\documentclass[a4paper, 11pt]{article}

\usepackage{jheppub}
\usepackage{abb}

\usepackage{amsmath, amsfonts}
\usepackage[normalem]{ulem}
\usepackage{bm}
\usepackage{physics}

\usepackage{soul}

\usepackage{comment}
\usepackage{color}

\usepackage{tikz}
\usetikzlibrary{decorations.markings}
\usetikzlibrary{shapes.misc}
\usetikzlibrary{arrows}
\definecolor{pyblue}{RGB}{31, 119, 180}
\definecolor{pyorange}{RGB}{255, 127, 14}
\definecolor{pygreen}{RGB}{44, 160, 44}
\definecolor{pyred}{RGB}{214, 39, 40}

\usepackage{tikz-cd}
\usepackage{circuitikz}

\definecolor{light-gray}{gray}{0.95}
\definecolor{lightgray}{gray}{0.9}

\newcommand{\Mp}{M_\mathrm{Pl}}
\renewcommand{\O}{\mathcal{O}}
\newcommand{\B}{\mathcal{B}}
\newcommand{\U}{\mathcal{U}}
\newcommand{\E}{\mathcal{E}}

\newcommand{\EOM}{\mathcal{O}^{\rm EoM}}

\makeatletter
\@addtoreset{equation}{section}

\makeatother

\allowdisplaybreaks[4]

\usepackage{cancel}

\usepackage{bbold}

\usepackage{empheq}
\usepackage{framed}
\usepackage[most]{tcolorbox}
\usepackage{xcolor}
\colorlet{shadecolor}{orange!15}
\usepackage{braket}

\usepackage{caption}
\usepackage{subcaption}

\usepackage{tikz}
\usetikzlibrary{intersections, calc, arrows.meta}
\usetikzlibrary{patterns}

\usepackage{array}
\usepackage{longtable}

\newcommand{\beq}{\begin{equation}}
\newcommand{\eeq}{\end{equation}}
\newcommand{\bal}{\begin{aligned}}
\newcommand{\eal}{\end{aligned}}
\newcommand{\bea}{\begin{eqnarray}}
\newcommand{\eea}{\end{eqnarray}}
\newcommand{\Lag}{{\mathcal{L}}}
\newcommand{\N}{{\mathcal{N}}}

\newcommand{\z}{{\zeta}}
\newcommand{\tz}{{\tilde{\zeta}}}
\newcommand{\tG}{{\tilde{G}}}
\newcommand{\p}{{p_\zeta}}
\newcommand{\tp}{{\tilde{p}_\zeta}}

\newcommand{\ttp}{{\tilde{\tilde{p}}_\zeta}}\newcommand{\ttz}{{\tilde{\tilde{\zeta}}}}

\newcommand{\zi}{\zeta_I}

\usepackage{adjustbox}
\usepackage{tikz-feynman}
\definecolor{rossocorsa}{rgb}{0.83, 0.0, 0.0}

\definecolor{gbcolor2}{rgb}{.9,.2,.6}
\definecolor{gbcolor3}{rgb}{.3,.2,.6}

\definecolor{verdechiaro}{rgb}{0.6,1,0.6}
\definecolor{giallochiaro}{rgb}{1,1,0.6}
\definecolor{bluscuro}{rgb}{0.15, 0.2, 0.9}
\definecolor{verdes}{rgb}{0.1, 0.5, 0.1}
\definecolor{tangerineyellow}{rgb}{1.0, 0.8, 0.0}
\definecolor{smokyblack}{rgb}{0.06, 0.05, 0.03}

\definecolor{americanrose}{rgb}{1.0, 0.01, 0.24}
\definecolor{cobalt}{rgb}{0.0, 0.28, 0.67}
\definecolor{brandeisblue}{rgb}{0.0, 0.44, 1.0}
\definecolor{mycolor}{rgb}{0.0, 0.0, 0.5}
\definecolor{oxfordblue}{rgb}{0.0, 0.13, 0.28}
\definecolor{azure}{rgb}{0.0, 0.5, 1.0}
\definecolor{turquoiseblue}{rgb}{0.0, 1.0, 0.94}

\definecolor{venetianred}{rgb}{0.78, 0.03, 0.08}
\newtcolorbox{mynamedbox1}[1]{colback=venetianred!5!white,colframe=venetianred!80!black,title=#1}
\usepackage{amssymb}
\usepackage{pifont}

\usepackage[symbol]{footmisc}

\newcommand{\ctext}[1]{\raise0.2ex\hbox{\textcircled{\scriptsize{#1}}}}

\newcommand{\highlight}[1]{%
  \colorbox{gray!50}{$\displaystyle#1$}}

\def\Mp{M_{{\rm Pl}}}

\def\k{\boldsymbol{k}}

\def\b{b}
\def\h{c}

\title{No time to derive: unraveling total time derivatives in in-in perturbation theory}

\author[a]{Matteo Braglia}
\author[b]{, Lucas Pinol}

\affiliation[a]{Center for Cosmology and Particle Physics, New York University, 726 Broadway, New York, NY 10003, USA}

\affiliation[b]{ Laboratoire de Physique de l’École Normale Supérieure, ENS, CNRS, Université PSL,\\ Sorbonne Université, Université Paris Cité, F-75005, Paris, France}

\emailAdd{mb9289@nyu.edu}
\emailAdd{lucas.pinol@phys.ens.fr}

\abstract{
The in-in formalism provides a way to systematically organize the calculation of primordial correlation functions.
Although its theoretical foundations are now firmly settled, the treatment of total time derivative interactions, incorrectly trivialized as ``boundary terms'', has been the subject of intense discussions and conceptual mistakes.
In this work, we demystify the use of total time derivatives---as well as terms proportional to the linear equations of motion---and show that they can lead to artificially large contributions cancelling at different orders of the in-in operator formalism.
We discuss the treatment of total time derivative interactions in the Lagrangian path integral formulation of the in-in perturbation theory, and we showcase the importance of interaction terms proportional to linear equations of motion.
We then provide a new route to the calculation of primordial correlation functions, which avoids the generation of total time derivatives, by working directly at the level of the full Hamiltonian in terms of phase-space variables.
Instead of integrating by parts, we perform canonical transformations to simplify interactions. 
We explain how to retrieve correlation functions of the initial phase-space variables from the knowledge of the ones after canonical transformations.
As an important first application, 
we find the explicit sizes of
Hamiltonian
cubic interactions in single-field inflation with canonical kinetic terms and for any background evolution, straight in terms of the primordial curvature perturbation and its canonical conjugate momentum, as well as the corresponding ones in the tensor sector, and the ones mixing scalars and tensors.
We also briefly comment on quartic interactions.
Our results are important for performing complete calculations of exchange diagrams in inflation, such as the (scalar and tensor) exchange trispectrum and the one-loop power spectrum.
Being already written in a form amenable to characterize quantum properties of primordial fluctuations, they also promise to shed light on the non-linear dynamics of quantum states during inflation.
}

\begin{document}

\maketitle

\section{Introduction}

\noindent

Originally designed to address the shortcomings of the Hot Big Bang scenario, the most fascinating aspect of the theory of cosmic inflation is its intriguing connection to the formation of the Large-Scale Structures that we now observe in our universe.
The exponential expansion stretches the physical wavelength of zero-point quantum fluctuations outside the Hubble radius, often dubbed ``horizon'' with a slight abuse of notation.
After inflation, the Hubble radius starts growing significantly and the large-scale fluctuations originating from inflation re-enter the horizon.
They are eventually  turned into perturbations to the matter and radiation fluids permeating the Universe.
Inflation thus inherently predicts the existence of primordial seeds for structure formation and provides a natural explanation for the observed anisotropies in the Cosmic Microwave Background (CMB).
Conversely, the fact that the distribution of matter and radiation that we routinely observe trace the earliest phase of the universe provides one with the tantalizing opportunity to test theories of high-energy physics likely to be at play during inflation.

Given the observational data that we currently have at hand, cosmic inflation is certainly the best theory to describe the earliest moments of our universe.
Yet, the precise mechanism responsible for it is yet to be confirmed.
The simplest model, canonical single-field slow-roll inflation, has successfully predicted the nearly scale-invariant spectrum of primordial scalar fluctuations, even before they were observed in the CMB.
Impressively, sixty years after the discovery of the CMB, inflation has withstood all observational tests, even though many specific realizations are now ruled out or in tension with latest data~\cite{Planck:2018jri}.
However, a good model does not make for a good theory, and canonical single-field slow-roll inflation arguably suffers from a lack of theoretical consistency.
Chief amongst its weaknesses is the required, though unnatural, flatness of the scalar potential, best known as the ``eta problem'' of inflation (see Ref.~\cite{McAllister:2007bg} for a review) and similar to the hierarchy problem in the Standard Model of particle physics.
Additionally, candidate theories for high-energy physics generically predict the presence of not only a scalar field, but several active degrees of freedom with various kinds of masses, spins and interactions~\cite{Baumann:2014nda}.

Given the stake---not less than probing new fundamental physics with cosmological observations---it appears crucial to pinpoint observable predictions that enable to decisively discriminate between inflationary scenarios.
Primordial Non-Gaussianities (PNGs) carry the hope to disentangle single-field from multifield scenarios, and hereby to advance our understanding of the laws of gravity and particle physics in a new regime.
Indeed, after first estimations of the smallness of PNGs in canonical single-field slow-roll inflation~\cite{Gangui:1993tt,Acquaviva:2002ud}, Maldacena computed the precise size and shape of the primordial three-point function, the bispectrum, definitively concluding that $f_{\rm NL}\sim\mathcal{O}(0.01)$ in this vanilla scenario~\cite{Maldacena:2002vr}.
This value, often dubbed as the ``gravitational floor'' as PNGs are then the result of the minimal, ever-present, gravitational interactions is---fortunately for inflation as a theory---well within the current upper bounds from the Planck satellite~\cite{Planck:2019kim} but it is also, unfortunately, well below the most optimistic projected sensitivities of cosmological surveys in the coming decade: $\sigma(f_{\rm NL})\sim\mathcal{O}(1)$~\cite{Meerburg:2019qqi,Achucarro:2022qrl}. On the other hand, inflationary scenarios breaking either of the assumptions of canonical single-field slow-roll may predict a substantial amount of PNGs.
More in detail, both the size and the shape of PNGs vary from model to model: single-field with non-canonical kinetic terms, with non-minimal coupling to gravity or higher-order derivatives, and all kinds of multifield models (see Ref.~\cite{Chen:2010xka} for a review).
At smaller scales, other cosmological observations could help pinpointing the correct mechanism behind inflation, or at least complete the picture to its latest stages.
In particular, gravitational-wave background phenomenology can be intimately related to non-linear interactions that are the subject of this work, such as gravitational waves non-linearly produced during inflation~\cite{Peng:2021zon,Cai:2021yvq,Inomata:2021zel,Fumagalli:2021mpc}, scalar-induced gravitational waves from non-Gaussian primordial fluctuations~\cite{Unal:2018yaa,Atal:2021jyo,Adshead:2021hnm,Garcia-Saenz:2022tzu,Garcia-Saenz:2023zue} or anisotropies originating from tensor and mixed scalar-tensor PNGs (see Ref.~\cite{Dimastrogiovanni:2022afr} for their incorporation within the in-in formalism).

Given the importance of both these predictions and their interpretations, it may seem surprising that ambiguities could persist in the formalism used for the calculation of primordial correlation functions, even in the simplest scenarios.
The now standard technique for computing PNGs is the quantum in-in formalism, first introduced in the context of inflation by Maldacena~\cite{Maldacena:2002vr}, later formalized by Weinberg~\cite{Weinberg:2005vy} and more generally known as the Schwinger-Keldysh formalism for out-of-equilibrium processes (see Ref.~\cite{Kamenev_2011} for a book on the topic, not related to cosmology).
However, even decades after its introduction, we found ourselves with no complete treatment---in the sense that it would be valid at any order of the in-in perturbation theory---
when the interaction Hamiltonian contains total time derivatives, as it is the case in single-field slow-roll inflation.
Total time derivative interactions arise as the byproduct of temporal integration by parts (other total time derivatives may appear from the Noether procedure to derive soft theorems, see Ref.~\cite{Hui:2022dnm}). 
In his seminal paper, Maldacena performed a large number of such integrations by parts to simplify the cubic order Lagrangian in the comoving gauge, where the dynamical scalar degree of freedom is identified with the curvature perturbation $\zeta$, in order to elucidate the true size of non-linear interactions.
As a result, the cubic interaction Hamiltonian (simply given by minus the cubic Lagrangian at this order) is composed by bulk slow-roll suppressed interactions, as well as total time derivative interactions, and terms proportional to the linear Equations of Motion (EoM) verified by free fields on shell.
Maldacena ignored total time derivatives and proposed a field redefinition of $\zeta\mapsto\zeta_n$ to remove the interactions proportional to the linear EoM. Through this field redefinition, he was able to compute correlation functions of $\zeta$ by relating them to those of $\zeta_n$, technically easier to compute and manifestly slow-roll suppressed.
While ultimately yielding the correct result for canonical single-field inflation, this procedure was later revisited in~\cite{Arroja:2011yj,Burrage:2011hd,Rigopoulos:2011eq,Renaux-Petel:2011zgy}.
The main argument in these works can be summarized as follows.
Since the upper integration limit of the in-in integrals lives at the finite-time boundary of the inflationary space-time, where interactions may not be negligible, overlooking total time derivatives may not always be justified.
On the contrary, the interaction Hamiltonian is evaluated on interaction picture fields and momenta which identically verify the equations of motion dictated by the free Hamiltonian, so terms proportional to the EoM can be safely evacuated.
The main reason why the procedure outlined by Maldacena gives the correct answer is that total time derivatives contributing to the bispectrum always come in pairs with terms proportional to the equations of motion, so by cancelling the latter, care was also taken about the former.
These conclusions in single-field inflation were later extended to multiple scalar fields~\cite{Garcia-Saenz:2019njm,Pinol:2020kvw} and tensor~\cite{Ning:2023ybc} fluctuations.
However, all these works specifically focus on the computation of the primordial bispectrum with cubic order interactions and at the one-vertex order of the in-in perturbation theory (see Ref.~\cite{Craig:2024qgy} for a notable exception where an exchange trispectrum from such cubic interactions is computed).

This paper aims at definitively addressing lingering ambiguities associated with total time derivative interactions and at systematizing their treatment at any order of perturbation theory, both for tree and loop processes.
After providing general formulae for the calculation of in-in correlation functions of theories with any number of fields, and which include total time derivative interactions in the interaction Hamiltonian, we show how they lead to a tedious perturbation theory, with cancellations between different orders.
Prompt by the need of a more straightforward method, we propose another route which avoids the use of integrations by parts altogether.
Our method  relies on the use of canonical transformations to simplify interactions in the full Hamiltonian, rather than the interaction one, which is therefore expressed in terms of canonical phase-space variables, instead of interaction picture fields and their time derivatives.
Applying our formalism to  canonical single-field inflation, we show how it offers clarity of interpretation and a computational simplification compared to the in-in method with total time derivative interactions.
The paper is structured as follows.

\begin{figure}
\centering
\resizebox{1\textwidth}{!}{%
\begin{circuitikz}
\tikzstyle{every node}=[font=\tiny]
\node [font=\small] at (8.25,12.25) {
$\mathcal{L}(\psi^a,\dot{\psi}^a_I)$};
\draw [->, >=Stealth] (6.75,12) -- (5,10.75)node[pos=0.5, fill=white]{Legendre transform};
\draw [->, >=Stealth] (9.5,12) -- (11,10.75)node[pos=0.5, fill=white]{integration by parts (Sec.~\ref{sec:ibp})};
\node [font=\small] at (4.5,10.25) {$\mathcal{H}(\psi^a,p_{\psi,a})$};
\node [font=\small] at (11.5,10.25) {$\mathcal{L}(\psi^a,\dot{\psi}^a_I)=-\U-\dd \B/\dd t - \E$};
\draw [->, >=Stealth] (4.5,9.75) -- (4.5,8.25)node[pos=0.5, fill=white]{canonical transformations (Secs.~\ref{sec:ct_generalities}--\ref{subsec: canonical transformations toy models})};
\draw [->, >=Stealth] (11.5,9.75) -- (11.5,8.25)node[pos=0.5, fill=white]{Legendre transform (Sec.~\ref{sec:interaction_picture})};
\node [font=\small] at (4.75,7.75) {$\tilde{\mathcal{H}}(\tilde{\psi}^a,\tilde{p}_{\psi,a})$};
\node [font=\small] at (11.5,7.75) {$\mathcal{H}(\psi^a,p_{\psi,a})$};
\draw [->, >=Stealth] (4.5,7.25) -- (4.5,5.75)node[pos=0.5, fill=white]{in-in perturbation theory (Sec.~\ref{subsec: canonical transformations toy models})};
\draw [->, >=Stealth] (11.5,7.25) -- (11.5,5.75)node[pos=0.5, fill=white]{interaction picture (Sec.~\ref{sec:interaction_picture})};
\node [font=\small] at (4.75,5.25) {$\braket{\mathcal{O}(\tilde{\psi}^a,\tilde{p}_{\psi,a})}$};
\node [font=\small] at (8.,3.25) {$\braket{\mathcal{O}(\psi^a)}$};
\draw [->, >=Stealth] (4.5,4.75) -- (6.75,3.5)node[pos=0.5, fill=white]{( diagrammatic rules (App.~\ref{app: diagrammatic rules}) )};
\node [font=\small] at (11.5,5.25) {$H^I= U + \dd B/\dd t$};
\draw [->, >=Stealth] (11.5,4.75) -- (9.5,3.5)node[pos=0.5, fill=white]{in-in perturbation theory (Secs.~\ref{sec:operator-formalism}--\ref{subsec: toy models})};
\draw [short] (9.25,10.25) -- (8.,10.25);
\draw [->, >=Stealth] (8.,10.25) -- (8.,3.75)node[pos=0.465, fill=white]{path integral formalism (Sec.~\ref{subsec: SK formalism})};
\end{circuitikz}
}%
\caption{Flowchart of the three possible ways considered in this work to compute the correlation functions $\braket{\mathcal{O}(\psi^a)}$ of fields $\psi^a$ dictated by an initial Lagrangian $\mathcal{L}(\psi^a,\dot{\psi}^a_I)$. We consider a procedure where one needs to manipulate interactions in order to break degeneracies amongst operators and render explicit their sizes, either with integration by parts in the Lagrangian (Sec.~\ref{sec:in-in}) or with canonical transformations in the Hamiltonian (Sec.~\ref{sec: canonical transformations}) obtained via Legendre transform of the initial Lagrangian. The different steps leading to correlators of $\psi^a$ are shown, with the corresponding sections of this paper to help the reader navigating through it.
For two toy models of cubic and quartic interactions, we prove explicitly that all relevant correlators exactly agree, thus showing agreement of the three distinct procedures.
Finally, Sec.~\ref{sec: single field inflation} consists in an application to single-field inflation with canonical kinetic terms of our new procedure with canonical transformations in the Hamiltonian.
}
\label{fig:flowchart}
\end{figure}
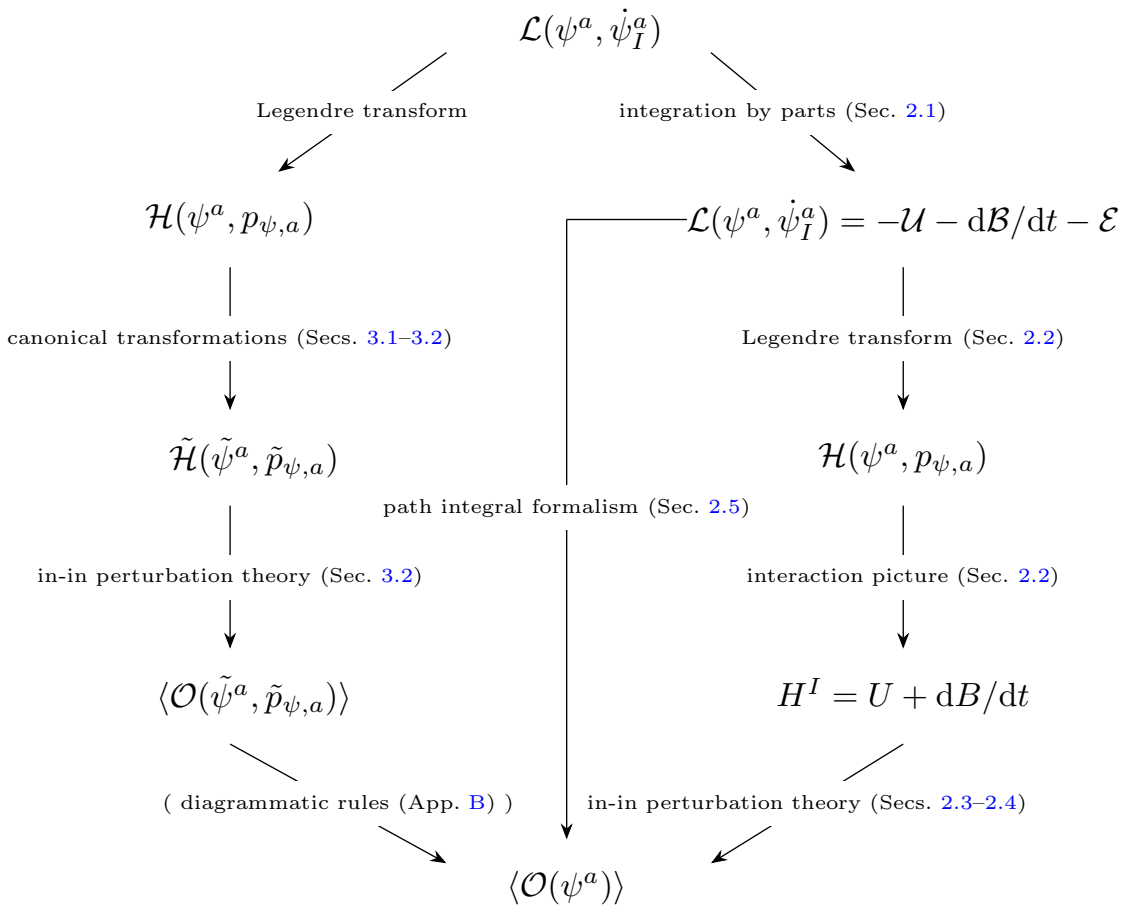

\paragraph{Structure of the paper.}

In Sec.~\ref{sec:in-in}, we solve the question of total time derivative interactions within the widely used operator in-in formalism, presenting generic formulae valid at any order in perturbation theory.
We also illustrate with two toy models how various superficially large contributions to correlation functions cancel each other at different vertex orders, making the treatment of total time derivative interactions non-trivial.
We show the ``conservation of trouble'' between the operator in-in formalism and the Lagrangian path integral one, where in the latter care must be taken concerning non-vanishing contributions coming from interactions proportional to the linear equations of motion.

In Sec.~\ref{sec: canonical transformations}, we introduce a new method, utilizing canonical transformations and the full Hamiltonian. 
After a brief pedagogical review of canonical transformations in classical mechanics, we apply this technique to calculate non-trivial correlation functions in the previously mentioned toy models, thus showcasing straightforward calculations without subtle cancellations and proving equivalence with the approach utilizing integrations by parts. Within this new method, the issue of total time derivative interactions is completely bypassed, as no such interactions appear in the Hamiltonian anymore.
We explain how to compute correlation functions of the initial phase-space variables in terms of the ones after the canonical transformation.

In Sec.~\ref{sec: single field inflation}, we apply those techniques to single-field inflation with canonical kinetic terms, focusing first on scalar fluctuations, but directly written in terms of the primordial curvature perturbation and its canonical conjugate momentum.
These variables are defined before expanding the Hamiltonian to a given order in perturbations, and we show how to perform a first non-linear canonical transformation that will simplify the calculation of interactions to all orders in perturbation theory.
We then dig into the perturbation theory, first defining the quadratic theory and solving the linear constraints in terms of phase-space variables.
After this, we go to cubic order and perform a series of canonical transformations to simplify the interactions, thus reducing the number of cubic operators and rendering manifest their sizes.
This is the equivalent of Maldacena's calculation, but at the level of the Hamiltonian, in terms of phase-space variables, and without introducing total time derivatives nor using the linear equations of motion.
Our final result is valid for any FLRW evolution, it does not assume a slow-roll dynamics and it needs not be expressed in the interaction picture.
It therefore extends the regime of application of already-known cubic interactions in single-field inflation to other contexts requiring a phase-space description, such as the non-linear evolution of the quantum properties of cosmological perturbations during inflation. 
We then turn to the equivalent calculations for non-linear tensor and mixed scalar-tensor perturbations, and we finally close this section with a few remarks regarding the calculation of the next, quartic order.
In particular, we comment on a non-trivial interplay of tensor and scalar fluctuations that makes it necessary to carefully take into account both of them to correctly derive quartic scalar interactions.
As a first application in this direction, we also show how our formalism allows to estimate the dominant quartic interactions in a regime of inflation with a large $\eta$.

In Sec.~\ref{sec: conclusions}, we close the paper with a summary of the main results and directions for future work, opened by our findings.

Our paper includes an Appendix, which offers additional material that complements the main text.
In App.~\ref{app:quantum_anomalies}, we briefly comment on
extensions of canonical transformations from classical to quantum field theories. In App.~\ref{app: diagrammatic rules}, we provide diagrammatic rules
 to compute correlation functions of the initial phase-space variables in terms of the ones after the canonical transformation. Finally,  in App.~\ref{app:tensors_L_and_H} we collect all Lagrangian interactions relevant to tensor and mixed scalar-tensor non-linearities.

\paragraph{Notations.}
Before starting, we set some notations for convenience. We denote derivatives with respect to the cosmic time $t$ with an over-dot $\dot{}\,$, and derivatives with respect to the conformal time $\dd t= a\,\dd \tau$, where $a$ is the scale factor.
We define the Hubble scale as $H=\dot{a}/a$, and sequentially define from it the first and higher order slow-roll parameter as $\epsilon_1\equiv-\dot{H}/H^2$  and $\epsilon_{i+1}=H^{-1} \dd\ln \epsilon_{i}/\dd t$ with $i=1,\,\cdots$. In this paper we will mainly be interested in $\epsilon\equiv\epsilon_1$, $\eta\equiv\epsilon_2$ and $\eta_2\equiv\epsilon_3$.  We use Latin indices $i, j,\ldots $ to denote spatial coordinates, unless otherwise stated. $\Mp$ is the Planck mass. Finally, we use italic fonts, respectively calligraphic ones, to indicate the Hamiltonian, Lagrangian,
respectively their densities.
For example, the Hamiltonian is related to the Hamiltonian density as $H=\int\,\dd^3x\,\mathcal{H}$.

\section{In-in formalism with total time derivatives}
\label{sec:in-in}
In this section, we explain how to consistently include the effect of total time derivatives in the interaction Hamiltonian, using  in-in perturbation theory.
We start by recalling the origin of those terms in the inflationary context, and remind how a sub-class of them are always generated together with terms proportional to the linear equations of motion.
We then present how the usual in-in integrals are affected by these terms, carefully taking into account subtle effects from the (anti)-time ordering operators and the non-commutativity of quantum operators.
Finally, we define two toy-model Lagrangians with different kinds of total time derivatives interactions, and terms proportional to linear equations of motion, and show how subtle cancellations among various terms affect their correlation functions.
We will use those toy models to showcase the technical difficulties with dealing with total time derivative interactions, and argue for the utility of a different approach, based on canonical transformations in phase space and exposed in the following section.

\subsection{The rise of total time derivatives in inflationary Lagrangians}
\label{sec:ibp}
After preliminary estimations of the (small) size of primordial non-Gaussianities in single-field, slow-roll inflation~\cite{Gangui:1993tt,Acquaviva:2002ud}, Maldacena provided the first complete calculation, including exact coefficients and the shape dependence of the primordial bispectrum~\cite{Maldacena:2002vr}. The calculation is conducted both in the flat gauge, where the only propagating scalar degree of freedom coincides with the fluctuation in the scalar field ($\delta \phi^\mathrm{flat}$), and in the comoving gauge, where it aligns with the curvature perturbation in the spacetime metric, denoted as $\zeta$ in this paper.
Each of the two gauges has its own advantages.
In the flat gauge, both the extrinsic curvature and the spatial curvature of three-dimensional hypersurfaces are trivial.
It is therefore easy to expand the action up to cubic order in fluctuations using the ADM formalism~\cite{Arnowitt:1962hi, Salopek:1990jq}, as relevant for the calculation of the primordial bispectrum, the three-point function.
Moreover, all cubic interactions are found to be proportional to at least two powers of the slow-roll parameters, showing explicitly the smallness of their sizes.
However, the variable used in the calculation, $\delta \phi^\mathrm{flat}$, is not conserved on super-horizon scales.
Actually, it is non-linearly related to the curvature perturbation $\zeta$ which is the conserved quantity on these scales~\cite{Lyth:2004gb}, and whose statistics need to be known to faithfully predict initial conditions in the radiation and matter eras through reheating.
It is therefore desirable to describe non-linearities directly in terms of the observable, adiabatic, fluctuation $\zeta$.
The calculation in the comoving gauge, however, is more involved as a direct expansion of the Lagrangian action does not yield manifestly slow-roll suppressed cubic interactions.
Moreover, several cubic operators are degenerate and their contributions to the three-point functions cancel at leading-order in the slow-roll expansion.
Maldacena proposed to perform a significant number of integrations by parts to explicitly cancel superficially large interactions and reduce the number of cubic operators.
For pedagogy, and as a warm-up for subsequent calculations, let us show two of such simplifications.

The Lagrangian of single-field inflation in the comoving gauge, after solving for the linear constraints and expanded at cubic order, contains the following contributions:
\begin{align}
\label{eq: 2 contrib to L3}
    \mathcal{L}^{(3)}(\zeta, \dot{\zeta}) &= \mathcal{L}^{(3)}_A(\zeta,\dot{\zeta}) +  \mathcal{L}^{(3)}_B(\zeta,\dot{\zeta}) + \ldots \,\, \,\, \,\, \text{with} \\
    \mathcal{L}^{(3)}_A(\zeta,\dot{\zeta}) &= a^3 \epsilon \Mp^2  \left(3\zeta - \frac{\dot{\zeta}}{H}\right)\dot{\zeta}^2  \,\, \,\, \,\, \,\,\,\, \,\, \,\, \,\, \,\, \,\,  \text{and}  \,\,\\ \mathcal{L}^{(3)}_B(\zeta,\dot{\zeta}) &= a \epsilon \Mp^2 \left(\frac{\dot{\zeta}}{H} - \zeta \right) \left(\partial\zeta\right)^2 \,. \nonumber
\end{align}
Both contributions can be simplified by the use of integrations by parts, but the first manipulation requires invoking the linear equation of motion for $\zeta$:
\begin{equation}
    \mathcal{E}_\zeta(\zeta,\dot{\zeta}) \equiv - \frac{1}{a^3} \frac{\delta S^{(2)}}{\delta \zeta} = 2 \Mp^2 \left[\frac{1}{a^3} \frac{\dd }{\dd t}\left(a^3 \epsilon \dot{\zeta} \right) - \frac{\epsilon}{a^2} \partial^2 \zeta \right] \,,
\end{equation}
which corresponds to the Euler-Lagrange equation computed from the quadratic action $S^{(2)} =  \int \dd^4 x \, \mathcal{L}^{(2)}(\zeta,\dot{\zeta})$ with
\begin{equation}
\label{eq: L2}
    \mathcal{L}^{(2)}(\zeta, \dot{\zeta}) =  a^3 \epsilon \Mp^2  \left[ \dot{\zeta}^2 - \frac{\left(\partial \zeta\right)^2}{a^2} \right] \,.
\end{equation}
Equipped with these notations, we can rewrite the ``$A$'' contribution  as
\begin{equation}
\label{eq:L_A_ibp}
    \mathcal{L}^{(3)}_A(\zeta,\dot{\zeta}) =a^3\Mp^2 \epsilon (\epsilon-\eta)\dot{\zeta}^2\zeta -\frac{\dd}{\dd t}\left[a^3 \Mp^2 \frac{\epsilon}{H}\dot{\zeta}^2 \zeta \right] +\frac{1}{H}\zeta \dot{\zeta} \mathcal{E}_{\zeta}(\zeta,\dot{\zeta}) + 2 a \frac{\epsilon}{H} \Mp^2\zeta \dot{\zeta}  \partial^2 \zeta \,.
\end{equation}
A few remarks are in order.
First, $\dot{\zeta}^3$ interactions have been removed and the explicit size of the $\dot{\zeta}^2 \zeta$ interaction has been reduced from order $\epsilon$ to order $\epsilon(\epsilon-\eta)$
, and is therefore further suppressed by slow-roll parameters.
Second, we see the appearance of a total time derivative term, as well as a term proportional to the linear equations of motion.
Actually, it is a general feature that total time derivatives of functions of fields and their first time derivatives---like what we have here---always appear along with a term proportional to $\mathcal{E}_{\zeta}(\zeta,\dot{\zeta})$.
The reason is simply that terms proportional to the second time derivatives of field must cancel out as they are absent from the initial Lagrangian.
This fact also ensures that no subtleties arise at the level of the variational principle and that no new \textit{ad hoc} degrees of freedom are added in the theory.
Therefore, they always come in pairs, from an integration by parts as in Eq.~\eqref{eq:L_A_ibp}. 
Third and finally, we seem to have generated a new interaction with the operator $\zeta \dot{\zeta} \partial^2 \zeta$, but this term can actually be combined with the ``$B$'' contribution:
\begin{equation}
\label{eq: combine with LB}
    2 a \frac{\epsilon}{H} \Mp^2 \zeta \dot{\zeta}  \partial^2 \zeta + \mathcal{L}^{(3)}_B(\zeta,\dot{\zeta}) = a \Mp^2  \epsilon(\epsilon+\eta) \zeta (\partial \zeta)^2 - \frac{\dd}{\dd t}\left[\frac{a\epsilon \Mp^2}{H} \zeta \left(\partial\zeta\right)^2 \right] \,,
\end{equation}
where we have not kept total \textit{spatial} derivatives.
Indeed, there is no physically well-defined notion of spatial boundary in cosmology, so we disregard them, which in practice amounts to neglecting interactions evaluated at infinite distances.
This is to be contrasted with total time derivatives: in the in-in formalism, the lower bound of the time integrals is still taken at (minus) infinity, but the upper one is living in the bulk of the inflationary spacetime or at its finite-time boundary, where interactions may not be negligible, as we will see soon.
At this stage, we can note that we have reduced the size of the $\zeta(\partial \zeta)^2$ interaction by one more order of slow-roll parameters, and we have generated another total time derivative.
However, no term proportional to the linear equation of motion appears from the manipulation of $\mathcal{L}^{(3)}_B$  because the total time derivative acts on a product of fields only, and not of their time derivatives.

After all simplifications, consisting in many integrations by parts and uses of linear equations of motion as in the two examples we showcased, one finds the total cubic Lagrangian~\cite{Maldacena:2002vr,Collins:2011mz,Burrage:2011hd}:
\begin{equation}
\label{eq:cubic action scalars}
    \mathcal{L}^{(3)}(\zeta,\dot{\zeta}) =  - \left( \U^{(3)}(\zeta,\dot{\zeta}) + \frac{\dd \B^{(3)}}{\dd t}(\zeta,\dot{\zeta}) + \E^{(3)}(\zeta,\dot{\zeta}) \right),
    \end{equation}
    with
    \begin{align}
    \U^{(3)}(\zeta,\dot{\zeta})
     = - \Mp^2 \,a^3
    \Biggl[&
    \epsilon (\epsilon-\eta) \dot{\zeta}^2\zeta+ \epsilon (\epsilon+\eta)\zeta \frac{\left(\partial \zeta \right)^2}{a^2}  + \epsilon^2 \left(\frac{\epsilon}{2}-2\right) \dot{\zeta} \partial_i \zeta \partial_i \partial^{-2} \dot{\zeta}
    \nonumber\\&+ \frac{\epsilon^3}{4} \partial^2 \zeta \left(\partial_i \partial^{-2} \dot{\zeta}\right)^2  \Biggr]  \,, \\
    \frac{\dd \B^{(3)}}{\dd t}(\zeta,\dot{\zeta}) =\Mp^2\frac{\dd}{\dd t}\bigg\{&9H  a^3 \zeta^3-\frac{a }{H}(1-\epsilon)\zeta(\partial\zeta)^2 +\frac{1 }{4aH^3} (\partial \zeta)^2 \partial^2 \zeta+\frac{a^3\epsilon}{  H  } \zeta \dot{\zeta}^2  \\
  &
+\frac{\epsilon^2 a^3}{2  H }\zeta\left( \partial^{-2} \dot{\zeta}_{, ij} \partial^{-2} \dot{\zeta}_{,ij}-\dot{\zeta}^2\right) -\frac{\epsilon a}{2 H^2}\zeta  \left(\zeta_{,ij} \partial^{-2} \dot{\zeta}_{,ij}-\partial^2\zeta \,\dot{\zeta}\right)\bigg\}\, \nonumber\\ 
    \E^{(3)}(\zeta,\dot{\zeta}) =\frac{a^3}{H}\,{\cal E}_{\zeta}\Biggl\{&-\dot{\zeta}\zeta+\frac{1}{4a^2H}\Biggl[ (\partial \zeta)^2-\partial^{-2}\partial_i\partial_j(\partial_i\zeta\partial_j\zeta) \nonumber\\&-2 a^2 \epsilon H\left(\partial \zeta\partial \chi-\partial^{-2}\partial_i\partial_j(\partial_i\zeta\partial_j \dot{\zeta})\right) \Biggr]\Biggr\}\nonumber \,,
\end{align}
where we defined $\partial^{-2}$ as the inverse Laplacian operator: $\partial^{-2}\partial^{2}=\partial^{2}\partial^{-2}=1$.
For later convenience, we have separated the final Lagrangian into usual interaction terms inside $\U$, total time derivatives inside $\dd \B/ \dd t$, and terms proportional to the linear equation of motion inside $\E$.
This result has then been extended to two-field inflation in Ref.~\cite{Garcia-Saenz:2019njm}, and then to any number of fields in Ref.~\cite{Pinol:2020kvw}, where all steps of the simplifications are shown explicitly.
Primordial tensor perturbations have also been added in single-field inflation up to cubic order, including their self-interactions and also their interactions with the scalar sector, see Ref.~\cite{Ning:2023ybc}.

\subsection{The interaction picture}
\label{sec:interaction_picture}
In perturbative calculations as described by the in-in formalism, one needs to define an interaction picture by separating the \textit{Hamiltonian} into a free part, defining the time evolution of the interaction picture fields and their propagators, and an interacting part, defining the vertices of the theory.
From $\mathcal{L}^{(2)}(\zeta,\dot{\zeta})$ in Eq.~\eqref{eq: L2}, we find the linear canonical conjugate momentum to be
\begin{equation}
    p_\zeta^\mathrm{lin} \equiv \frac{\partial \mathcal{L}^{(2)}(\zeta,\dot{\zeta})}{\partial \dot{\zeta}} = 2 \epsilon a^3 \Mp^2 \dot{\zeta} \,,
\end{equation}
from which we deduce, after performing a Legendre transformation, that the quadratic Hamiltonian density, which we \textit{define} to be the free Hamiltonian $\mathcal{H}_\mathrm{free}$, reads:
\begin{equation}
    \mathcal{H}_\mathrm{free}(\zeta, p_\zeta) \equiv \mathcal{H}^{(2)}(\zeta, p_\zeta) = \frac{p_\zeta^2}{4a^3 \epsilon \Mp^2} + a \epsilon \Mp^2 \left(\partial\zeta\right)^2
\end{equation}
The interaction picture fields and momenta therefore verify the linear equations of motion
\begin{align}
    \dot{\zeta}_I &= \frac{p_\zeta^I}{2\epsilon a^3 \Mp^2} \\
    \dot{p}_\zeta^I &=  2\epsilon a \Mp^2 \partial^2 \zeta_I
\end{align}
Combining those two equations, we find:
\begin{equation}
    \mathcal{E}_{\zeta}(\zi, \dot{\zeta}_I) = 0 \,,
\end{equation}
meaning that the interaction picture fields identically verify the second-order linear equations of motion dictated by $\mathcal{L}^{(2)}$, as expected.
Now that interaction picture fields are defined, vertices of the theory need to be written down.
Those are found from the interaction Hamiltonian  $\mathcal{H}_\mathrm{int}(\zeta, p_\zeta)= \mathcal{H}(\zeta, p_\zeta)- \mathcal{H}_\mathrm{free}(\zeta, p_\zeta)$, but expressed in terms of interaction picture fields and momenta, $\mathcal{H}_\mathrm{int}(\zi, p_\zeta^I) $.
Using the linear equation of motion relating $\dot{\zeta}^I$ to $p_\zeta^I$, one may express the interaction Hamiltonian in terms of fields and their time derivatives in the interaction picture.

In general, going from the Lagrangian to the full Hamiltonian $\mathcal{H}(\zeta, p_\zeta)$ requires defining the non-linear momenta and performing the Legendre transformation consistently order by order in fields and momenta.
We find the cubic order interaction Hamiltonian of single-field inflation, in the interaction picture, to read:

\begin{equation}
\label{eq: Hint3 SFSR}
    \mathcal{H}_\mathrm{int}^{(3)}\left(\zi, p_\zeta^I(\zi,\dot{\zeta}^I)\right) = \mathcal{U}^{(3)}(\zi, \dot{\zeta}^I) + \frac{\dd \mathcal{B}^{(3)}}{\dd t}(\zi, \dot{\zeta}^I)  \,.
\end{equation}
Note that an intermediate step of the calculation is the interaction Hamiltonian in terms of the general phase-space variables $(\zeta,p_\zeta)$.
In order to get it, it is crucial to keep all three contributions to the Lagrangian.
In particular, terms proportional to the equations of motion in $\mathcal{E}$ cancel second order derivatives of the fields in the subset of the $\dd \mathcal{B}/ \dd t$ terms where $\mathcal{B}$ does contain time derivatives of the fields.
Only in the final expression $\mathcal{H}_\mathrm{int}(\zeta, p_\zeta)$ did we express the Hamiltonian in terms of interaction picture fields, for which one can use $\E^{(3)}(\zi, \dot{\zeta}^I)\propto  \mathcal{E}_{\zeta}(\zi, \dot{\zeta}^I) =0$.
We will show more details of this procedure in a toy model with simpler interactions in Sec.~\ref{subsec: toy models}.
Eq.~\eqref{eq: Hint3 SFSR} confirms that $\mathcal{H}_\mathrm{int}^{(3)}\left(\zi, p_\zeta^I(\zi,\dot{\zeta}^I)\right) = - \mathcal{L}^{(3)}(\zi, \dot{\zeta}^I)$, 
as already well-known 
for Lagrangians without total time derivatives nor terms proportional to the equations of motion.
But a cubic Lagrangian may also generate quartic (and sometimes higher) order interactions in the interaction Hamiltonian, which is therefore not simply given by minus the Lagrangian in general ~\cite{Chen:2006dfn} (see also~\cite{Wang:2013zva} for a review).
Curiously, contributions of this kind, i.e. terms of order four and more in the interaction Hamiltonian and arising from cubic total time derivative interactions in the Lagrangian, have been overlooked in the literature on single-field slow-roll inflation with canonical kinetic terms.

In particular, let us look at a subset of these interactions: $\dd \B^{(3)} / \dd t \supset \dd ( 9 a^3 H  \Mp^2  \zeta^3)/\dd t$ (we will consider them again and in greater detail in the insert page 13).
It is easy to show that, through a non-linear correction to the momentum, this cubic Lagrangian interaction results in a quartic Hamiltonian interaction\footnote{In the Lagrangian path integral approach to the in-in formalism as presented in Ref.~\cite{Chen:2017ryl}, an equivalent statement holds.
Exchange diagrams with two Lagrangian cubic vertices corresponding to $\dd \B^{(3)} / \dd t $ contain a permutation for which the time derivatives of both vertices hit the internal propagator.
For this permutation, there exists a term $\propto \delta(\tau_1-\tau_2)$ where $\tau_{1,2}$ are the conformal times at which the two vertices are, leading to an effective quartic unique vertex with strength exactly equal to Eq.~\eqref{eq: large H4 interaction}.
However, other contributions will eventually cancel out this particular permutation, as we show in Sec.~\ref{subsec: SK formalism}. 
}
\begin{equation}
\label{eq: large H4 interaction}
    \mathcal{H}_\mathrm{int}^{(4)}(\zi, p_\zeta^I) \supset  \frac{729}{4\epsilon} a^3 H^2 \Mp^2 \zi^{4} \,, 
\end{equation}
whose size is inversely proportional to the slow-roll parameter $\epsilon$.
This term cannot be cancelled by other contributions to $\mathcal{H}_\mathrm{int}^{(4)}$, neither from $\mathcal{L}^{(3)}$ as one can check explicitly, nor from $\mathcal{L}^{(4)}$ that we have not written.
Indeed, it is possible to prove non-perturbatively that $\mathcal{L}(\zeta,\dot{\zeta})$ does not contain any non-derivative interactions like $\zeta^n$~\cite{Maldacena:2002vr}; such interactions in the Lagrangian would also violate the theorem of constancy of $\zeta$ on super-horizon scales~\cite{Lyth:2004gb,Langlois:2005qp,Senatore:2012ya}.
This is a rather troubling fact, that single-field inflation in slow roll predicts very large interactions like in Eq.~\eqref{eq: large H4 interaction}.
Actually, we will show that there are subtle cancellations: in-in diagrams including this large quartic interaction are exactly cancelled by a subset of the same diagrams where the quartic vertex is replaced by two cubic vertices $\B^{(3)}$ and a ``collapsed'' internal propagator.
To see that, we first need to develop a perturbative in-in formalism in the presence of total time derivatives.

\subsection{Total time derivatives in the interaction Hamiltonian}
\label{sec:operator-formalism}
Let us now take a step back and neither specify a particular field content nor the cosmological background.
We do not even specify particular kinds of non-linear interactions.
We simply consider fields $\psi^a$ labelled by $a, b, \ldots$
and their canonically conjugate momenta $p_{\psi,a}$.
We suppose the total Hamiltonian has already been split into a free and an interaction part, as $H=H_\mathrm{free}+H_\mathrm{int}$.
The interaction picture has been defined as the one in which fields and momenta verify the equations of motion dictated by $H_\mathrm{free}$ that we assume is quadratic, so there exists a linear relation $p_{\psi,a}^I(\psi^b_I,\dot{\psi}^b_I)$.
We also assume that the interaction Hamiltonian, once written in terms of interaction picture fields and momenta, reads
\begin{equation}
    H^I_\mathrm{int} \equiv H_\mathrm{int}\left(\psi^a_I,p_{\psi,a}^I(\psi^b_I,\dot{\psi}^b_I)\right) =  U(\psi^a_I,\dot{\psi}^a_I)  + \frac{\dd B  }{\dd t}(\psi^a_I,\dot{\psi}^a_I)  \,,
\end{equation}
where $U$ encodes the usual interactions and $\dd B/\dd t$ is a total time derivative.
Each of these two terms is generally a sum of different operators with various powers of the fields and their time (and spatial) derivatives.

In the operator approach of the in-in formalism, the vacuum expectation value of an operator $\mathcal{O}$ is written:
\begin{equation}
\label{eq: in-in general formula}
    \Braket{\hat{\O}}(t) = \Braket{0|\Bar{T}\left[\exp{i \int_{-\infty^+}^t \dd t^\prime \hat{H}_\mathrm{int}^I (t^\prime)}\right]\hat{\O}^I(t)T\left[\exp{-i \int_{-\infty^-}^t \dd t^\prime \hat{H}_\mathrm{int}^I (t^\prime)}\right]|0} \,.
\end{equation}
where the superscript in $-\infty^\pm$ denotes the $i \epsilon$-prescription needed to regularize ambiguous phases in the infinite past and project the vacuum of the full theory onto the one of the free theory: $\ket{\mathrm{in}} \rightarrow \ket{0}$.
$T$ and $\bar{T}$ are the time and anti-time ordering operators respectively.
An important comment is that Eq.~\eqref{eq: in-in general formula} is nothing but a useful, compact formula to encode perturbation theory at any order with the in-in formalism.
Rigorously though, the time-ordered exponential should really be understood as the re-summed series (in the following we drop the superscript $I$ in order to avoid cluttered notations)
\begin{equation}
\label{eq: time-ordered exponential explicitly}
 \lim_{N\rightarrow \infty} \sum_{n=0}^N (-i)^n \int_{-\infty^-}^t \dd t_1 \int_{-\infty^-}^{t_1} \dd t_2 \ldots \int_{-\infty^-}^{t_{n-1}} \dd t_n \hat{H}_\mathrm{int}(t_1)\hat{H}_\mathrm{int}(t_2)\ldots \hat{H}_\mathrm{int}(t_n) \,,
\end{equation}
and similarly for the anti-time ordered one.
Forgetting for the moment the $i \epsilon$-prescriptions, it will turn useful to use the so-called commutator form~\cite{Weinberg:2005vy}
of the in-in formalism at the $n$-vertices order:
\begin{align}
     \Braket{\hat{\O}}^{(n)}(t) = i^n & \int_{-\infty}^t \dd t_1 \int_{-\infty}^{t_1} \dd t_2 \ldots \int_{-\infty}^{t_{n-1}} \dd t_n  \times \\ & \Braket{ 0| \left[ \hat{H}_\mathrm{int}(t_n ), \left[ \hat{H}_\mathrm{int}(t_{n-1} ), \ldots, \left[\hat{H}_\mathrm{int}(t_1), \hat{\O}(t) \right] \ldots \right] \right]|0 } \,. \nonumber
\end{align}
Let us investigate more concretely the effects of the $U$ and $B$ terms composing $H_\mathrm{int}$.
We will do so explicitly up to two vertices, but later we will generalize to any order.

\paragraph{One-vertex order.} We get:
\begin{align}
     \Braket{\hat{\O}}^{(1)}(t) &= i  \int_{-\infty}^t \dd t_1 \Braket{0|\left[\hat{U}(t_1),\hat{\O}(t)\right]|0}  + i \Braket{0|\left[\hat{B}(t),\hat{\O}(t)\right]|0} \,.
\end{align}
First, any type of interactions encoded in $U$ would contribute to $\Braket{\hat{\O}}^{(1)}(t)$, as even the commutator of fields only, but at different times ($t_1$ for the ones in $U$ and $t$ for the ones in $\mathcal{O}$) is always non-vanishing.
Moreover, they are all integrated over time from the birth of interactions on sub-Hubble scales to the external time $t$ at which expectation values are calculated.
We now turn to the contribution from the total time derivative term $B$.
Clearly, the usual nomenclature {\em boundary} terms for total time derivative interactions carries, at first order in vertices, a meaningful name: they only result in a local-in-time contribution evaluated at the external time $t$.\footnote{For in-out perturbation theory like in $S$-matrix amplitudes calculations in flat spacetime, the upper limit of the time integrals is taken to be $+\infty$.
In that case, total time derivatives are always boundary terms evaluated at infinity, where interactions are supposed to shut down adiabatically, so their contribution is always vanishing.
In cosmology, breaking of Lorentz invariance forbids one to take time integrals up to future infinity and connect to a free theory with unambiguous vacuum.
A consequence of this fact is the necessity to use the more involved in-in formalism, but more generally this symmetry breaking can be considered as the fundamental difficulty in computing cosmological correlation functions and understanding their analytic structures.}
Note also that they only appear in a commutator with the external operator $\mathcal{O}$.
For definiteness, we now focus on cases where this external operator contains only powers of fields and not of their conjugate momenta, $\mathcal{O}(\psi_a)$: scalar or tensor power spectra, bispectra, trispectra, etc.
We can then make use of the canonical commutation relations
\begin{equation}
    \left[\hat{\psi}^a(t,\vec{x}),\hat{p}_{\psi,b}(t,\vec{y})\right]=i \delta^a_{b} \delta^{(3)}(\vec{x}-\vec{y}) \iff \left[\hat{\psi}^a_{\vec{k}}(t),\hat{p}_{\psi,b}^{\vec{k}^\prime}(t)\right]=i  \delta^a_b\delta^{(3)}(\vec{k}+\vec{k}^\prime) \,.
\end{equation}
Let us consider different cases:
\begin{itemize}
    \item If a $B$-term contains only fields and no time derivatives of the fields---i.e. no momenta---then it commutes with $\mathcal{O}$ and its contribution to correlation functions vanishes identically at this one-vertex order.
    \item If a $B$-term contains two or more powers of momenta, then at least one of them will survive the operation consisting in taking the commutator.
    The corresponding contribution, if not identically vanishing, may still be completely negligible in the inflationary context, i.e. for an external time $t$ at the end of inflation and scales of cosmological interest.
    Indeed, if an adiabatic limit is reached on super-horizon scales, which is both a necessary condition for inflation to remain predictive without a full description of reheating and a fairly generic feature of inflationary models, then conjugate momenta decay exponentially on super-horizon scales.
    Note however that this contribution may not be negligible at external times $t$ not taken at the end of inflation, or during or slightly after non-attractor phases of inflation like ultra-slow-roll.
    \item  If a $B$-term contains exactly one power of momenta, then its contribution may be important if it is not suppressed by spatial gradients, which are as negligible as time derivatives at the end of inflation for scales of cosmological interest.
\end{itemize}
The way we displayed usual interactions and total time derivative terms in the cubic interactions in Eqs.~\eqref{eq:cubic action scalars} is optimized to compute primordial bispectra as only few boundary terms can contribute to correlation functions.
For the scalar bispectrum specifically, only the last boundary term in $B^{(3)}$ contains a single power of the momentum $p_\zeta$, however it is suppressed by spatial gradients that are exponentially small on super-horizon scales, and it results in a completely negligible contribution~\cite{Arroja:2011yj,Burrage:2011hd}.
This is the main reason why one usually disregards the contribution of total time derivative terms after having chosen them appropriately.
But, this intuition is based on one-vertex in-in perturbation theory in the context of bispectrum calculations.
Let us now investigate the two-vertices order.

\paragraph{Two-vertices order.} The effect of total time derivatives interactions becomes more subtle at the two-vertices order as they can mix, both together and with usual interactions.
Diagramatically, the two-vertices order also corresponds to the appearance of internal, so-called ``bulk-to-bulk'' propagators.
We find:
\begin{align}
     \Braket{\hat{\O}}^{(2)}(t) &= \sum_{a,b\,\in\{B,U\}} \Braket{\hat{\O}}^{(2)}_{ab}(t) \,, \\  
     \Braket{\hat{\O}}^{(2)}_{BB}(t) & = - \int_{-\infty}^t \dd t_1  \int_{-\infty}^{t_1} \dd t_2  \Braket{0|\left[\frac{\dd \hat{B}}{\dd t_2}(t_2),\left[\frac{\dd \hat{B}}{\dd t_1}(t_1),\hat{\O}(t)\right]\right]|0}\,, \nonumber \\
     \Braket{\hat{\O}}^{(2)}_{BU}(t) & = - \int_{-\infty}^t \dd t_1  \int_{-\infty}^{t_1} \dd t_2  \Braket{0|\left[\frac{\dd \hat{B}}{\dd t_2}(t_2),\left[\hat{U}(t_1),\hat{\O}(t)\right]\right]|0}\,, \nonumber \\
     \Braket{\hat{\O}}^{(2)}_{UB}(t) & = - \int_{-\infty}^t \dd t_1  \int_{-\infty}^{t_1} \dd t_2  \Braket{0|\left[\hat{U}(t_2),\left[\frac{\dd \hat{B}}{\dd t_1}(t_1),\hat{\O}(t)\right]\right]|0}\,, \nonumber \\
     \Braket{\hat{\O}}^{(2)}_{UU}(t) & = - \int_{-\infty}^t \dd t_1  \int_{-\infty}^{t_1} \dd t_2  \Braket{0|\left[\hat{U}(t_2),\left[\hat{U}(t_1),\hat{\O}(t)\right]\right]|0}\,, \nonumber 
\end{align}
where care needs to be taken due to the nested time integrals.
We investigate each contribution separately.

The $BB$ contribution is rather simple, as the integral over $t_2$ can be carried explicitly, yielding a function of $t_1$ and $t$ only (we disregard contributions from $t \rightarrow -\infty$ as those are consistently shut down by the $i \epsilon$-prescription, even if not explicited any more in the above expressions),
\begin{equation}
    \hat{B}(t_1) \hat{B}^\prime(t_1) \hat{\mathcal{O}}(t) - \hat{B}(t_1) \hat{\mathcal{O}}(t) \hat{B}^\prime(t_1)  -  \hat{B}^\prime(t_1) \hat{\mathcal{O}}(t) \hat{B}(t_1) + \hat{\mathcal{O}}(t) \hat{B}^\prime(t_1)  \hat{B}(t_1)  \nonumber \,,
\end{equation}
which can be recast as a total derivative of $t_1$ plus a remaining term, finally yielding
\begin{align}
     \Braket{\hat{\O}}^{(2)}_{BB}(t) =& -  \frac{1}{2}\Braket{0|\left[\hat{B}(t),\left[\hat{B}(t),\hat{\O}(t)\right]\right]|0} \notag\\
     &- \frac{1}{2} \int^t_{-\infty} \dd t_1  \Braket{0|\left[\left[\hat{B}(t_1),\frac{\dd \hat{B}}{\dd t_1}(t_1)\right],\hat{\O}(t)\right]|0} \,.
\end{align}
The first term shares some features with the first-order equivalent one: it is a local contribution at the final time $t$ and it only contributes to field correlation functions if it has at least one term with one momentum.
However, this time, different $B$-terms may mix non-trivially.
For example, consider $B=B_1+B_2$, schematically with $B_1 \sim \int \psi^3$ and $B_2 \sim \int \psi p_\psi^2$, and $\O=\psi^4$, then $[\hat{B}(t),[\hat{B}(t),\hat{\O}(t)]] \sim \psi^6(t)$ where no momenta survived the commutators.
The second term above, however, is still integrating over time, showing that the nomenclature ``boundary term'' is not justified once one considers interactions involving at least two vertices, and, therefore, internal propagators.
As we will see, this term generally cancels with the contribution from one-vertex diagrams with usual interactions but of higher order, that come from switching from total time derivative interactions in the Lagrangian to the Hamiltonian.
Actually, such cancellation is precisely of the kind already mentioned at the end of the previous section, needed to remove large contributions in single-field slow-roll inflation from the large quartic interaction in Eq.~\eqref{eq: large H4 interaction}, as we show in the following insert.

\begin{framed}
{\small \noindent
Here, we consider only one cubic interaction with a total time derivative, and the corresponding quartic interaction generated from going from the Lagrangian to the Hamiltonian, in real space:
\begin{align}
    \mathcal{L}^{(3)}(\zeta,\dot{\zeta}) &\supset - \frac{\dd \B^{(3)}_A }{\dd t}  (\zeta,\dot{\zeta}) = -\frac{\dd}{\dd t} \left( 9 a^3 H \Mp^2 \zeta^3  \right) \\
    \implies H_\mathrm{int}^I & \supset \int \dd^3\vec{x} \left[\frac{\dd}{\dd t}\left( 9 a^3 H \Mp^2 \zeta_I^3 \right) + \frac{729}{4 \epsilon}a^3 H^2 \Mp^2 \zeta_I^4\right] \,.
\end{align}
The first term is a total time derivative, let us call it $B^{(3)}_A$, and the second one is a usual interaction, let us call it $U^{(4)}_A=U^{(4)}_{\mathrm{from\,}-\dd \B^{(3)}_A  / \dd t}$.
Let us add the contribution of the former at the two-vertices order to the one of the latter at the one-vertex order.
For this, we compute the following commutator appearing at the two-vertices order:
\begin{align}
     -\frac{1}{2} \left[\hat{B}^{(3)}_A(t_1),\frac{\dd \hat{B}^{(3)}_A}{\dd t_1}(t_1)\right] &= -\frac{1}{2}\int \dd^3 \vec{x} \int \dd^3 \vec{y} \left[9a^3 H \Mp^2 \zeta_I^3(\vec{x},t_1) , 27 a^3 H \Mp^2 \zeta_I^2(\vec{y},t_1) \dot{\zeta}_I(\vec{y},t_1)\right] \nonumber \\
     &= -  \frac{1}{2} \times 9 \times 27 \times \frac{3 i}{2 \epsilon} \int \dd^3 \vec{x} \,   a^3 H^2 \Mp^2 \zeta_I^4(\vec{x},t_1) \,, 
\end{align}
where we used the linear relation verified by interaction picture fields $\dot{\zeta}_I = p_\zeta^I / (2\epsilon a^3 \Mp^2 )$, and the equal-time commutation relation with appropriate symmetry factors $[\zi^3(\vec{x},t_1),\zi^2(\vec{y},t_1) p_\zeta^I(\vec{y},t_1)] = 3 i \times \delta^{(3)}(\vec{x}-\vec{y}) \times \zi^4(\vec{x},t_1)$. 
Collecting the pre-factors, we find the overall coefficient to be $-729/(4 \epsilon)$.
Adding this contribution to the one-vertex contribution from $U^{(4)}_A$, we find:
\begin{equation}
    i  \int_{-\infty}^t \dd t_1 \Braket{0|\left[\hat{U}^{(4)}_A(t_1),\hat{\O}(t)\right]|0}
    - \frac{1}{2} \int^t_{-\infty} \dd t_1  \Braket{0|\left[\left[\hat{B}^{(3)}_A(t_1),\frac{\dd \hat{B}^{(3)}_A}{\dd t_1}(t_1)\right],\hat{\O}(t)\right]|0} = 0 \,,
\end{equation}
thus proving the exact cancellation.
Note that the cancellation happens even in the integrand of the $t_1$-integral, so that there is a notion of \textit{effective quartic} interaction, which exactly vanishes in the example given above.
We will come back to this notion soon.
This example shows also that it is inconsistent to take only the cubic Hamiltonian to compute a given correlation function: one could wrongly interpret the result to be large from the $[B^{(3)}, \dd B^{(3)} /\dd t]$ contribution, while it gets cancelled by carefully computing, and taking into account, $U^{(4)}$ terms generated from the $-\B^{(3)}$ term in the cubic Lagrangian density.
}
\end{framed}

The $BU$ contribution is found trivially after integrating explicitly over $t_2$:
\begin{equation}
     \Braket{\hat{\O}}^{(2)}_{BU}(t) = - \int_{-\infty}^t \dd t_1  \Braket{0|\left[\hat{B}(t_1),\left[\hat{U}(t_1),\hat{\O}(t)\right]\right]|0}\,,
\end{equation}
and constitutes yet another example where the total time derivative term is ``integrated over the history'', and not a ``boundary term''.

The $UB$ contribution is more subtle.
We define $\hat{F}(t_1) = \int_{-\infty}^{t_1} \dd t_2 \, \hat{U}(t_2)$, in terms of which we have:
\begin{equation}
    \hat{F}(t_1) \hat{B}^\prime(t_1) \hat{\mathcal{O}}(t) - \hat{F}(t_1) \hat{\mathcal{O}}(t) \hat{B}^\prime(t_1)  -  \hat{B}^\prime(t_1) \hat{\mathcal{O}}(t) \hat{F}(t_1) + \hat{\mathcal{O}}(t) \hat{B}^\prime(t_1)  \hat{F}(t_1)  \nonumber \,.
\end{equation}
We then integrate by parts over $t_1$ and we find two terms:
\begin{equation}
     \Braket{\hat{\O}}^{(2)}_{UB}(t) = - \int_{-\infty}^t \dd t_1  \Braket{0|\left\{\left[\hat{U}(t_1),\left[\hat{B}(t),\hat{\O}(t)\right]\right] - \left[\hat{U}(t_1),\left[\hat{B}(t_1),\hat{\O}(t)\right]\right]  \right\}|0} \,,
\end{equation}
with the first one verifying the usual statement about total time derivatives being boundary terms, and with the second one which does not.
Actually, the two terms $BU$ and $UB$ may be rewritten together in a useful way as:
\begin{align}
     \Braket{\hat{\O}}^{(2)}_{BU}(t)+ \Braket{\hat{\O}}^{(2)}_{UB}(t)  = & \, - \int_{-\infty}^t \dd t_1  \times  \\
     & \Braket{0|\left\{\left[\hat{U}(t_1),\left[\hat{B}(t),\hat{\O}(t)\right]\right] + \left[\left[\hat{B}(t_1),\hat{U}(t_1)\right],\hat{\O}(t)\right] \right\}|0}\,.
     \nonumber
\end{align}
We argue that this form is enlightening as it enables one to conclude that total time derivatives may also contribute to the bulk evolution (via $B(t_1)$ above) when mixing with usual interactions $U(t_1)$, but only provided they do not commute with them.

The $UU$ contribution, from usual interactions only, is the one that has always been considered, and cannot be written more explicitly.
We close this paragraph by re-arranging in-in perturbation theory up to second order and in the commutator form, including both total time derivatives ($ \dd B / \dd t$) and usual interactions ($U$) in the interaction Hamiltonian explicitly.
This extends the in-in perturbative formalism to theories with total time derivatives:

\begin{align}
\label{eq: in-in recast for boundary and bulk}
    \Braket{\hat{\O}}(t) =& \Braket{0|\hat{\O}(t)|0}+\highlight{i \Braket{0|\left[\hat{B}(t),\hat{\O}(t)\right]|0} -  \frac{1}{2}\Braket{0|\left[\hat{B}(t),\left[\hat{B}(t),\hat{\O}(t)\right]\right]|0}}  \\
    &+  i  \int_{-\infty}^t \dd t_1 \Braket{0|\left\{\left[\hat{U}(t_1),\hat{\O}(t)\right]+\highlight{  \left[i[\hat{B}(t_1),\hat{U}(t_1)],\hat{\O}(t)\right] + \left[\hat{U}(t_1), i[\hat{B}(t),\hat{\O}(t)]\right]} \right\}|0} \nonumber \\
    & + \highlight{i \int^t_{-\infty} \dd t_1  \Braket{0|\left[\frac{i}{2} \left[\hat{B}(t_1),\frac{\dd \hat{B}}{\dd t_1}(t_1)\right],\hat{\O}(t)\right]|0}} \nonumber  \\
    & - \int_{-\infty}^t \dd t_1  \int_{-\infty}^{t_1} \dd t_2  \Braket{0|\left[\hat{U}(t_2),\left[\hat{U}(t_1),\hat{\O}(t)\right]\right]|0}
    + \ldots  \,, \nonumber
\end{align}
Contributions including total time derivatives interactions are highlighted.

Interestingly, all these new contributions, shown here explicitly up to two-vertices order, may be recast as either a redefinition of the external operator $\O$ as an effective external operator $\tilde{\O}$, or a redefinition of the usual interactions $U$ as effective interactions $\tilde{U}$:
\begin{align}
\label{eq: effective O and U 2-vertices order first line}
    \hat{\tilde{\O}} &\equiv \hat{\O} + i \left[\hat{B},\hat{\O}\right]  -  \frac{1}{2}\left[\hat{B},\left[\hat{B},\hat{\O}\right]\right] + \ldots \\
    \label{eq: effective O and U 2-vertices order}
    \hat{\tilde{U}} &\equiv \hat{U} + i \left[\hat{B},\hat{U}\right] + \frac{i}{2} \left[\hat{B},\frac{\dd \hat{B}}{\dd t}\right] + \ldots \,.
\end{align}
As we are going to see, this statement actually holds at any order in perturbation theory and considerably simplifies the in-in formalism when total time derivatives interactions are present.

\paragraph{Any vertex order.}
We would like to generalize Eqs.~\eqref{eq: effective O and U 2-vertices order first line}--\eqref{eq: effective O and U 2-vertices order} beyond the two-vertices order.
The first line of Eq.~\eqref{eq: in-in recast for boundary and bulk} seems straightforward to extend at any order as
\begin{align}
\label{eq: guess eff O}
    \sum_{n=0}^\infty \frac{i^n}{n!} \Braket{0|\underbrace{[\hat{B}(t)[\hat{B}(t)[...,[\hat{B}(t),\hat{\O}(t)]...]]}_{n \,\, \text{commutators}}|0}\,.
\end{align}
If true, this would mean that we could redefine \textit{non-perturbatively} the external operator $\O$ to take into account this (infinite) subset of contributions.
To build physical intuition, we first show that this property indeed holds, and then only we will prove the general form of Eqs.~\eqref{eq: effective O and U 2-vertices order first line}--\eqref{eq: effective O and U 2-vertices order} at any order including both redefinitions of $\mathcal{O}$ and $U$ at once.

The effect $\mathcal{O} \rightarrow \tilde{\mathcal{O}}$ from total time derivatives interactions corresponds to the only one that one would find by wrongly interpreting the ``non-perturbative'' expression of the in-in perturbation theory, Eq.~\eqref{eq: in-in general formula}, and performing explicitly the time integral inside the exponential, inside the (anti-)time-ordering operators.
Clearly, this procedure is wrong as it misses all other contributions from total time derivatives that are still integrated over time.
It has however been used in the past, in particular to justify that total time derivatives without momenta of the fields inside could not participate to correlation functions made of fields only.
Although our findings will eventually agree with this statement, those proofs using the in-in operator formalism are clearly incomplete as we show now.
One can use the (left-ordered) Zassenhaus formula for the exponential of the sum of two operators:
\begin{equation}
    e^{X+Y}=e^X e^Y \prod_{n=2}^\infty e^{C_n(X,Y)} \,, \nonumber
\end{equation}
with $C_n$ living in the Lie algebra of $(X,Y)$, i.e. they can all be written only in terms of $n$ nested commutators of $X$ and $Y$.
For example, $C_2(X,Y) = - [X,Y]/2\,, \,\, C_3(X,Y) = \left(2[Y,[X,Y]]+[X,[X,Y]\right)/6\ldots$
Applying this left-ordered formula to the time-ordered exponential with $X=- i \int \dd B/\dd t$ and $Y=- i \int U $, we are able to take the single $B$ term outside the operator $T$ as
\begin{align}
\label{eq: time ordered expansion left}
    T& \left[\exp \left\{ - i\int^t_{-\infty^-} \dd t^\prime \left(U + \dd B / \dd t^\prime\right) \right\} \right]= \exp \left\{ - i B(t) \right\} T \Biggl[ \exp \left\{ - i \int^t_{-\infty^-} \dd t^\prime \, U\right\}\\ &\times\prod_{n=2}^\infty \exp \left\{C_n\left(- i \int^t_{-\infty^-} \dd t^\prime \, \dd B/\dd t^\prime,- i\int^t_{-\infty^-} \dd t^\prime \, U \right)\right\} \Biggr] \,. \nonumber
\end{align}
Similarly, the anti-time-ordered product may be rewritten by using the right-ordered  Zassenhaus formula 
\begin{equation}
    e^{-X-Y}= \prod_{n=\infty}^2 e^{\tilde{C}_n(-X,-Y)}  e^{-Y} e^{-X}\,, \nonumber
\end{equation} 
with $\tilde{C}_n(X,Y)=(-1)^{n+1}C_n(X,Y)$, giving
\begin{align}
\label{eq: time ordered expansion right}
    \bar{T} \Biggl[&\exp \left\{ i\int^t_{-\infty^+} \dd t^\prime \left(U + \dd B / \dd t^\prime\right) \right\} \Biggr]
    \\ =& \bar{T} \Biggl[ \prod_{n=\infty}^2 \exp \left\{(-1)^{n+1} C_n\left(i \int^t_{-\infty^+} \dd t^\prime \, \dd B/\dd t^\prime,i\int^t_{-\infty^+} \dd t^\prime \, U \right)\right\} \notag\\&\exp \left\{ - i \int^t_{-\infty^+} \dd t^\prime \, U\right\}\Biggr]\exp \left\{ i B(t) \right\}  \,. \nonumber
\end{align}
Unfortunately, there is no closed formula for the $C_n$'s (see however Ref.~\cite{Casas_2012} for an efficient algorithm to compute them recursively up to a given order $n$ and Ref.~\cite{Kimura:2017xxz} for an explicit expression for $e^{-X} e^{X+Y}$ although not in terms of commutators only).
Therefore, this formula is not particularly useful to perform concrete calculations in perturbation theory.
However, it does prove the non-perturbative form of $\tilde{\mathcal{O}}$ in terms of $\mathcal{O}$ as:
\begin{equation}
\label{eq: non-pert eff O}
    \hat{\tilde{\mathcal{O}}}(t)= e^{ i \hat{B}(t)} \hat{\mathcal{O}}(t) e^{ - i \hat{B}(t)} \,,
\end{equation}
which exactly coincides with the infinite series expansion that we guessed in Eq.~\eqref{eq: guess eff O}. Expressions~\eqref{eq: time ordered expansion left}--\eqref{eq: time ordered expansion right}, although not explicit, also prove the presence of additional non-zero contributions from $B$ integrated over time when mixing either with itself or with ordinary interactions in $U$.
Importantly, this proves also that these contributions are always in the form of commutators and start at the two-vertices order only, given the properties of the $C_n$'s.

We are now ready for the general proof, which is rather short and elegant.
We define the usual interaction-picture evolution operator as
\begin{equation}
    \hat{F}(t, t_0) = T \left[\exp \left\{ - i\int^t_{t_0} \dd t^\prime \hat{H}_\mathrm{int} (t^\prime) \right\} \right] \,.
\end{equation}
$\hat{F}$ is a unitary operator and its Hermitian conjugate defines the inverse evolution along the anti-time ordered path.
The general in-in formula~\eqref{eq: in-in general formula} is precisely found by separating the full Hamiltonian into a free part dictating the dynamics of the interaction picture fields and momenta, and an interaction part that dictates the evolution of $\hat{F}$ as
\begin{equation}
     i \frac{\partial \hat{F}(t,t_0)}{\partial t} = \hat{H}_\mathrm{int}(t) \hat{F}(t,t_0) \,.
\end{equation}
We now split $H_\mathrm{int} =U + \dd B/\dd t$, and we define a new evolution operator $\tilde{F}$ as 
\begin{equation}
\label{eq: F to tilde(F)}
    \hat{F}(t,t_0)= e^{- i \hat{B}(t) }  \hat{\tilde{F}}(t,t_0) \,.
\end{equation}
One can then find the equation of evolution of $\tilde{F}$ using the one for $F$ and their relation through Eq.~\eqref{eq: F to tilde(F)}, being careful about the fact that $B$ does not commute with $\dd B/\dd t$:
\begin{equation}
\label{eq: evolution of tilde(F)}
   i \frac{\partial \hat{\tilde{F}}(t,t_0)}{\partial t} = \left\{ e^{i \hat{B}(t)}\left[ \hat{U}(t) + \frac{\dd \hat{B}(t)}{\dd t} + i  e^{-i \hat{B}(t)} \frac{\dd}{\dd t}  \left(e^{i \hat{B}(t)}\right)\right]  e^{-i \hat{B}(t)} \right\} \hat{\tilde{F}}(t,t_0) \,.
\end{equation}
This equation can be formally solved with a time-ordered product and we can retrieve the initial evolution operator as:
\begin{equation}
    \hat{F}(t,t_0)= e^{- i \hat{B}(t) } \, T \left[\exp \left\{ - i\int^t_{t_0} \dd t^\prime \hat{\tilde{U}}(t^\prime) \right\} \right] \,,
\end{equation}
where, remarkably, we were able to define the effective interactions $\tilde{U}$ non-perturbatively as:
\begin{equation}
\label{eq: non-pert eff U}
    \hat{\tilde{U}}(t) = e^{i \hat{B}(t)}\left[ \hat{U}(t) + \frac{\dd \hat{B}(t)}{\dd t} + i  e^{-i \hat{B}(t)} \frac{\dd}{\dd t}  \left(e^{i \hat{B}(t)}\right)\right]  e^{-i \hat{B}(t)} \,.
\end{equation}
This expression was quoted in a different context and without proof by Weinberg in Ref.~\cite{Weinberg:2005vy}.
Following similar steps for the anti-time-ordered path of the in-in, finally setting $t_0 \rightarrow -\infty^\pm$\footnote{
Strictly speaking, the $i \epsilon$-prescription to adiabatically shut down interactions in the infinite past time breaks the unitarity property of the evolution.
See~\cite{Kaya:2018jdo} for an interesting discussion on this topic, as well as~\cite{Baumgart:2020oby} for the proposition of a different prescription that respects unitarity and allows for a manifestly causal in-in perturbation theory.
}, we eventually find
\begin{equation}
\label{eq: in-in general formula our version}
    \Braket{\hat{\O}}(t) = \braket{0|\Bar{T}\left[\exp{i \int_{-\infty^+}^t \dd t^\prime \hat{\tilde{U}}(t^\prime)}\right] \underbrace{e^{i \hat{B}(t)} \hat{\O}^I(t)  e^{-i \hat{B}(t)}}_{\hat{\tilde{\O}}^I(t)  } T\left[\exp{-i \int_{-\infty^-}^t \dd t^\prime\hat{\tilde{U}}(t^\prime)}\right]|0} \,.
\end{equation}
This proves non-perturbatively that one can write the expectation value of interest in the theory of interest including both usual interactions and total time derivatives $\braket{\hat{\O}}_{\hat{U},\hat{B}}$ in terms of an expectation value of another operator in another theory without total time derivatives interactions:
\begin{equation}
\label{eq: effective O and U}
    \Braket{\hat{\O}}_{\hat{U},\hat{B}} = \Braket{\hat{\tilde{\O}}}_{\hat{\tilde{U}}} \,.
\end{equation}
Expressions for $\tilde{\mathcal{O}}$ and $\tilde{U}$ have been found first perturbatively up to two-vertices orders in Eqs.~\eqref{eq: effective O and U 2-vertices order first line}--\eqref{eq: effective O and U 2-vertices order}, then non-perturbatively in Eqs.~\eqref{eq: non-pert eff O}--\eqref{eq: non-pert eff U}.
For performing concrete perturbative calculations, it will prove useful to also expand these non-perturbative formulas at arbitrary finite vertex order $n$ as:
\begin{align}
    \label{eq: eff O arbitrary order}
    \hat{\tilde{\mathcal{O}}}&=\sum_{n=0}^\infty \frac{i^n}{n!} \left(\mathcal{L}_{\hat{B}}\right)^n \cdot \hat{\mathcal{O}} \equiv e^{i \mathcal{L}_{\hat{B}}} \cdot \hat{\mathcal{O}}   \\
    \label{eq: eff U arbitrary order}
    \hat{\tilde{U}}&=\sum_{n=0}^\infty \frac{i^n}{n!} \left(\mathcal{L}_{\hat{B}}\right)^n \cdot \hat{U} +  \sum_{n=0}^\infty \frac{n \, i^n}{(n+1)!} \left(\mathcal{L}_{\hat{B}}\right)^n \cdot \frac{\dd \hat{B}}{\dd t} \\
    & \equiv e^{i \mathcal{L}_{\hat{B}}}\cdot \hat{U} + \frac{e^{i \mathcal{L}_{\hat{B}}}(i\mathcal{L}_{\hat{B}}-1)+1}{i \mathcal{L}_{\hat{B}}} \cdot \frac{\dd \hat{B}}{\dd t} \nonumber \,,
\end{align}
where we have defined the operation ``taking the commutator of $\ldots$ with $\hat{B}$'' as $\mathcal{L}_{\hat{B}}$ with
\begin{equation}
    \mathcal{L}_{\hat{B}} \cdot \hat{X} \equiv \left[\hat{B},\hat{X}\right]\implies \left(\mathcal{L}_{\hat{B}}\right)^n \cdot \hat{X} = \underbrace{\left[\hat{B},\left[\hat{B},\left[...,\left[\hat{B},\hat{X}\right]...\right]\right]\right]}_{n \,\, \text{commutators}}
\end{equation}
and where in the second lines above we have defined a non-perturbative version of the infinite sums of such commutators.
Looking at $n \leqslant 2$ in those equations, we consistently recover Eqs.~\eqref{eq: effective O and U 2-vertices order first line}--\eqref{eq: effective O and U 2-vertices order}.

\vspace{10pt}
We conclude this section by summarizing what we have learnt about total time derivative terms in the Lagrangian, and all possible ways they can affect expectation values of an operator:
\begin{itemize}
    \item If $\mathcal{L} \supset - \dd \B /\dd t$, we have seen that it generates the same interaction with an opposite sign in the interaction Hamiltonian in terms of interaction picture fields and momenta, plus a new contribution as a usual interaction but of higher-order in powers of fields and momenta, $H^I_\mathrm{int} \supset \dd B /\dd t + U_{\mathrm{from\,}-\dd \B /\dd t} $;
    \item The resulting total-time derivative interaction in the Hamiltonian redefines the external operator $\O$ whose expectation values are sought for. This effect really corresponds to a ``boundary term'', and it is the only one that was considered before this work, as it is the only relevant one for the computation of the primordial bispectrum at tree level;
    \item $B$ terms in the interaction Hamiltonian also contribute as integrated over time, i.e. as redefining effective interactions, which cannot be understood as a ``boundary term'' effect.
    \item 
    For $B$-terms that do not contain powers of the momenta inside the total time derivative, we have proved that this last contribution at the two cubic vertices order exactly cancels the one from the quartic $U_{\mathrm{from\,}-\dd \B /\dd t}$ at the one-vertex order.
    Furthermore, in the particular case where the Lagrangian interactions are made \textit{only} of total time derivatives of fields as $\mathcal{L}-\mathcal{L}^{(2)}=-\dd \mathcal{B}/\dd t$ (and no usual interactions), it is straightforward to prove that these cancellations happen at every order in vertices and powers of the fields.
    Therefore, these interactions do not affect correlation functions $\mathcal{O}$ made of fields only.
    Indeed, one finds $\tilde{\mathcal{O}}=\O$ and $\tilde{U}= \dd B/\dd t +  U_{\mathrm{from\,}-\dd \B /\dd t} + i [B,\dd B/\dd t]/2 = 0$, which is an exact expression since $B$ commutes with $[B,\dd B/\dd t]$ and with $U_{\mathrm{from\,}-\dd \B /\dd t} $.
    We find the general proof of this statement when the Lagrangian interactions \textit{also} contain usual interactions $-\mathcal{U}$ rather cumbersome in this formalism, as many cancellations should be sought for in $\tilde{U}$ between different vertex orders.
    Instead, we will prove the equivalent statement with a different approach in Sec.~\ref{sec: canonical transformations}.
\end{itemize}
We now show concrete calculations of correlation functions for two toy models, in order to get more intuition on the aforementioned cancellations within the in-in perturbation theory with total time derivative interactions.
The reader not interested in these calculations may directly jump to Sec.~\ref{subsec: SK formalism} and Sec.~\ref{sec: canonical transformations}, where we present alternative methods to compute the same expectation values in a more direct way.

\subsection{Toy models}
\label{subsec: toy models}

We here define two toy models of a single scalar degree of freedom $\psi(\vec{x},t)$ and its momentum $p_\psi(\vec{x},t)$.
In both cases, we consider the following quadratic Lagrangian:
\begin{equation}
\label{eq: Lfree toy models}
    \mathcal{L}^{(2)}(\psi,\dot{\psi}) =  \frac{c(t)}{2} \left[\dot{\psi}^2 - (\partial \psi)^2 \right]  \,,
\end{equation}
with $c(t)$ a time-varying parameter.
For $t \rightarrow \tau$ the conformal time and $\dot{\psi} = \dd \psi/\dd t \rightarrow \psi^\prime= \dd \psi/\dd \tau$, with $S=\int \dd \tau \mathcal{L}$, $\psi = \zeta$ and $ c = 2 a^2 \epsilon \Mp^2$, one recovers the quadratic Lagrangian of single-field inflation with canonical kinetic terms.
However in the following we remain generic and do not specify neither a cosmology nor a particular field content.
This Lagrangian results in the following quadratic Hamiltonian, which we define as the free Hamiltonian:
\begin{equation}
\label{eq: free Hamiltonian toy model}
    \mathcal{H}_\mathrm{free} (\psi, p_\psi) \equiv \mathcal{H}^{(2)}(\psi, p_\psi) = \frac{p_\psi^2}{2 c} + \frac{c}{2}  (\partial \psi)^2  \,.
\end{equation}
In particular, interaction picture fields and momenta verify:
\begin{align}
    \dot{\psi}_I &= \frac{p_\psi^I}{c}\,, \\
    \dot{p}_{\psi}^I &=  c \,\partial^2 \psi_I \,. \nonumber
\end{align}
By combining Hamilton's equations, we find $\dd(c  \dot{\psi}_I)/\dd t - c \,\partial^2 \psi_I = 0$, which corresponds to the linear equations of motion one would derive from the quadratic Lagrangian.
We now specify to two different interacting theories.

\subsubsection{No time derivatives inside the total time derivative interaction}

The first toy model contains the following cubic order Lagrangian density, with both a total time derivative and a usual interaction:
\begin{equation}
\label{eq: toy model 1}
    \mathcal{L}^{(3)}(\psi,\dot{\psi}) =  \underbrace{- \frac{\dd}{\dd t}\left(\alpha(t) \psi^3\right)}_{- \dd \mathcal{B}^{(3)} /\dd t} \underbrace{- \beta(t) \dot{\psi}\psi^2}_{-\mathcal{U}^{(3)}} \,,
\end{equation}
with $\alpha(t), \beta(t)$ two time-varying parameters.
Together with the quadratic Lagrangian, it can be used to derive the full Hamiltonian density of the theory, $\mathcal{H}$, from the full momentum $p_\psi =  c \dot{\psi} - (\beta + 3 \alpha) \psi^2 $:
\begin{equation}
\label{eq: full Hamiltonian toy model 1}
\mathcal{H}(\psi,p_\psi)= \mathcal{H}^{(2)}(\psi,p_\psi) + \frac{(\beta + 3 \alpha ) p_\psi \psi^2}{c} +\dot{\alpha} \psi^3 + \frac{1}{2 c}(\beta^2 + 6 \alpha \beta + 9 \alpha^2) \psi^4 \,,
\end{equation}
which is an exact expression.
Having done so, one can evaluate the interaction Hamiltonian $\mathcal{H}_\mathrm{int} = \mathcal{H}-\mathcal{H}_\mathrm{free}$ in terms of interaction picture fields, and replace the momentum using the linear equations of motion, giving:
\begin{align}
    H^I &\equiv \int \dd^3 \vec{x}\, \mathcal{H}_\mathrm{int}(\psi_I,p_\psi^I(\psi_I,\dot{\psi}_I))\nonumber \\
    &= \int \dd^3 \vec{x} \left[ \underbrace{\frac{\dd}{\dd t}\left(\alpha\psi_I^3 \right)}_{\rightarrow \,\, \dd B/\dd t} + \underbrace{\beta \dot{\psi}_I \psi_I^2}_{\rightarrow \,\, U^{(3)}}  + \underbrace{\frac{1}{2 c}(\beta^2 + 6 \alpha \beta + 9 \alpha^2) \psi_I^4 }_{\rightarrow \,\, U^{(4)}_{\mathrm{from}\,\mathcal{L}^{(3)}}} \right] \,,
\end{align}
where we were able to re-form the total time derivative at cubic order, as in the Lagrangian but with an opposite sign.
We identify this term as $\dd B/\dd t$ in the notations of the previous section.
Once more, we also note the appearance of new quartic interaction, with a size inversely proportional to the normalisation of the quadratic order Lagrangian, $c$.
In single-field inflation where $c \propto \epsilon^1$ and $\alpha \propto \epsilon^0$, this leads in particular to the large quartic interaction inversely proportional to $\epsilon$ that we have already encountered.
There are also two other quartic terms, including one that mixes the two Lagrangian cubic interactions non-trivially and is $\propto \alpha\times \beta$.
We identify the cubic interaction $\propto \beta$ plus the new quartic interactions as the term $U=U^{(3)}+U^{(4)}_{\mathrm{from}\,\mathcal{L}^{(3)}}$ in the notations of the previous section.

Using this interaction Hamiltonian, we can now compute correlation functions of the theory.
For this, we will take the shortest route, using the lessons learnt in the previous section, and compute the effective external operator $\tilde{O}$ and usual interactions $\tilde{U}$.
We find:
\begin{equation}
    \tilde{U}  =  U  + i \left[B, U \right]  + \frac{i}{2} \left[B, \frac{\dd B}{\dd t}\right] = \int \dd^3 \vec{x} \left( \beta \dot{\psi}_I \psi_I^2  + \frac{\beta^2}{2 c} \psi_I^4 \right)  \,, 
\end{equation}
where we used the derivative of an exponential operator and the fact that $B$ commutes with $[B, \dd B/\dd t]$ and with $[B, U]$ to write an exact expression for $\tilde{U}$ at all orders.
This proves the cancellation between two of the possible effects from cubic total time derivative terms $\dd \mathcal{B}/\dd t$ in the Lagrangian: induce in the Hamiltonian usual interactions $U_{\mathrm{from}\,-\dd \mathcal{B}/\dd t}$ of higher order, and modify interactions into new effective ones $U \rightarrow \tilde{U}$.
Evidently, the cancellation happens even in the presence of additional interactions (here $\propto \beta$) which mix non-trivially.
Now, we want to compute $\tilde{\O}$, a step for which we first need to specify the correlation function of interest, i.e. $\mathcal{O}$.
In cosmology, we are typically interested in the bispectrum, the trispectrum, the one-loop power spectrum, etc.
For all these correlation functions, the external operator $\O$ is only made of fields.
Since in this example, $B$ is also made of fields only, it is straightforward to show that:
\begin{equation}
    \tilde{\O} = \O \,.
\end{equation}
We have therefore proved, without even calculating correlation functions explicitly, that the total time derivative cubic interaction $\propto \alpha$ in the Lagrangian, Eq.~\eqref{eq: toy model 1}, does not generate \textit{at all} a bispectrum, nor a trispectrum, nor a one-loop correction to the power spectrum, etc.
Note however that the statement is non-trivial in the sense that we used cancellations between different vertex orders of the in-in perturbation theory.
Of course, the usual interaction $\propto \beta$ in Eq.~\eqref{eq: toy model 1} does correspond to non-linearities that will generate those (corrections to) correlation functions, but in a way that is not affected by the presence of the total time derivative interaction, as can be seen from the absence of the parameter $\alpha$ in $(\tilde{U}, \tilde{O})$.
In particular the Hamiltonian quartic interaction mixing the two Lagrangian cubic non-linearities and $\propto \alpha \times \beta$, was cancelled by the term $i [B,U]$ in the expression of the effective interaction $\tilde{U}$.
This is the first proof of this paper that total time derivative interactions made of fields only do not contribute to correlation functions of fields, at any order in vertex theory and including mixed diagrams with other vertices from usual interactions.
We show a second proof of this in the next paragraph below, and a third proof using the Lagrangian path integral approach is proposed in Sec.~\ref{subsec: SK formalism}.
Note also that the Lagrangian interaction $\propto \beta$ can almost be written as a total time derivative interaction without time derivatives of the fields inside, indeed $\beta \dot{\psi} \psi^2 = \dd(\beta \psi^3/3)/\dd t - \dot{\beta} \psi^3 /3$.
Therefore, from the current discussion, we already know that the final contributions to correlation functions from this interaction must be proportional to $\dot{\beta}$, and not $\beta$.
Indeed, one can write $\tilde{U} = \dd B^\prime/\dd t + U^\prime $ with $B^\prime= \beta \psi_I^3/3$ and where $U^\prime$ contains the cubic interaction $\propto \dot{\beta}$ and the quartic interaction $\propto \beta^2$.
This procedure indeed results in a new effective operator $\tilde{\tilde{U}} = \tilde{U} - \beta^2 \psi_I^4/(2c) = -\dot{\beta} \psi_I^3 /3 $, and with $\tilde{\tilde{O}}=O$ for correlation functions of fields only, thus proving our point.

Actually, it would have been straightforward to reach the same conclusions by avoiding the definition of the perturbative theory with interaction picture fields to begin with, and simply write the a priori non-linear equations of motion for the full fields:
\begin{align}
\label{eq: second order EoM toy model}
    0 &= \frac{\delta \left( \mathcal{L}^{(2)}+ \mathcal{L}^{(3)}\right)}{\delta \psi} \\&
    = \frac{\dd}{\dd t}\left[ c \dot{\psi} - (\beta + 3 \alpha) \psi^2\right] - c \partial^2 \psi +  3 \dot{\alpha} \psi^2 + 2 (\beta + 3 \alpha )\dot{\psi} \psi
    \nonumber 
    \\ &= \frac{\dd}{\dd t}\left( c \dot{\psi}\right) - c \partial^2 \psi  - \dot{\beta} \psi^2 \,, \nonumber
\end{align}
where the two contributions from the total time derivative cubic term cancel each other.
Note also how the interactions $\propto \beta$ combine to result in a contribution $\propto \dot{\beta}$ only, as already shown above.
This shows that at the non-perturbative level, the total time derivative interaction in Eq.~\eqref{eq: toy model 1} does not affect the dynamics of the field $\psi$, making for the second proof.
The lesson of this short paragraph is that the perturbative in-in formalism with total time derivatives may be unnecessary complicated, with cancellations to be sought for between different orders of the perturbation theory.

Also, if we write the Hamilton equations from the full Hamiltonian in Eq.~\eqref{eq: full Hamiltonian toy model 1},
\begin{align}
    \dot{\psi} &= \frac{p_\psi}{c} + \frac{(\beta + 3 \alpha) \psi^2}{c}  \,, \\ 
    \dot{p}_\psi &= c \partial^2 \psi  - \frac{2}{c} (\beta + 3 \alpha) \psi p_\psi - 3 \dot{\alpha} \psi^2 - \frac{2}{c}(\beta^2 + 6 \alpha \beta + 9 \alpha^2) \psi^3 \,, \nonumber
\end{align} 
we do find that those are non-linear even when $\dot{\beta}=0$.
Of course, by combining them we retrieve the second-order equations of motion~\eqref{eq: second order EoM toy model}, which are much simpler. The conclusion is that a total time derivative term in the Lagrangian $-\dd \B/\dd t$ where $\B$ is a function of fields only, and not of time derivatives of the fields---i.e. momenta---, does not contribute to correlation functions of fields, although they do affect the correlation functions of momenta.
Indeed, if e.g. $\O = \psi^2 p_\psi$, then $\tilde{O} \neq \O$.
This can also be seen from the comparison between the second-order equation of motion for $\psi$---linear when $\dot{\beta}=0$---and the first-order equations for $(\psi,p_\psi)$---non-linear even when $\dot{\beta}=0$.
This simple example also already points towards the idea that there may be a ``better'' notion of conjugate momentum.
We will come back to this in Sec.~\ref{sec: canonical transformations}.

\subsubsection{Time derivatives inside the total time derivative interaction}
\label{sec:toymodel2_standardinin}

The second toy model contains the following cubic order Lagrangian density, with both a total time derivative of a function of a time derivative, and a term proportional to the linear equations of motion, as they always appear in pairs:
\begin{align}
\label{eq: toy model 2}
    \mathcal{L}^{(3)}(\psi,\dot{\psi}) &= \underbrace{\frac{\dd}{\dd t}\left(c (t) \dot{\psi} \psi^2\right)}_{- \dd \mathcal{B}^{(3)}/\dd t} - \underbrace{\left[\frac{\dd}{\dd t} \left(c(t) \dot{\psi} \right)- c(t) \partial^2 \psi \right] \psi^2}_{- \mathcal{E}^{(3)}}  \\
    &=  2 c \dot{\psi}^{2} \psi + c \psi^2 \partial^2 \psi\,. \nonumber 
\end{align}
For simplicity we have assumed that the size of this interaction is proportional to the same homogeneous function $c(t)$ appearing in the free Lagrangian.
Developing the total time derivatives as we did in the second line of this expression, one can check the explicit cancellation of terms proportional to $\ddot{\psi}$, as expected.
There is therefore a single degree of freedom in the system, characterized by a momentum $p_\psi = c \dot{\psi} + 4 c \dot{\psi} \psi$.
This expression needs to be inverted, $\dot{\psi} = (p_\psi/c)\times (1+ 4 \psi )^{-1}$ in order to derive the Hamiltonian density as the Legendre transform of the Lagrangian.
Clearly, the expression becomes non-perturbative with inverse powers of the field $\psi$.
We therefore consistently expand the Hamiltonian up to quartic order in fields and momenta, which requires the expression of $\dot{\psi}(\psi,p_\psi)$ up to quadratic order only, giving:
\begin{equation}
\label{eq: full Hamiltonian toy model 2}
\mathcal{H}(\psi,p_\psi)= \mathcal{H}^{(2)}(\psi,p_\psi) - 2 \frac{p^2_\psi \psi}{c} - c \psi^2 \partial^2 \psi + 8 \frac{p^2_\psi \psi^2 }{c} + \ldots \,,
\end{equation}
where we omitted terms of order five or more.
Having done so, one can evaluate the interaction Hamiltonian $\mathcal{H}_\mathrm{int} = \mathcal{H}-\mathcal{H}_\mathrm{free}$ in terms of interaction picture fields, and replace the momentum using the linear equations of motion, giving:
\begin{equation}
    H^I \equiv \int \dd^3 \vec{x}\, \mathcal{H}_\mathrm{int}(\psi_I,p_\psi^I(\psi_I,\dot{\psi}_I)) = \int \dd^3 \vec{x} \left[ \underbrace{-\frac{\dd}{\dd t}\left(c\dot{\psi}_I \psi_I^2 \right)}_{\rightarrow \,\, \dd B^{(3)}/\dd t} + \underbrace{ 8 c \dot{\psi}_I ^2 \psi_I^2 }_{\rightarrow \,\, U^{(4)}_{\mathrm{from}\,\mathcal{L}^{(3)}}} \right] \,.
\end{equation}
In particular, in this expression, we have used that $E^I = \int \mathcal{E}(\psi_I,\dot{\psi}_I) =0$.
Once more, we see that at cubic order the interaction Hamiltonian is given by minus the Lagrangian expressed in terms of interaction picture fields, while a new interaction is generated at quartic order.
Note also that keeping in the initial cubic Lagrangian density, the terms proportional to the linear equation of motion, was crucial to correctly define the momentum and the Hamiltonian of the theory.
Moreover in this example, one can observe the qualitative differences between the full and the interaction Hamiltonians.

\paragraph{Effective interactions.}
Having identified a total time derivative, $\dd B/\dd t$, and usual interactions, $U$, we can proceed with the computation of the effective interactions $\tilde{U}$ in an equivalent theory without total time derivatives.
The derivation is complicated by the fact that $B$ does not commute neither with $U$, nor with $\dd B/\dd t$, nor with their commutators with itself, $[B, U]$ and $[B,\dd B / \dd t]$.
However, we have already truncated interactions at the quartic order in the full Hamiltonian, so it would be inconsistent to proceed more generally at a later stage.
Therefore, we consistently compute effective interactions up to quartic order in interaction picture fields and their time derivatives.
Given the interactions at hand, this calculation only requires the two-vertices order expression for $\tilde{U}$ of Eq.~\eqref{eq: effective O and U 2-vertices order}.
We find $i [\hat{B},\hat{U}] $ to yield a quintic order interaction, and therefore to be negligible.
We also find
\begin{align}
    \frac{i}{2} [B, \dd B/\dd t](t) &= \frac{i c^2(t) }{2}\int \dd^3 \vec{x}   \int \dd^3 \vec{y}   \left[ \dot{\psi}_I(t,\vec{x}) \psi^2_I(t,\vec{x})  , 2 \dot{\psi}^2_I(t,\vec{y})  \psi_I(t,\vec{y})  + \psi^2_I(t,\vec{y}) \partial^2 \psi_I(t,\vec{y})\right] \nonumber \\
    & = - 3 c(t) \int \dd^3 \vec{x} \, \dot{\psi}^2_I(t,\vec{x})  \psi^2_I(t,\vec{x}) + \frac{5}{3} c(t)\int \dd^3 \vec{x} \, \psi^3_I(t,\vec{x})\partial^2 \psi_I(t,\vec{x})  \,.
\end{align}
Up to quartic order, we therefore have:
\begin{equation}
    \tilde{U} =  U + \frac{i}{2} [B, \dd B/\dd t] =  \int \dd^3 \vec{x} \,\left(5 c   \dot{\psi}^2_I \psi^2_I  + \frac{5c}{3}\psi_I^3\partial^2 \psi_I \right)\,.
\end{equation}
Note that, quite strikingly, these two interactions can be further simplified as a total time derivative using the linear equation of motion for $\psi_I$:
\begin{equation}
    \tilde{U} = \int \dd^3 \vec{x} \left[  \frac{\dd}{\dd t}\left(\frac{5c}{3} \dot{\psi}_I \psi_I^3 \right)\right]\,,
\end{equation}
Identifying this term as total time derivative interaction $\dd B^{(4)}/\dd t$, we can iterate the calculation of effective interactions without it, and we simply find
\begin{equation}
\label{eq: tilde tilde U toy model 2}
    \tilde{\tilde{U}} = 0 + \ldots\,,
\end{equation}
where, once more, we truncated at quartic order.

We now want to compute the effective external operator $\tilde{\mathcal{O}}$---respectively $\tilde{\tilde{\mathcal{O}}}$---corresponding to the theory without the cubic total time derivative---respectively without cubic nor quartic total time derivatives.
This time, since $B$-terms involve time derivatives of the fields, their commutators with $\mathcal{O}$ are non-trivial even when the latter is only composed of fields.
We therefore separate the calculations for different correlation functions.
\paragraph{Bispectrum.} We first compute the tree-level three-point function of the theory, in Fourier space, i.e.
\begin{equation}
    \mathcal{O} = \psi_{\vec{k}_1} \psi_{\vec{k}_2} \psi_{\vec{k}_3}\,.  
\end{equation}
Only the interaction Hamiltonian up to cubic order is relevant.
Since $U^{(3)} = 0 $, only the effect of $\dd B^{(3)}/\dd t$ is relevant.
Using Eq.~\eqref{eq: effective O and U 2-vertices order} it is straightforward to find that the tree-level bispectrum can be found from an equivalent theory without interactions at all ($\tilde{U}^{(3)}=0$), but with a modified external operator
\begin{align}
    \Braket{\mathcal{O}}_{B^{(3)}} &= \Braket{\tilde{\mathcal{O}}}_{0}  \\
    &= i \Braket{0|\int \left(\prod_{i=1}^3 \frac{\dd^3 \vec{q}_i}{(2\pi)^3} \right)\delta^{(3)}\left(\sum_{i=1}^3\vec{q}_i\right) \left[ - c \dot{\psi}_{I}^{\vec{q}_1} \psi_{I}^{\vec{q}_2} \psi_{I}^{\vec{q}_3} , \psi_{I}^{\vec{k}_1} \psi_{I}^{\vec{k}_2} \psi_{I}^{\vec{k}_3} \right]|0}\,.
\end{align}
Taking into account permutations, and defining the bispectrum as
\begin{equation}
   \Braket{\psi_{\vec{k}_1} \psi_{\vec{k}_2} \psi_{\vec{k}_3}} = (2\pi)^3\delta^{(3)}\left(\sum_{i=1}^3\vec{k}_i\right)   B_\psi(k_1,k_2,k_3) \,,
\end{equation}
we find:
\begin{equation}
\label{eq: bispectrum toy model 2}
    B_\psi(k_1,k_2,k_3)  = - 2 \left[ P_\psi(k_1)P_\psi(k_2) + P_\psi(k_2)P_\psi(k_3)  + P_\psi(k_3)P_\psi(k_1) \right] \,,
\end{equation}
where we have defined $P_\psi(k)$ the two-point function of interaction picture fields, $\Braket{\psi_I^{\vec{k}} \psi_I ^{\vec{k}^\prime}}=(2\pi)^3 \delta^{(3)}(\vec{k}+\vec{k}^\prime) P_\psi(k)$.
This bispectrum corresponds to a \textit{local} shape, with an amplitude $f_\mathrm{NL}^\mathrm{loc} = -5/3$.
Note that we used only the 1-vertex order of the effective external operator $\tilde{O}$, even though in principle higher orders like $[B,[B,\mathcal{O}]]$, etc., also generate non-zero contributions.
One can check, however, that those additional terms all contribute at a higher loop level.
It is therefore consistent to neglect them at tree level, which we have assumed any way to consider only cubic (and not quintic, etc.) interactions to start with.

\paragraph{Trispectrum.}
We now turn to the computation of the trispectrum, the connected four-point correlation function.
It is the lowest $n$-point function for which non-trivial effects of the total time derivatives, and Lagrangian versus Hamiltonian interactions, can be seen at tree level.
We consider
\begin{equation}
    \mathcal{O} = \psi_{\vec{k}_1} \psi_{\vec{k}_2} \psi_{\vec{k}_3} \psi_{\vec{k}_4}\,,
\end{equation}
and we remind that we are interested in the connected piece only:
\begin{align}
    \Braket{\psi_{\vec{k}_1} \psi_{\vec{k}_2} \psi_{\vec{k}_3} \psi_{\vec{k}_4}}_c &= \Braket{\psi_{\vec{k}_1} \psi_{\vec{k}_2} \psi_{\vec{k}_3} \psi_{\vec{k}_4}} - \left[P_\psi(k_1) P_\psi(k_2) \delta^{(3)}(\vec{k}_1+\vec{k}_3)\delta^{(3)}(\vec{k}_2+\vec{k}_4) + \text{2 perm.}\right]  \nonumber \\
    & = (2\pi)^3 \delta^{(3)}\left(\sum_{i=1}^4\vec{k}_i\right)   T_\psi(\vec{k}_1,\vec{k}_2,\vec{k}_3,\vec{k}_4) \,, 
\end{align}
where the second line serves as a definition for the trispectrum, which in general can be a function of the four vectors $\vec{k}_i$, and not just their norms.
The shortest route to the final result consists in considering the effective interaction $\tilde{\tilde{U}}$ in Eq.~\eqref{eq: tilde tilde U toy model 2}, coming from both cubic and quartic total time derivatives $B = B^{(3)} + B^{(4)} = - c \dot{\psi}_I \psi_I^2 + 5  c \dot{\psi} \psi^3 /3 $.
In that case, we read from Eq.~\eqref{eq: effective O and U 2-vertices order} that $\tilde{\tilde{O}}$ receives two corrections, and therefore that
\begin{align}
    \Braket{\mathcal{O}}_{U, B^{(3)}} =& \Braket{\tilde{\mathcal{O}}}_{\tilde{U}=\dd B^{(4)}/\dd t}=  \Braket{\tilde{\tilde{\mathcal{O}}}}_{0} \\
    =&\,\Braket{0|\mathcal{O}|0} +  i \Braket{0|[B^{(4)},\mathcal{O}]|0} - \frac{1}{2}\Braket{0|[B^{(3)},[B^{(3)},\mathcal{O}]]|0} \nonumber \\ &  + \ldots \nonumber \,
\end{align}
where we have consistently truncated the result at tree level.

The first term, $\Braket{0|\psi_{\vec{k}_1} \psi_{\vec{k}_2} \psi_{\vec{k}_3} \psi_{\vec{k}_4}|0}$, corresponds precisely to the disconnected piece of the four-point function, $P_\psi(k_1) P_\psi(k_2) \delta^{(3)}(\vec{k}_1+\vec{k}_3)\delta^{(3)}(\vec{k}_2+\vec{k}_4) + \text{2 perm}$.

The second term gives a connected contribution (we factor out the $(2\pi)^3  \delta^{(3)}(\sum_{i=1}^4\vec{k}_i)  $):
\begin{align}
    & i \Braket{0|\int \left(\prod_{i=1}^4 \frac{\dd^3 \vec{q}_i}{(2\pi)^3} \right)\delta^{(3)}\left(\sum_{i=1}^4\vec{q}_i\right) \left[  \frac{5c}{3} \dot{\psi}_{I}^{\vec{q}_1} \psi_{I}^{\vec{q}_2} \psi_{I}^{\vec{q}_3}
    \psi_{I}^{\vec{q}_4}, \psi_{I}^{\vec{k}_1} \psi_{I}^{\vec{k}_2} \psi_{I}^{\vec{k}_3}
    \psi_{I}^{\vec{k}_4}\right]|0}^\prime \nonumber \\
    = & 10 \left[P_\psi(k_1)P_\psi(k_2)P_\psi(k_3)  + \text{3 perm.} \right] \,.
\end{align}
The total of four permutations comes from choosing which of the $\vec{k}_i$'s is taken to commute with $\dot{\psi}$, also giving a factor $-i/c$.
A symmetry factor of $6=3!$ also arises from equivalent permutations of the connected contractions amongst the remaining six $\psi$, giving the prefactor $i \times (5 c/3) \times  (-i/c) \times  = 10$.
This contribution consists in a local trispectrum shape of the $g_\mathrm{NL}$ kind, as can be found from a local parameterization of non-linearities as $\psi(t,\vec{x}) = \psi_\mathrm{Gaussian}(t,\vec{x})+ (9 g_\mathrm{NL}^\mathrm{loc}/25)\psi_\mathrm{Gaussian}^3(t,\vec{x})$.

The third term has more complicated combinatorics, but it is straightforward to find that it gives
\begin{align}
    & - \frac{1}{2} \int \left(\prod_{i=1}^3 \frac{\dd^3 \vec{q}_i}{(2\pi)^3} \right)\delta^{(3)}\left(\sum_{i=1}^3\vec{q}_i\right)
    \int \left(\prod_{i=1}^3 \frac{\dd^3 \vec{p}_i}{(2\pi)^3} \right)\delta^{(3)}\left(\sum_{i=1}^3\vec{p}_i\right)  \nonumber \\
    & \quad \times\Braket{0|\left[ c
    \dot{\psi}_{I}^{\vec{q}_1} \psi_{I}^{\vec{q}_2} \psi_{I}^{\vec{q}_3}, [c\dot{\psi}_{I}^{\vec{p}_1} \psi_{I}^{\vec{p}_2} \psi_{I}^{\vec{p}_3},\psi_{I}^{\vec{k}_1} \psi_{I}^{\vec{k}_2} \psi_{I}^{\vec{k}_3}
    \psi_{I}^{\vec{k}_4}]
    \right]|0}^\prime \nonumber \\
    = &  6 \left[P_\psi(k_1)P_\psi(k_2)P_\psi(k_3)  + \text{3 perm.} \right] + 4  \left[P_\psi(k_1)P_\psi(k_2)P_\psi(k_{13})  + \text{11 perm.} \right] \,,
\end{align}
where $k_{13}=|\vec{k}_1+\vec{k}_3|$, etc., and the 12 permutations correspond to the $4!$ possible permutations of the ordered four $\vec{k}_i$, with half of them being explicitly equal.
This corresponds to a local trispectrum shape, with both a $g_\mathrm{NL}$ contribution and a $\tau_\mathrm{NL}$ contribution, as can be found from a local parameterization of non-linearities as $\psi(t,\vec{x}) = \psi_\mathrm{Gaussian}(t,\vec{x})+ (3 f_\mathrm{NL}^\mathrm{loc} /5) \left(\psi_\mathrm{Gaussian}^2(t,\vec{x})-\Braket{\psi_\mathrm{Gaussian}^2(t,\vec{x})}\right) + (9 g_\mathrm{NL}^\mathrm{loc}/25)\psi_\mathrm{Gaussian}^3(t,\vec{x})$.
Together with the previous contribution, we find the total shape of the trispectrum to be local, $T_\psi = T_\psi^\mathrm{loc} $, with
\begin{align}
\label{eq: trispectrum toy model 2}
     T_\psi^\mathrm{loc}  &= \tau_\mathrm{NL}  \left[P_\psi(k_1)P_\psi(k_2)P_\psi(k_{13})  + \text{11 perm.} \right] +\frac{54}{25} g_\mathrm{NL} \left[P_\psi(k_1)P_\psi(k_2)P_\psi(k_3)  + \text{3 perm.} \right] \nonumber  \\
    \text{with} \quad \tau_\mathrm{NL} & = \left( \frac{6 f_\mathrm{NL}^\mathrm{loc}}{5}\right)^2= 4 \quad \text{and} \quad g_\mathrm{NL} =  g_\mathrm{NL}^\mathrm{loc} = \frac{200}{27} \,. 
\end{align}
Note that, as expected, we recovered the single-field consistency relation for local non-Gaussianities, relating the exchange trispectrum to the bispectrum through $\tau_\mathrm{NL} = \left(6 f_\mathrm{NL}^\mathrm{loc}/5\right)^2$.

The conclusion is that an interaction in the form of a total time derivative of a function including a time derivative of the field does contribute to correlation functions, a priori both as a boundary term and as effective interactions to be integrated over the cosmic history.
We have shown in practice how to compute correlation functions in theories involving these interactions, where considering different vertices-orders of the in-in perturbation theory was crucial to get a consistent result.
Although suitable to perform concrete calculations, it would be desirable to have a simpler framework to deal with these interactions.
This is the point of Section~\ref{sec: canonical transformations}.
Before that, we comment on another version of the in-in formalism with total time derivative interactions.

\subsection{Lagrangian path integral formulation}
\label{subsec: SK formalism}

Another formulation of the in-in, equivalent to the operator formalism that is the main focus of this work, is provided by the in-in path integral approach~\cite{Schwinger:1960qe,Keldysh:1964ud}, sometimes dubbed ``Schwinger-Keldysh'' to honor the authors of these references.
In this approach, one manipulates classical field configurations instead of quantum operators.
However, as we quickly review in the following, interactions are still specified by operators acting as derivatives with respect to external currents, which also leads to a number of subtleties.
The Schwinger-Keldysh, in-in, partition function with external currents can be derived from first principles; in perturbation theory it reads
\begin{align}
\label{eq: in-in partition function}
    \mathcal{Z}_{\mathrm{in}-\mathrm{in}}\left[J^{\pm},K^{\pm}\right] = \exp&{\left\{- i \int \dd t \int  \dd^3 \vec{x}  \left(\mathcal{H}_\mathrm{int}\left[\psi^+ \rightarrow\frac{\delta}{i \delta J_+},p_{\psi}^+ \rightarrow\frac{\delta}{i \delta K+}\right]-(+ \leftrightarrow -)\right)\right\} } \nonumber \\
    & \times \mathcal{Z}_\mathrm{in-in}^\mathrm{free}\left[J^{\pm},K^{\pm}\right]  \nonumber \,,\\
    \text{with}\,\, \mathcal{Z}_\mathrm{in-in}^\mathrm{free}\left[J^{\pm},K^{\pm}\right] =  & \int_{\mathcal{C}(t,-\infty^\pm)} \mathcal{D} \psi^\pm \mathcal{D} p_{\psi}^\pm  
    \delta(\psi^+(t)-\psi^-(t)) \\ 
    &  \times \exp\left\{i \int \dd t \int  \dd^3 \vec{x}  \left[p_\psi^+\dot{\psi}^+ - \mathcal{H}_\mathrm{free}^+ +J^+\psi^++K^+p_\psi^+-(+\leftrightarrow -)\right]\right\} \nonumber \,. \nonumber
\end{align}
The path integral follows a closed time path from $-\infty$ (with slight deformations in the complex plane in order to implement the $i \epsilon$-prescriptions) to the external time $t$ at which one wants to compute observables, and back to $-\infty$, as denoted by $\mathcal{C}(t,-\infty^\pm)$.
Fields and momenta in each of the two branches, forward time evolution and backwards one, are independent in the bulk of the closed time path, hence we have defined them with a superscript $\pm$.
Importantly, at the external time $t$, the two branches of the path integral are sewn; this means that field configurations $\psi^\pm$ (and not momenta) need to coincide at $t$ and that the degrees of freedom $+$ and $-$ are actually not completely independent.
We have already defined the path integral in a perturbation theory, with the total Hamiltonian divided into a free part and an interacting one, as in the operator formalism.
The free partition function can be used to define propagators of the free theory, and derivatives with respect to the external currents $J^\pm$ and $K^\pm$, generated by the interaction Hamiltonian, bring down respectively powers of $\psi^\pm$ and $p_\psi^\pm$ in the path integral.
These operations can be neatly encompassed by a diagrammatic representation with vertices and propagators.

But before that, and importantly for the matter of this work given the aforementioned complications with the Hamiltonian operator formalism when the Lagrangian contains total time derivative interactions, we define a simplified path integral approach at the level of the interaction Lagrangian.
When the (full) Hamiltonian is only quadratic in the momenta $p_\psi$, one can explicitly integrate over them in the closed-time-path integral, giving schematically $\int \mathcal{D}\psi \mathcal{D}p_\psi e^{i \int \left( p_\psi \dot{\psi} - \mathcal{H}\right)}  \rightarrow \int \mathcal{D}^\prime \psi e^{i \int \mathcal{L}}$, where $\mathcal{L}$ is the Lagrangian density of the theory.
Although this procedure can be performed exactly, we are interested in situations where the Hamiltonian density may contain cubic or quartic interactions with momenta.
Fortunately for us, it was proved that up to quartic order in the momenta, the path integral over momenta can still be performed perturbatively, leading to~\cite{Chen:2017ryl}
(see also~\cite{Prokopec:2010be})
\begin{align}
\label{eq:path integral with Lint}
    \mathcal{Z}_\mathrm{in-in}\left[J^{\pm}\right] = \exp&{\left\{i \int \dd t \int  \dd^3 \vec{x}  \left(\mathcal{L}_\mathrm{int}\left[\psi^+ \rightarrow\frac{\delta}{i \delta J_+}\right]-(+ \leftrightarrow -)\right)\right\} }  \mathcal{Z}_\mathrm{in-in}^\mathrm{free}\left[J^{\pm}\right]  \nonumber\,,\\
    \text{with}\,\, \mathcal{Z}_\mathrm{in-in}^\mathrm{free}\left[J^{\pm}\right] =  & \int_{\mathcal{C}(t,-\infty^\pm)} \mathcal{D}^\prime \psi^\pm  
    \delta(\psi^+(t)-\psi^-(t))  \\ 
    &  \times 
    \exp\left\{i \int \dd t \int  \dd^3 \vec{x}  \left[\mathcal{L}_\mathrm{free}^+ +J^+\psi^+-(+\leftrightarrow -)\right]\right\} \nonumber \,, \nonumber
\end{align}
where $\mathcal{L}_\mathrm{free}$ is the free Lagrangian and $\mathcal{L}_\mathrm{int}$ the interaction one.
In the following, we will closely follow the notations of Ref.~\cite{Chen:2017ryl}.
In particular, from now on, we switch to the notation $t\rightarrow \tau$ for the time variable, $\dot{f}\rightarrow f^\prime$ for the time derivative and we consider $\tau=0$ as the time at which observables are sought for, having in mind conformal time in the cosmological context and in order to match the literature.
Propagators of the free theory are then defined as Fourier transforms of the (anti) time-ordered two point functions in real space:
\begin{equation}
    \forall \mathsf{a}, \mathsf{b} \in \{+,-\}\,,\,\,\, - i \Delta_{\mathsf{a}\mathsf{b}}(\tau_1,\vec{x}_1;\tau_2,\vec{x}_2) = \frac{\delta}{i\mathsf{a} \delta J^\mathsf{a}(\tau_1,\vec{x}_1)}\frac{\delta}{i\mathsf{b} \delta J^\mathsf{b}(\tau_2,\vec{x}_2)} \mathcal{Z}^\mathrm{free}_\mathrm{in-in}[J^\pm],
\end{equation}
as
\begin{equation}
    G_{\mathsf{a}\mathsf{b}}(k;\tau_1,\tau_2)= - i \int \dd^3 \vec{x} e^{-i \vec{k} \cdot \vec{x}}\Delta_{\mathsf{a}\mathsf{b}}(\tau_1,\vec{x};\tau_2,\vec{0}).
\end{equation}
These $\pm$ propagators can then be expressed in terms of the causal bulk-to-bulk propagators
\begin{equation}
    \label{eq: causal propag}
    G_>(k;\tau_1,\tau_2) = \psi_k(\tau_1)\psi_k^*(\tau_2) \,,\,\, G_<(k;\tau_1,\tau_2) = \psi_k^*(\tau_1)\psi_k(\tau_2)\,,
\end{equation}
where $\psi_k(\tau)$ are the Fourier-space mode functions whose dynamics are dictated by the free Lagrangian $\mathcal{L}_\mathrm{free}$, as
\begin{align}
    G_{++}(k;\tau_1,\tau_2) &=\theta(\tau_1-\tau_2)G_>(k;\tau_1,\tau_2)+\theta(\tau_2-\tau_1)G_<(k;\tau_1,\tau_2) \\
    G_{+-}(k;\tau_1,\tau_2) &=G_<(k;\tau_1,\tau_2) \nonumber \\
    G_{-+}(k;\tau_1,\tau_2) &=G_>(k;\tau_1,\tau_2)\nonumber \\
    G_{--}(k;\tau_1,\tau_2) &=\theta(\tau_1-\tau_2)G_<(k;\tau_1,\tau_2)+\theta(\tau_2-\tau_1)G_>(k;\tau_1,\tau_2)\nonumber \,.
\end{align}
We find it useful to also define bulk-to-boundary propagators $\tilde{G}$ as 
\begin{equation}
    G_{\pm\pm}(k;\tau_1,0)=G_{\pm\mp}(k;\tau_1,0)=G_{\lessgtr}(k;\tau_1,0)\equiv \tilde{G}_\pm(k,\tau_1) \,.
\end{equation}
Note that when both times are evaluated at the boundary all these propagators simply reduce to the final two-point function of $\psi$, $P_\psi(k)$.
Vertices of the perturbation theory arise from $\mathcal{L}_\mathrm{int}$.
A vertex on the branch $\mathsf{a}\in\{+,-\}$ at $\tau_i$ with one power of $\psi$ in it will bring down a power of $\psi^\mathsf{a}(\tau_i)$, to be contracted with an other $\psi^\mathsf{b}(\tau_j)$, either from another vertex at $\tau_j$ on the branch $\mathsf{b}$ or with an external field with $\tau_j = 0$, leading to respectively a bulk-to-bulk propagator $G_{\mathsf{a}\mathsf{b}}(k;\tau_i,\tau_j)$ or a bulk-to-boundary one $\tilde{G}_{\mathsf{a}}(k,\tau_i)$.
Interactions with spatial derivatives are simply given by suitable multiplications of the propagators by wavenumbers $\vec{k}$.
Interactions with time derivatives are slightly more subtle (strictly speaking they should be taken care of in a Hamiltonian path integral approach), however they can be understood as the operator multiplication $\psi^{\pm\prime} \rightarrow \partial_\tau \cdot \left(- i \delta / \delta J^\pm \right) \cdot $, to be applied to the left of a functional.
Therefore a vertex at $\tau_i$ on the branch $\mathsf{a}$ with one power of $\psi^\prime$ in it will bring down a power of $\psi^{\mathsf{a} \prime}(\tau_i)$ giving, after suitable contraction with another field, a partial time derivative of a propagator: $\partial_{\tau_i} G_{\mathsf{a}\mathsf{b}}(k;\tau_i,\tau_j)$ or $\partial_{\tau_i} \tilde{G}_{\mathsf{a}}(k,\tau_i)$.
One should note, however, that although time derivatives of bulk-to-bulk propagators $\mp\pm$ reduce to time derivatives of the causal propagators $G_{\gtrless}$ (as do the bulk-to-boundary ones by definition), time derivatives of the $\pm\pm$ bulk-to-bulk propagators are more subtle as one needs to take into account derivatives of the Heaviside distributions $\theta$. 
For example, one can show that 
the following equalities hold
in the sense of distributions (i.e. once integrated over the two time variables with test functions $f(\tau_1)$ and $g(\tau_2)$),
\begin{align}
\label{eq: derivatives of internal propagators}
  \partial_{\tau_i}G_{\pm\pm}(k;\tau_1,\tau_2) = & \, \theta(\tau_1-\tau_2) \partial_{\tau_i }G_{\gtrless}(k;\tau_1,\tau_2)+\theta(\tau_2-\tau_1) \partial_{\tau_i }G_{\lessgtr}(k;\tau_1,\tau_2)  \\
  & + (-1)^{i-1} \delta(\tau_1-\tau_2) (G_{\gtrless}(k;\tau_1,\tau_2)-G_{\lessgtr}(k;\tau_1,\tau_2))\,, \nonumber \\
   \partial^2_{\tau_i\tau_j}G_{\pm\pm}(k;\tau_1,\tau_2) = & \,\theta(\tau_1-\tau_2) \partial^2_{\tau_i \tau_j }G_{\gtrless}(k;\tau_1,\tau_2) \nonumber +\theta(\tau_2-\tau_1) \partial^2_{\tau_i  \tau_j }G_{\lessgtr}(k;\tau_1,\tau_2) \nonumber \\ & \pm (-1)^{i+j-1} \frac{i}{c(\tau_1)}\delta(\tau_1-\tau_2)\nonumber \,,
\end{align}
valid $\forall i,j \in \{1,2\}$.
Note that, if no other derivative operator is acting on them, the second line in single-derivatives proportional to $\delta(\tau_1-\tau_2)$ vanishes identically, since $G_<(k;\tau,\tau) = G_>(k;\tau,\tau)$.
Moreover, as before $c(\tau)$ denotes the normalisation of $\psi$ in the free Lagrangian, see Eq.~\eqref{eq: Lfree toy models}, and for which we used the Wronskian condition $W(k,\tau)=\psi_k^{\prime*}(\tau)\psi_k(\tau) - \psi_k^\prime(\tau) \psi_k^*(\tau) = i / c(\tau) $ from the canonical commutation relation and the linear momentum definition valid for the free theory defining the propagators. 

There is one last ingredient we need in order to compute, in the path integral approach, correlation functions for our theory with total time derivative interactions that include powers of the time derivatives of the fields.
As we have seen above, a general feature of such interactions, arising from integration by parts, is that they always come in pairs with interactions proportional to the linear equations of motion.
While, as previously discussed, vertices from such interactions are zero when evaluated on interaction picture fields in the in-in operator formalism, they must be carefully taken into account in the path integral approach, as we now explain.
Let us define the Fourier space equation of motion operator as
\begin{equation}
\label{eq:EOM_operator}
\EOM(k,\,\tau) f(k,\,\tau)\equiv\partial_\tau\left[ c (\tau) \partial_\tau f(k,\tau)\right] + c(\tau) k^2 f(k,\tau).
\end{equation}
By definition of the free theory in the path integral approach, the causal propagators identically verify these equations of motion:
\begin{equation}
    \forall i \in \{1,2\}\,,\,\, \EOM(k,\,\tau_i) G_{\gtrless}(k;\tau_1,\tau_2) = 0 \,.
\end{equation}
Therefore, when this operator acts on a bulk-to-boundary propagator, the corresponding contribution vanishes.
However, when the operator acts on a bulk-to-bulk propagator, its effect is highly non-trivial.
Indeed, once more in the sense of distributions, the following properties hold:
\begin{align}
\label{eq: EOM of internal propagators}
  \forall i \in \{1,2\} \,,\,\, \EOM(k,\,\tau_i)G_{\pm\mp}(k;\tau_1,\tau_2) = & \,0 \,, \\ 
  \forall i \in \{1,2\} \,,\,\,  \EOM(k,\,\tau_i)G_{\pm\pm}(k;\tau_1,\tau_2) =& \mp i\,\frac{\delta(\tau_1-\tau_2)}{c(\tau_1)}  \,, \nonumber \\
  \int \dd \tau_1  f(\tau_1)\int  \dd \tau_2 g(\tau_2) \EOM(k,\,\tau_2)\EOM(k,\,\tau_1)G_{\pm\pm}(k;\tau_1,\tau_2) = & \nonumber \\  \mp i  \int \dd \tau_1  f(\tau_1)\int  \dd \tau_2 \delta(\tau_1-\tau_2) & \EOM(k,\,\tau_2) g(\tau_2)  \,. \nonumber
\end{align}
In the last equality, we made explicit the time integration with test functions, in order to show the effect of the EOM operator.
Keeping track of this subtle contribution from interactions proportional to the linear equations of motion is crucial to obtain the correct result in the path integral form of the in-in formalism, as we will prove explicitly for our toy model that contains them. 
To our knowledge, this is the first time that the importance of interaction terms proportional to linear equations of motion is stressed.
We explained their roles both in the operator formalism in order to consistently derive the Hamiltonian from the Lagrangian, and here in the path integral approach not to miss any contribution.

Let us introduce here some diagrammatic rules to compute correlation functions. 
Following the notations of Ref.~\cite{Chen:2017ryl}, we use a black and a white dot to denote $+$ and $-$ vertices respectively. The bulk-to-bulk propagators are represented as:
\begin{align}
&\begin{tikzpicture}[line width=1.pt, scale=2]
\vspace*{-2cm}
\draw[black] (0.5, 0) -- (1, 0);
\draw[fill=black] (0.5, 0) circle (.03cm);
\draw[fill=black] (1, 0) circle (.03cm);
\node at (0.5, 0.2) {$\tau_1$};
\node at (1, 0.2) {$\tau_2$};
\end{tikzpicture}
\,=\,
G_{++}(k;\tau_1,\tau_2)\,, \nonumber\\
&\begin{tikzpicture}[line width=1.pt, scale=2]
\vspace*{-2cm}
\draw[black] (0.5, 0) -- (1, 0);
\draw[fill=black] (0.5, 0) circle (.03cm);
\draw[fill=white] (1, 0) circle (.03cm);
\node at (0.5, 0.2) {$\tau_1$};
\node at (1, 0.2) {$\tau_2$};
\end{tikzpicture}
\,=\,
G_{+-}(k;\tau_1,\tau_2)\,, \nonumber\\&\begin{tikzpicture}[line width=1.pt, scale=2]
\vspace*{-2cm}
\draw[black] (0.5, 0) -- (1, 0);
\draw[fill=white] (0.5, 0) circle (.03cm);
\draw[fill=black] (1, 0) circle (.03cm);
\node at (0.5, 0.2) {$\tau_1$};
\node at (1, 0.2) {$\tau_2$};
\end{tikzpicture}
\,=\,
G_{-+}(k;\tau_1,\tau_2)\,, \nonumber\\&\begin{tikzpicture}[line width=1.pt, scale=2]
\vspace*{-2cm}
\draw[black] (0.5, 0) -- (1, 0);
\draw[fill=white] (0.5, 0) circle (.03cm);
\draw[fill=white] (1, 0) circle (.03cm);
\node at (0.5, 0.2) {$\tau_1$};
\node at (1, 0.2) {$\tau_2$};
\end{tikzpicture}
\,=\,
G_{--}(k;\tau_1,\tau_2)\,. \nonumber
\end{align}
Furthermore, denoting the boundary with a square, we can also write the bulk-to-boundary propagators
\begin{align}
&\begin{tikzpicture}[line width=1.pt, scale=2]
\vspace*{-2cm}
\draw[black] (0.5, 0) -- (1, 0);
\draw[fill=black] (0.5, 0) circle (.03cm);
\draw[fill=white] (1, 0-0.03) rectangle +(0.06,0.06);
\node at (0.5, 0.2) {$\tau$};
\end{tikzpicture}
\,=\,
\tilde{G}_{+}(k,\,\tau)\,, \nonumber\\
&\begin{tikzpicture}[line width=1.pt, scale=2]
\vspace*{-2cm}
\draw[black] (0.5, 0) -- (1, 0);
\draw[fill=white] (0.5, 0) circle (.03cm);
\draw[fill=white] (1, 0-0.03) rectangle +(0.06,0.06);
\node at (0.5, 0.2) {$\tau$};
\end{tikzpicture}
\,=\,
\tilde{G}_{-}(k,\tau)\,.
\end{align}
Diagrams with circles filled in gray denote that the sum over $\pm$ values of the vertex is taken: $\begin{tikzpicture}[line width=1.pt, scale=2]
\draw[pattern=horizontal lines gray] (0.5, 0) circle (.03cm);
\end{tikzpicture} = \begin{tikzpicture}[line width=1.pt, scale=2]
\draw[fill=black] (0.5, 0) circle (.03cm);
\end{tikzpicture} + \begin{tikzpicture}[line width=1.pt, scale=2]
\draw[fill=white] (0.5, 0) circle (.03cm);
\end{tikzpicture}$ .

Equipped with the tools just introduced, we now compute correlation functions for the two toy models presented in Sec.~\ref{subsec: toy models}.

\subsubsection{Toy model 1}

The calculation of correlation functions made of fields only in the first toy model, whose interactions were specified in Eq.~\eqref{eq: toy model 1}, can be carried relatively easily in this approach.
Indeed, one needs not define the Hamiltonian nor look for cancellations between different orders of the perturbation theory.

First, as already stressed in the Hamiltonian operator formalism, the easiest way to show that interactions in the form of total time derivatives of functions of fields only, do not contribute to correlation functions of fields, is to treat them in the free theory.
In the Lagrangian path integral formulation, this can be done by using Eq.~\eqref{eq:path integral with Lint} and incorporating the total time derivative in $\mathcal{L}_\mathrm{free}$ in the free partition function with external currents.
This way, those interactions completely disappear from the perturbation theory and cannot be used to form vertices, and their only possible effect is through the mode functions $\psi_k(\tau)$ appearing in the causal propagators~\eqref{eq: causal propag}.
However, these total time derivatives do not affect the equations of motion, as already shown explicitly in Eq.~\eqref{eq: second order EoM toy model}.
This concludes this proof, which is by far the simplest one.

Would one insist in treating those total time derivatives as interactions in the perturbation theory, they could still be shown to give vanishing contributions.
An argument sketched in~\cite{Fumagalli:2023loc} can already give us this intuition.
Here we make this proof explicit, though underlining its implicit assumptions.
We use again Eq.~\eqref{eq:path integral with Lint} but with the total time derivative interaction in $\mathcal{L}_\mathrm{int}$.
The effect of such a vertex, when combined with other vertices left implicit, is found from the diagrammatic rules of Ref.~\cite{Chen:2017ryl} $$ - i \sum_{\mathsf{d}\in\{+,-\}}  \mathsf{a} \mathsf{b}  \mathsf{c}\mathsf{d}  \int \dd \tau \partial_\tau \left[- \alpha(\tau) G_{\mathsf{d}\mathsf{a}}(k_1;\tau,\tau_1) G_{\mathsf{d}\mathsf{b}}(k_2;\tau,\tau_2) G_{\mathsf{d}\mathsf{c}}(k_3;\tau,\tau_3) \right]\,,$$ as can be seen by acting on the free partition function with the operator $\partial_\tau \left[ i \alpha(\tau) \left(\delta/\delta J^\mathsf{d}(\tau,\vec{x})\right)^3\right]$ and going to Fourier space.
By integrating this operator over time in the path integral of the perturbation theory, using vanishing of interactions at $-\infty^\pm$ as prescribed by the deformed time contour, one finds:
\begin{equation}
    \mathrm{exp}\left\{ i \,  \alpha(0) \int \dd^3 \vec{x}  \left[ \left(\frac{\delta}{\delta J^{+}(0,\vec{x})}\right)^3 - \left(\frac{\delta}{\delta J^{-}(0,\vec{x})}\right)^3 \right]\right\} \mathcal{Z}^\mathrm{free}_\mathrm{in-in}\left[J^\pm\right] \,,
\end{equation}
which can only bring down powers of $i\,\alpha(0)\left[\left(\psi^+(0,\vec{x})\right)^3-\left(\psi^-(0,\vec{x})\right)^3\right]$ in the path integral.
Since $\pm$ fields coincide at the external time, here $\tau=0$, where the time path closes, this proves the vanishing of all diagrams including at least one power of the total time derivative of a function made of fields only.
Importantly, we stress that the proof above relies on the implicit assumption that the action of the derivative operator at time $\tau$ commutes with the time integral over $\tau$.
This hypothesis is not obviously correct, as we already stressed that internal $\pm\pm$ propagators are not strictly speaking functions of time, but distributions.
We have shown that their time derivatives must be taken with care as they involve derivatives of the Heaviside distributions, making for non-trivial contributions.
In order to reinforce the intuition that we have just developed, and to incorporate the effect of mixing total time derivative interactions with strength $\alpha$ to normal ones with $\beta$ in Eq.~\eqref{eq: toy model 1}, we now turn to the explicit calculation of the first two relevant correlation functions for this toy model: the bispectrum and the exchange trispectrum.
The latter case makes for an important check as it involves permutations with two time derivatives of the internal propagator.

At one-vertex order, only the three-point function of $\psi$ is relevant. 
Following the diagrammatic rules of~\cite{Chen:2017ryl}, it is immediate to realize that the result, proportional to $$- i \sum_{\mathsf{a}\in \{+,-\}} \mathsf{a} \int \dd \tau \partial_\tau \left(\alpha(\tau) \tilde{G}_{\mathsf{a}}(k_1,\tau) \tilde{G}_{\mathsf{a}}(k_2,\tau)\tilde{G}_{\mathsf{a}}(k_3,\tau)\right)$$ and expressible in terms of the causal propagators $G_{\gtrless}$ only, vanishes after performing the time integral and summing over $\pm$ vertices.
As for the second interaction $\propto \beta$, one concludes that it can only contribute as $\propto \dot{\beta}$, following a similar reasoning.
Subtleties start arising at the two-vertices order at which internal propagators $\pm\pm$ with Heaviside distributions appear.
There, for the reasons already stressed above, it is less obvious that time integrals can be performed without expanding the total time derivatives first.
We first focus on the four-point function of $\psi$ generated by two total time derivative interactions.
Overlooking symmetry factors, the result is
\begin{align*}
    - \sum_{\mathsf{a},\mathsf{b}\in \{+,-\}} \mathsf{a} \mathsf{b}  & \int_{-\infty}^0 \dd \tau_1  \int_{-\infty}^0  \dd \tau_2 \partial_{\tau_1} \left[\alpha(\tau_1) \tilde{G}_{\mathsf{a}}(k_1,\tau_1) \tilde{G}_{\mathsf{a}}(k_2,\tau_1) \partial_{\tau_2} \Bigl( \alpha(\tau_2)\right.  \\ 
    & \left.
    G_{\mathsf{a}\mathsf{b}}(|\vec{k}_1+\vec{k}_2|;\tau_1,\tau_2) \tilde{G}_{\mathsf{b}}(k_3,\tau_2)\tilde{G}_{\mathsf{b}}(k_4,\tau_2)\Bigr) \right] + \text{perm.}\,,
\end{align*}
where ``perm.'' means all other permutations of the external wavevectors $\vec{k}_i$.
The $\pm\mp$ contributions being given in terms of causal propagators $G_{\gtrless}$ only, they can be straight integrated over time.
Being real, they are equal and their sum is proportional to $$+2 \alpha^2(0) P(k_1) P(k_2) P(k_3) P(k_4) P(|\vec{k}_1+\vec{k}_2|)+ \text{perm.}$$
We turn to the $\pm\pm$ diagrams and we develop the total time derivatives being careful that, whenever they hit the internal propagator, formulas of Eq.~\eqref{eq: derivatives of internal propagators} should be used.
By rendering explicit the causal structure, one finds the integrand of the $\pm\pm$ diagrams to read:
\begin{align*}
    &\theta(\tau_1-\tau_2) \partial_{\tau_1}
    \left[\alpha(\tau_1) G_{\lessgtr}(k_1,\tau_1) G_{\lessgtr}(k_2,\tau_1) \partial_{\tau_2} \Bigl( \alpha(\tau_2)G_{\gtrless}(|\vec{k}_1+\vec{k}_2|;\tau_1,\tau_2) G_{\lessgtr}(k_3,\tau_2) G_{\lessgtr}(k_4,\tau_2)\Bigr) \right]\\
    +&\theta(\tau_2-\tau_1) \partial_{\tau_1}  \left[\alpha(\tau_1) G_{\lessgtr}(k_1,\tau_1) G_{\lessgtr}(k_2,\tau_1) \partial_{\tau_2} \Bigl( \alpha(\tau_2)G_{\lessgtr}(|\vec{k}_1+\vec{k}_2|;\tau_1,\tau_2) G_{\lessgtr}(k_3,\tau_2) G_{\lessgtr}(k_4,\tau_2)\Bigr) \right]\\
    \pm & \frac{i}{c(\tau_1)} \delta(\tau_1-\tau_2) 
    \alpha(\tau_1)
    \alpha(\tau_2)
    G_{\lessgtr}(k_1,\tau_1) G_{\lessgtr}(k_2,\tau_1) G_{\lessgtr}(k_3,\tau_2) G_{\lessgtr}(k_4,\tau_2) + \text{perm.} \,,
\end{align*}
where we wrote $G_{\lessgtr}(k_i,\tau)\equiv G_{\lessgtr}(k_i,\tau, 0)$ to avoid cluttered expressions.
Fixing $\tau_1$ as the outermost integral, integration over $\tau_2$ can be performed explicitly on the domain defined by the Heaviside distributions. 
The contribution from the first line above (with $\tau_2 \in (-\infty,\tau_1]$) cancels with the sum of the lower bound evaluation from the second line (with $\tau_2 \in [\tau_1,0]$) and the third line.
The only surviving term comes from the upper bound evaluation of the second line and is already expressed as a total time derivative with respect to $\tau_1$, leading to a result proportional to $$-2\alpha^2(0) P(k_1) P(k_2) P(k_3) P(k_4) P(|\vec{k}_1+\vec{k}_2|)+ \text{perm.}$$ after explicit integration. The factor of two comes from including both $++$ and $--$ diagrams.
Taking into account all diagrams, we proved that the four-point function of $\psi$ is not affected by two insertions of total time derivative interactions made of fields only.
We now turn to the mixed diagram with one $\alpha$-type and one $\beta$-type cubic interaction.
First, the permutations for which the time derivative of the $\beta$-diagram hits a bulk-to-boundary propagator do not contain second-order derivative of the internal propagator; the total time derivative interaction can be safely integrated over time and gives $-i\sum_{\mathsf{a}}\mathsf{a} \, \alpha(0) P(k_1) P(k_2) \times(\ldots) $ with ``$\ldots$'' being independent of $\mathsf{a}$ and they therefore vanish.
The only potentially dangerous contribution therefore comes from diagrams with the time derivative of the $\beta$-vertex hitting the internal propagator.
The $\pm\mp$ diagrams give:
\begin{equation*}
    +\partial_{\tau_1}\left[G_{\lessgtr}(k_1,\tau_1) G_{\lessgtr}(k_2,\tau_1)  \partial_{\tau_2}(G_{\lessgtr}(|\vec{k}_1+\vec{k}_2|;\tau_1, \tau_2) ) G_{\gtrless}(k_3,\tau_2) G_{\gtrless}(k_4,\tau_2) \right].
\end{equation*}
The most complicated diagrams are the $\pm\pm$ ones, and they give, after developing the total time derivatives:
\begin{align}
   & -\partial_{\tau_1}\left[G_{\lessgtr}(k_1,\tau_1) G_{\lessgtr}(k_2,\tau_1)  \partial_{\tau_2}(G_{\gtrless}(|\vec{k}_1+\vec{k}_2|;\tau_1, \tau_2) ) G_{\lessgtr}(k_3,\tau_2) G_{\lessgtr}(k_4,\tau_2) \right]\notag\\
     &   -\partial_{\tau_1}\left[G_{\lessgtr}(k_1,\tau_1) G_{\lessgtr}(k_2,\tau_1)  \partial_{\tau_2}(G_{\lessgtr}(|\vec{k}_1+\vec{k}_2|;\tau_1, \tau_2) ) G_{\lessgtr}(k_3,\tau_2) G_{\lessgtr}(k_4,\tau_2) \right]\theta(\tau_2-\tau_1)\notag\\
   & +\partial_{\tau_1}\left[G_{\lessgtr}(k_1,\tau_1) G_{\lessgtr}(k_2,\tau_1)  \partial_{\tau_2}(G_{\gtrless}(|\vec{k}_1+\vec{k}_2|;\tau_1, \tau_2) ) G_{\lessgtr}(k_3,\tau_2) G_{\lessgtr}(k_4,\tau_2) \right] \theta(\tau_2-\tau_1)\notag\\
     & \mp \frac{i}{c(\tau_2)}\delta(\tau_1-\tau_2) G_{\lessgtr}(k_1,\tau_1) G_{\lessgtr}(k_2,\tau_1)  G_{\lessgtr}(k_3,\tau_2) G_{\lessgtr}(k_4,\tau_2).
\end{align}
The first line cancels with the $\mp\pm$ diagrams upon integrating over $\dd\tau_1$. The second and third lines are total time derivatives with respect to $\tau_1$. The integral over $\tau_1$ can thus be performed easily. Upon using the Wronskian condition, the result cancels together with the fourth line.
This finishes the explicit proof that total time derivative interactions that are functions of fields only do not contribute to correlation functions of fields at all, even in the presence of internal propagators and including mixed diagrams with other interactions.
We believe the result applies for any $n$-point function of $\psi$, at any order in vertex theory and also including loop diagrams, but we find the general proof using explicitly the perturbation theory and developing the total time derivatives to be cumbersome.

\subsubsection{Toy model 2}
Let us now move to toy model 2. The corresponding Lagrangian contains the two interactions in Eq.~\eqref{eq: toy model 2}. The first one is a total time derivative, and the second one is proportional to the linear equations of motion for $\psi$, and was thus vanishing in the interaction Hamiltonian used for the calculation in section~\ref{sec:toymodel2_standardinin}. However, as explained above, we cannot neglect it a priori in the Lagrangian path integral approach. We can associate the following vertices to each of these two interactions, respectively
\begin{align}
V_1=&\vcenter{\hbox{\begin{tikzpicture}[line width=1. pt, scale=2]
\vspace*{-2cm}
\draw[black,dashed] (0.2, 0) -- (0.6, 0);
\draw[black] (0.6, 0.) -- (0.8, 0.346) ;
\draw[black]  (0.6, 0.)  -- (0.8, -0.346);
\draw[fill=black] (0.6, 0) circle (.03cm);
\node at (0.6-0.07, 0.07) {$\tau$};
\node at (0.2-0.1, 0.) {$\tau_1$};
\node at (0.8, 0.346+0.1) {$\tau_2$};
\node at (0.8, -0.346-0.1) {$\tau_3$};
\end{tikzpicture}}}\,=\,-i\mathsf{a}\mathsf{b}\mathsf{c} \int \dd\tau \partial_\tau\left\{c(\tau)\left[\partial_\tau 
G_{+\mathsf{a}}(k_1;\tau,\tau_1)\right] G_{+\mathsf{b}}(k_1;\tau,\tau_2)  G_{+\mathsf{c}}(k_3;\tau,\tau_3) \right\}\,,\label{vertex:TM2_1}\\
V_2=&\vcenter{\hbox{\begin{tikzpicture}[line width=1. pt, scale=2]
\vspace*{-2cm}
\draw[black,decorate, decoration={snake}] (0.2, 0) -- (0.6, 0);
\draw[black] (0.6, 0.) -- (0.8, 0.346) ;
\draw[black]  (0.6, 0.)  -- (0.8, -0.346);
\draw[fill=black] (0.6, 0) circle (.03cm);
\node at (0.6-0.07, 0.07) {$\tau$};
\node at (0.2-0.1, 0.) {$\tau_1$};
\node at (0.8, 0.346+0.1) {$\tau_2$};
\node at (0.8, -0.346-0.1) {$\tau_3$};
\end{tikzpicture}}}
\,=\,-i \mathsf{a}\mathsf{b}\mathsf{c} \int \dd\tau \, c(\tau)\left[\EOM(k_1,\,\tau) 
G_{+\mathsf{a}}(k_1;\tau,\tau_1)\right] G_{+\mathsf{b}}(k_1;\tau,\,\tau_2)  G_{+\mathsf{c}}(k_3;\tau,\tau_3)\label{vertex:TM2_2} \,,
\end{align}
and vertices with a white circle are related to the ones above by complex conjugation.
In these vertices, dashed or wiggly lines are here to distinguish non-equivalent permutations and denote the propagator to which is acting respectively the time derivative and $\EOM$ operators.

Since the wiggly line in the vertex~\eqref{vertex:TM2_2} is only non-zero when acting on an internal line, the calculation of the bispectrum from this vertex is just 0. The bispectrum is thus calculated solely from the first vertex~\eqref{vertex:TM2_1}, which can easily be shown to be equal to that computed in the previous section, see Eq.~\eqref{eq: bispectrum toy model 2}.

We therefore move to the calculation of the trispectrum, which is far less trivial.
In the Lagrangian path integral approach that we are embracing, only interactions in the Lagrangian are relevant.
Since our toy model has an exactly cubic Lagrangian, the trispectrum can only be given by an exchange channel made of two insertions of cubic interactions with an internal propagator.
In particular, in contrast with the operator formalism, there is no quartic contact interaction.
We now calculate all contributions.

$\bullet$ {\bf $V_1$-$V_1$.} This contribution corresponds to a diagram from two total time derivative vertices. There are 3  contributions to this diagram. The first  one can be represented schematically as
\begin{align}
&\vcenter{\hbox{\begin{tikzpicture}[line width=1. pt, scale=2]
\vspace*{-2cm}
\draw[black,dashed] (-0.2, 0) -- (0.6, 0);
\draw[black] (-0.2, 0) -- (-0.4, 0.346) ;
\draw[black]  (-0.2, 0)  -- (-0.4, -0.346);
\draw[black] (0.6, 0.) -- (0.8, 0.346) ;
\draw[black]  (0.6, 0.)  -- (0.8, -0.346);
\draw[pattern =horizontal lines gray] (-0.2, 0) circle (.03cm);
\draw[pattern =horizontal lines gray] (0.6, 0) circle (.03cm);
\draw[fill=white] (-0.4-0.03, 0.346-0.03) rectangle +(0.06,0.06);
\draw[fill=white] (-0.4-0.03, -0.346-0.03) rectangle +(0.06,0.06);
\draw[fill=white] (0.8-0.03, -0.346-0.03) rectangle +(0.06,0.06);
\draw[fill=white] (0.8-0.03, 0.346-0.03) rectangle +(0.06,0.06);
\node at (0.6-0.07, 0.12) {$\tau_2$};
\node at (-0.2+0.07, 0.12) {$\tau_1$};
\end{tikzpicture}}}
= 0 \,,
\end{align}
and is seen to vanish when summing the gray circles over $+$ and $-$. 

The second contribution is:
\begin{align}
&\vcenter{\hbox{\begin{tikzpicture}[line width=1. pt, scale=2]
\vspace*{-2cm}
\draw[black,dashed] (-0.2, 0) -- (0.2, 0);
\draw[black] (-0.2, 0) -- (-0.4, 0.346) ;
\draw[black]  (-0.2, 0)  -- (-0.4, -0.346);
\draw[black] (0.2, 0) -- (0.6, 0);
\draw[black,dashed] (0.6, 0.) -- (0.8, 0.346) ;
\draw[black]  (0.6, 0.)  -- (0.8, -0.346);
\draw[pattern =horizontal lines gray] (-0.2, 0) circle (.03cm);
\draw[pattern =horizontal lines gray] (0.6, 0) circle (.03cm);
\draw[fill=white] (-0.4-0.03, 0.346-0.03) rectangle +(0.06,0.06);
\draw[fill=white] (-0.4-0.03, -0.346-0.03) rectangle +(0.06,0.06);
\draw[fill=white] (0.8-0.03, -0.346-0.03) rectangle +(0.06,0.06);
\draw[fill=white] (0.8-0.03, 0.346-0.03) rectangle +(0.06,0.06);
\node at (0.6-0.07, 0.12) {$\tau_2$};
\node at (-0.2+0.07, 0.12) {$\tau_1$};
\end{tikzpicture}}}+\tau_1\leftrightarrow\tau_2\,\nonumber=\,-\sum_{\mathsf{a},\mathsf{b}\in\{+,-\}} \mathsf{a}\mathsf{b} \int \dd\tau_1\dd\tau_2\partial^2_{\tau_1\tau_2}\Bigl[c(\tau_1)c(\tau_2)
\end{align}
\vspace{-0.6cm}
\begin{align}
  \times  \Bigl( &
\tG_\mathsf{a}(k_1,\tau_1) 
\tG_\mathsf{a}(k_2,\tau_1) \partial_{\tau_1} G_{\mathsf{a}\mathsf{b}}(|\vec{k}_1+\vec{k}_2|;\tau_1,\tau_2) \partial_{\tau_2} \tG_\mathsf{b}(k_3,\,\tau_2)  \tG_\mathsf{b}(k_4,\tau_2) \,\,\nonumber\\
 + \, &\tG_\mathsf{a}(k_1,\tau_2) 
\tG_\mathsf{a}(k_2,\tau_2) \partial_{\tau_2} G_{\mathsf{a}\mathsf{b}}(|\vec{k}_1+\vec{k}_2|;\tau_2,\tau_1) \partial_{\tau_1} \tG_\mathsf{b}(k_3,\tau_1)  \tG_\mathsf{b}(k_4,\tau_1) +\,\,{\rm perm.} \Bigr)\Bigr]\,,\nonumber\\
&=24 \left[P_\psi(k_1)P_\psi(k_2)P_\psi(k_3)  + \text{3 perm.} \right] \,,
\end{align}
where we used the relations in Eqs.~\eqref{eq: derivatives of internal propagators}, and where ``perm.'' in the last line represent nonequivalent permutations of the external wavevectors.

The last contraction is given by the following diagram:
\begin{align}
&\vcenter{\hbox{\begin{tikzpicture}[line width=1. pt, scale=2]
\vspace*{-2cm}
\draw[black] (-0.2, 0) -- (0.2, 0);
\draw[black, dashed] (-0.2, 0) -- (-0.4, 0.346) ;
\draw[black]  (-0.2, 0)  -- (-0.4, -0.346);
\draw[black] (0.2, 0) -- (0.6, 0);
\draw[black,dashed] (0.6, 0.) -- (0.8, 0.346) ;
\draw[black]  (0.6, 0.)  -- (0.8, -0.346);
\draw[pattern =horizontal lines gray] (-0.2, 0) circle (.03cm);
\draw[pattern =horizontal lines gray] (0.6, 0) circle (.03cm);
\draw[fill=white] (-0.4-0.03, 0.346-0.03) rectangle +(0.06,0.06);
\draw[fill=white] (-0.4-0.03, -0.346-0.03) rectangle +(0.06,0.06);
\draw[fill=white] (0.8-0.03, -0.346-0.03) rectangle +(0.06,0.06);
\draw[fill=white] (0.8-0.03, 0.346-0.03) rectangle +(0.06,0.06);
\node at (0.6-0.07, 0.12) {$\tau_2$};
\node at (-0.2+0.07, 0.12) {$\tau_1$};
\end{tikzpicture}}}\,\nonumber=\,-\sum_{\mathsf{a},\mathsf{b}\in\{+,-\}}\mathsf{a}\mathsf{b} \int \dd\tau_1\dd\tau_2\partial^2_{\tau_1 \tau_2}\Bigl[c(\tau_1)c(\tau_2)
\end{align}
\vspace{-0.6cm}
\begin{align}
\times  \Bigl(  &  
\partial_{\tau_1}\tG_\mathsf{a}(k_1,\tau_1) 
\tG_\mathsf{a}(k_2,\tau_1)  G_{\mathsf{a}\mathsf{b}}(|\vec{k}_1+\vec{k}_2|;\tau_1,\tau_2) \partial_{\tau_2} \tG_\mathsf{b}(k_3,\tau_2)  \tG_\mathsf{b}(k_4,\tau_2)\,\,+\,\,{\rm perm.} \Bigr)\Bigr]\,,\nonumber\\
&=4 \left[P_\psi(k_1)P_\psi(k_2)P_\psi(k_{13})  + \text{11 perm.} \right]
\end{align}
The result from summing the contributions from these diagrams is different from the one obtained in Section~\ref{sec:toymodel2_standardinin}, which confirms the need to include diagrams with vertices $V_2$, which contain the EOM operator $\EOM$.

$\bullet$ {\bf $V_1$-$V_2$.} Contributions from mixed interactions are vanishing.

$\bullet$ {\bf $V_2$-$V_2$.} Finally, we compute the contribution to the trispectrum from two insertions of the EOM operator. Before we draw the corresponding diagram, let us mention that the calculation can be simplified using the property that, as mentioned at the beginning of this section, the operator $\EOM$ gives 0 when applied to a bulk-to-boundary propagator. For this reason, only one contraction is contributing. Furthermore, we have that $\EOM G_{\pm\mp}=0$, which implies that the only non-vanishing diagram must have both vertices on $+$, or both on $-$. Taking these considerations into account, the trispectrum is given by:
\begin{align}
&\vcenter{\hbox{\begin{tikzpicture}[line width=1. pt, scale=2]
\vspace*{-2cm}
\draw[black,decorate, decoration={snake}] (-0.2, 0) -- (0.6, 0);
\draw[black] (-0.2, 0) -- (-0.4, 0.346) ;
\draw[black]  (-0.2, 0)  -- (-0.4, -0.346);
\draw[black] (0.6, 0.) -- (0.8, 0.346) ;
\draw[black]  (0.6, 0.)  -- (0.8, -0.346);
\draw[pattern =horizontal lines gray] (-0.2, 0) circle (.03cm);
\draw[pattern =horizontal lines gray] (0.6, 0) circle (.03cm);
\draw[fill=white] (-0.4-0.03, 0.346-0.03) rectangle +(0.06,0.06);
\draw[fill=white] (-0.4-0.03, -0.346-0.03) rectangle +(0.06,0.06);
\draw[fill=white] (0.8-0.03, -0.346-0.03) rectangle +(0.06,0.06);
\draw[fill=white] (0.8-0.03, 0.346-0.03) rectangle +(0.06,0.06);
\node at (0.6-0.07, 0.12) {$\tau_2$};
\node at (-0.2+0.07, 0.12) {$\tau_1$};
\end{tikzpicture}}}\,\nonumber=\,-\sum_{\mathsf{a}\in\{+,-\}} \int \dd\tau_1\dd\tau_2 \Bigl[\tG_\mathsf{a}(k_1,\tau_1) 
\tG_\mathsf{a}(k_2,\tau_1)
\end{align}
\vspace{-0.6cm}
\begin{align}
    &  
\left(\EOM(|\vec{k}_1+\vec{k}_2|,\,\tau_2)\EOM(|\vec{k}_1+\vec{k}_2|,\tau_1)  G_{\mathsf{a}\mathsf{a}}(|\vec{k}_1+\vec{k}_2|;\,\tau_1,\tau_2)\right) \tG_\mathsf{a}(k_3,\tau_2)  \tG_\mathsf{a}(k_4,\tau_2)\,\,+\,\,{\rm perm.}\Bigr]\,,\nonumber\\
&=-8 \left[P_\psi(k_1)P_\psi(k_2)P_\psi(k_3)  + \text{3 perm.} \right].
\end{align}
In order to arrive at this result, we have used the identities derived in Eqs.~\eqref{eq: EOM of internal propagators} to express the action of two EOM operators. We have used the linear equations of motion  for the $\tilde{G}_\lessgtr$ to simplify the integrand, and performed the integral over the Dirac delta from Eq.~\eqref{eq: EOM of internal propagators}. 
The resulting integrand turns out to be a total time derivative and can be integrated straight, giving the result above.

Summing all contributions, we obtain exactly Eq.~\eqref{eq: trispectrum toy model 2}.
Our findings may not look surprising.
Indeed, the equivalence of the Lagrangian path integral approach with the Hamiltonian operator formulation of the in-in formalism was already proved in Ref.~\cite{Chen:2017ryl} up to quartic interactions and two-vertices order.
Here we extended it explicitly to total time derivatives and interactions proportional to the linear equations of motion.
Although the two versions of the in-in formalism are consistent and are both suitable for performing calculations, we finish this section with a few comments. 

First of all, the apparent simplicity with which total time derivative interactions of functions of fields only are seen to give vanishing contributions to correlation functions, should be contrasted with the absence of complete and explicit proof---valid at any vertex order and for any $n$-point function both at tree and loop levels---in the literature before our work.
Here we have shown this result with three different methods: incorporating these ``interactions'' in the free theory which is by far the simplest way (equations of motion are not affected), defining effective external operator and interactions (after cancellations, they are not affected) in the operator in-in formalism, and in the Lagrangian path integral approach by using the fact that $\pm$ fields coincide at the external time (showing the vanishing of all possible diagrams after carefully accounting for non-trivial time derivatives of internal propagators), an argument already sketched in~\cite{Fumagalli:2023loc}.
Second, time derivatives of functions that contain time derivatives of the fields contribute to correlation functions.
We have shown that these contributions are not particularly easy to take into account.
In particular, they always come in pairs with interactions proportional to the linear equations of motion that should crucially be taken into account.
In the operator formalism, we showed that they are needed to even define the Hamiltonian interactions from the Lagrangian ones, before turning to the calculation of the effective external operator and interactions; in the Lagrangian path integral approach they crucially give non-zero contributions to correlation functions, which are generally harder to compute.
To our knowledge, this is the first time that total time derivative interactions of both kinds, as well as terms proportional to the linear equations of motion, are consistently taken into account, also with both versions of the in-in formalism.
Besides having shed light on these aspects, our calculation makes it clear that
either procedures are not particularly easy to implement.
We therefore move on to present a more flexible method to deal with redundant interactions without introducing total time derivative interactions.
Indeed, as we will see, there will be no time to derive with.

\section{Canonical transformations}
\label{sec: canonical transformations}

In this section, we start by briefly recalling the notion of canonical transformations, and how they can be used to solve a Hamiltonian system in terms of new phase-space variables whose dynamics are dictated by a simpler Hamiltonian.
We then explain how to use them in practice to simplify interactions in the Hamiltonian, just like what can be done with integration by parts in the Lagrangian.
This procedure does not introduce total time derivative interactions nor ones proportional to the linear equations of motion.
We showcase the simplicity of the calculation of correlation functions for the two toy models introduced in the previous section, and we propose some diagrammatic rules to list all possible contributions from the canonical transformation.

\subsection{Generalities}
\label{sec:ct_generalities}
Canonical transformations in the context of a classical field theory consist in a transformation of canonical phase-space variables that preserves the Poisson bracket---denoted as $\left\{\cdot,\cdot\right\}$---relations.
Consider a field $\psi(t, \vec{x})$ and its canonically conjugate momentum $p_\psi(t, \vec{x})$ describing the system dynamics.
They verify:
\begin{equation}
    \forall t \,, \quad \left\{\psi(t, \vec{x}), p_\psi(t, \vec{y})\right\} = \delta^{(3)}(\vec{x}-\vec{y}) \,.
\end{equation}
The initial Hamiltonian density is a function of the original variables, and possibly time explicitly, $\mathcal{H}(\psi, p_\psi,t)$.
This Hamiltonian defines the equations of motion of the system through:
\begin{equation}
    \dot{\psi} = \frac{\partial \mathcal{H}}{\partial p_\psi} \,, \quad \dot{p}_\psi= - \frac{\partial \mathcal{H}}{\partial \psi} \,.
\end{equation}

\paragraph{New phase-space variables.}
After a canonical transformation, new canonical variables $\tilde{\psi}$ and $\tilde{p}_\psi$ are introduced.
By construction, they are enforced to verify the same Poisson brackets as the original variables:
\begin{equation}
    \forall t \,, \quad \left\{\tilde{\psi}(t, \vec{x}), \tilde{p}_\psi(t, \vec{y})\right\} = \delta^{(3)}(\vec{x}-\vec{y}) \,.
\end{equation}
After the transformation, a new Hamiltonian $\tilde{\mathcal{H}}$ is introduced and dictates the dynamics of the new variables via the following equations:
\begin{equation}
    \dot{\tilde{\psi}} = \frac{\partial \tilde{\mathcal{H}}}{\partial \tilde{p}_\psi} \,, \quad \dot{\tilde{p}}_\psi= - \frac{\partial \tilde{\mathcal{H}}}{\partial \tilde{\psi}} \,.
\end{equation}

\paragraph{Generating functions.}
How to find in practice the new Hamiltonian density $\tilde{\mathcal{H}}$ expressed in terms of the new phase-space variables, depends on the kind of canonical transformations.
Indeed, there are four possible types of canonical transformations, depending on which of the new or old phase-space variables are expressed explicitly in terms of the other ones.
Those four types are characterized by their so-called generating functions $F$, which correspond to the following cases:
\begin{enumerate}
    \item \textbf{Type I:}
    \begin{align}
        F_1(\psi,\tilde{\psi},t) \,\,\rightarrow\,\,
        p_\psi &= \frac{\partial F_1}{\partial \psi}, \quad \tilde{p}_\psi = -\frac{\partial F_1}{\partial \tilde{\psi}}
    \end{align}
    
    \item \textbf{Type II:}
    \begin{align}
    \label{eq: generating function properties}
    F_2(\psi,\tilde{p}_\psi,t) \,\,\rightarrow\,\,
        p_\psi
        &= \frac{\partial F_2}{\partial \psi}, \quad \tilde{\psi} = \frac{\partial F_2}{\partial \tilde{p}_\psi}
    \end{align}
    
    \item \textbf{Type III:}
    \begin{align}
    \label{eq: type III generating function}
    F_3(p_\psi,\tilde{\psi},t) \,\,\rightarrow\,\,
        \psi
        &=- \frac{\partial F_3}{\partial p_\psi}, \quad \tilde{p}_\psi = -\frac{\partial F_3}{\partial \tilde{\psi}}
    \end{align}
    
    \item \textbf{Type IV:}
    \begin{align}
     F_4(p_\psi,\tilde{p}_\psi,t) \,\,\rightarrow\,\,
        \psi
        &=- \frac{\partial F_4}{\partial p_\psi}, \quad \tilde{\psi} = -\frac{\partial F_4}{\partial \tilde{p}_\psi}
    \end{align}
\end{enumerate}
For example, for type II generating functions, the old momentum $p_\psi$ and the new position $\tilde{\psi}$ are explicitly expressed in terms of the old position $\psi$ and new momentum $\tilde{p}_\psi$, while the converse expressions have to be found by inversion.
Canonical transformations are actually restricted to the class of generating functions with invertible Hessian matrices, so that one can always define any old or new variables in terms of the other ones.
In the generating function, the time $t$ serves as an external parameter not involved in the phase-space structure.
However it does play an important role, as the Poisson bracket structure needs to be preserved for all values of such an external parameter.
Indeed, the new Hamiltonian can be found by imposing invariance of the least action principle derived from the Hamiltonian action $I$ under the canonical transformation, which we assume to be of the type II here for definiteness:
\begin{align}
    I\left[\psi,p_\psi\right] &\equiv \int \dd t \int \dd^3 \vec{x}  \left[p_\psi \dot{\psi} - \mathcal{H}(\psi,p_\psi)\right] \\
    & = \int \dd^3 \vec{x} \left[ F_2\left(\psi,\tilde{p}_\psi, \infty \right) - \tilde{\psi}\tilde{p}_\psi\right] + \tilde{I}\left[\tilde{\psi},\tilde{p}_\psi\right] \nonumber \,, \\
    \text{with} \quad \tilde{I}\left[\tilde{\psi},\tilde{p}_\psi\right] &=  \int \dd t \int \dd^3 \vec{x}  \left[\tilde{p}_\psi \dot{\tilde{\psi}} -  \tilde{\mathcal{H}}\left(\tilde{\psi},\tilde{p}_\psi\right) \right]\,,
\end{align}
where the surface term at infinity in the second line can be evacuated at this non-perturbative level, i.e. in the full path integral of the theory, and with
\begin{equation}
    \label{eq: effective Hamiltonian after canonical transformation}
    \tilde{\mathcal{H}}\left(\tilde{\psi},\tilde{p}_\psi\right) = \mathcal{H}\left(\psi(\tilde{\psi},\tilde{p}_\psi),p_\psi(\tilde{\psi},\tilde{p}_\psi)\right) + \left.\frac{\partial F_2}{\partial t}\right|_{\psi(\tilde{\psi},\tilde{p}_\psi),\tilde{p}_\psi,t} \,.
\end{equation}
The same relation between the old and new Hamiltonians actually holds for any of the four types of generating function, where after the partial time derivative any old variable must be expressed in terms of the new ones.
Finally, the preservation of the Poisson bracket imposes that the corresponding symplectic form on the phase space is invariant:
\begin{equation}
    \dd \Omega \equiv \dd \psi \wedge \dd p_\psi =  \dd \tilde{\psi} \wedge \dd \tilde{p}_\psi
    \,,
\end{equation}
a consequence of which being to enforce unity of the Jacobian determinant of the transformation:
\begin{equation}
\label{eq: Jac condition}
    J=\frac{\partial \psi}{\partial \tilde{\psi}}\frac{\partial p_\psi}{\partial \tilde{p}_\psi}-\frac{\partial p_\psi}{\partial \tilde{\psi}}\frac{\partial\psi}{\partial \tilde{p}_\psi} = 1 \,.
\end{equation}
This imposes further restrictions on the Hessian matrix of the generating function $F$.

\paragraph{Properties of the initial system in terms of the new one.}
If the system in terms of the new phase-space variables is ``solved'', be it its full dynamics, or the statistical properties of interest, one can retrieve the corresponding information about the initial variables by inverting the canonical transformation.
Specifically in cosmology, we are interested in correlation functions of fields and momenta.
In a classical field theory, with a phase-space path integral approach, expectation values of operators $\mathcal{O}(\psi,p_\psi)$ can be found from those of the same operator in terms of the new variables after the canonical transformation and calculated under the new Hamiltonian:
\begin{align}
\label{eq: relating correlation functions by canonical transformations}
    \Braket{\mathcal{O}(\psi,p_\psi)}_\mathcal{H} & \equiv \int \mathcal{D} \psi \mathcal{D} p_\psi \mathcal{O}(\psi,p_\psi) e^{i I\left[\psi,p_\psi\right]} \nonumber \\
    &= \int \mathcal{D} \tilde{\psi} \mathcal{D} \tilde{p}_\psi J(\tilde{\psi},\tilde{p}_\psi)\mathcal{O}\left(\psi(\tilde{\psi},\tilde{p}_\psi),p_\psi(\tilde{\psi},\tilde{p}_\psi)\right) e^{i \tilde{I}\left[\tilde{\psi},\tilde{p}_\psi\right]} \nonumber \\
    &= \Braket{\mathcal{O}\left(\psi(\tilde{\psi},\tilde{p}_\psi),p_\psi(\tilde{\psi},\tilde{p}_\psi)\right)}_{\tilde{\mathcal{H}}}\,,
\end{align}
where $\tilde{I}$ was defined above and where we used unity of the Jacobian determinant as already proved.
We will show concrete examples of non-linear canonical transformations and their possible uses to simplify interactions in perturbation theory, as well as compute correlation functions of the initial theory, in Sec.~\ref{subsec: canonical transformations toy models}.
In Appendix~\ref{app:quantum_anomalies}, we briefly comment on extensions of canonical transformations from classical to quantum field theories.

\subsection{Use to simplify Hamiltonian interactions in the toy models}
\label{subsec: canonical transformations toy models}

We now apply the general techniques presented in the previous paragraph, first to the two toy models presented in the previous section, see Sec.~\ref{subsec: toy models}, and then to generic theories with a  Lagrangian that would lead to total time derivative interactions in the interaction Hamiltonian and the in-in perturbation theory.

\subsubsection{Toy model 1}

In the first toy model, the Hamiltonian is given by Eq.~\eqref{eq: full Hamiltonian toy model 1}.
We would like to simplify the interactions, but before defining the interaction picture and without introducing total time derivatives.
Let us find the canonical transformation needed to remove the cubic Hamiltonian interaction, in the form:
\begin{equation}
    p_\psi \rightarrow \tilde{p}_\psi \quad \text{with} \quad p_\psi = \tilde{p}_\psi + A(t) \psi^2 + B(t)\tilde{p}_\psi \psi + C(t) \tilde{p}_\psi^2 \,,
\end{equation}
with $A, B, C$ three real functions of time.
The generating function must be of the type II, $F(\psi,\tilde{p}_\psi,t)$ and verify the corresponding equations~\eqref{eq: generating function properties}.
Integrating the first equation, we find
\begin{equation}
    F(\psi,\tilde{p}_\psi,t) = \tilde{p}_\psi \psi + \frac{A}{3} \psi^3 + \frac{B}{2} \tilde{p}_\psi \psi^2 + C \tilde{p}_\psi^2 \psi + f(\tilde{p}_\psi,t) \,.
\end{equation}
The second equation, by requiring consistency, gives
\begin{equation}
    \tilde{\psi} = \psi + \frac{B}{2} \psi^2 + 2 C \tilde{p}_\psi \psi + \frac{\partial f}{\partial \tilde{p}_\psi} \,.  
\end{equation}
After the canonical transformation $(\psi,p_\psi)\rightarrow (\tilde{\psi},\tilde{p}_\psi)$, the new Hamiltonian of the theory is given by Eq.~\eqref{eq: effective Hamiltonian after canonical transformation}.
For general $B$ and $C$ functions of time, as well as a general $f$ function of $(\tilde{p}_\psi,t)$, unwanted new cubic interactions will be generated, so we look for a canonical transformation with these three functions put to zero: $B=C=f=0$.
After this simplification, leading in particular to $\tilde{\psi} = \psi$ and $\partial F/\partial t = \dot{A} \psi^3/3$, we find up to quartic order in fields and momenta:
\begin{align}
    \tilde{\mathcal{H}} = &\,  \mathcal{H}^{(2)}(\tilde{\psi},\tilde{p}_\psi) + \frac{(\beta + 3 \alpha + A ) \tilde{p}_\psi \tilde{\psi}^2}{c} +\dot{\alpha} \tilde{\psi}^3 + \frac{\dot{A}}{3}\tilde{\psi}^3 \\
   & + \frac{1}{2 c}(\beta^2 + 6 \alpha \beta + 9 \alpha^2 + 2 A (\beta+3\alpha)+ A^2 ) \tilde{\psi}^4  +\ldots \,, \nonumber
\end{align}
where we have neglected terms of order five and more in the new phase-space variables.
Choosing $A=-\beta-3\alpha$ to cancel the $\tilde{p}_\psi \psi^2 $ coefficient, results in a number of additional simplifications:
\begin{equation}
    \tilde{\mathcal{H}} = \mathcal{H}^{(2)}(\tilde{\psi},\tilde{p}_\psi) - \frac{\dot{\beta}}{3}\tilde{\psi}^3 \,.
\end{equation}
In the new theory after canonical transformation, only a single cubic interaction is present, and it is proportional to $\dot{\beta}$.
After going to the interaction picture, one finds exactly the effective interactions $\tilde{\tilde{U}}$ found after two iterations in Sec.~\ref{subsec: toy models}.
In particular, we have avoided defining a perturbation theory with several cubic and quartic interactions leading to diagrams at different vertices-order and that cancel each other.
Indeed, all interactions involving $\alpha$ have been taken into account simply as a redefinition of the momentum of the theory:
\begin{equation}
    \tilde{p}_\psi = p_\psi - (3 \alpha + \beta) \psi^2 \,.
\end{equation}
Clearly, this change does not affect correlation functions of fields only.
From Eq.~\eqref{eq: relating correlation functions by canonical transformations}, we indeed find 
\begin{equation}
    \Braket{\mathcal{O}(\psi)}_\mathcal{H} =  \Braket{\mathcal{O}\left(\psi=\tilde{\psi}\right)}_{\tilde{\mathcal{H}}}
\end{equation}
where we also used unity of the Jacobian.
This confirms our conclusions from the in-in perturbation theory including total time derivative interactions, but in a much more straightforward way: interactions as total time derivatives of functions of fields only, do not affect correlation functions of fields.
Moreover, we found again that the other cubic interaction $\propto \beta$ contributes only as proportional to $\dot{\beta}$.
Lastly, correlation functions of the initial momentum $p_\psi$ are non-trivially affected by the presence of the interaction $\propto \alpha$ and can be found from the correlation functions of the new phase-space variables after canonical transformation $(\tilde{\psi},\tilde{p}_\psi)$.
Note, however, that the initial momentum may or may not carry a physical meaning, so retrieving its correlation functions may or may not be interesting, depending on what can be measured by experiments.

\subsubsection{Toy model 2}

In the second toy model, the Hamiltonian is perturbatively given by Eq.~\eqref{eq: full Hamiltonian toy model 2} up to quartic order.
To simplify interactions, we want to include the terms $p_\psi^2 \psi$ and $p_\psi^2 \psi^2$ in the kinetic term.
We try the following canonical transformation:
\begin{equation}
    F(\psi,\tilde{p}_\psi,t) = \tilde{p}_\psi \psi + \frac{A}{2} \tilde{p}_\psi \psi^2 + \frac{B}{3} \tilde{p}_\psi \psi^3 \,,
\end{equation}
which results in
\begin{align}
    p_\psi &= \tilde{p}_\psi + A \tilde{p}_\psi \psi + B \tilde{p}_\psi \psi^2 \,, \\
    \tilde{\psi} &=  \psi + \frac{A}{2} \psi^2 + \frac{B}{3} \psi^3 \,,
\end{align}
and, after inversion:
\begin{align}
    p_\psi &= \tilde{p}_\psi +A \tilde{p}_\psi \tilde{\psi} + \left( B - \frac{A^2}{2}\right) \tilde{p}_\psi \tilde{\psi}^2 +\ldots \,, \\
    \psi &= \tilde{\psi} - \frac{A}{2} \tilde{\psi}^2 + \left(\frac{A^2}{2}-\frac{B}{3}\right)\tilde{\psi}^3 + \ldots \,,
\end{align}
where we truncated at cubic order in fields and momenta.
The coefficients of the interactions that we want to remove are now
\begin{equation}
 \left(A+\frac{c\dot{A}}{2}-2\right) \frac{\tilde{p}_\psi \tilde{\psi}^2}{c} \,,\left(B- 3A + 8 + \frac{c\dot{B}}{3} \right)\frac{\tilde{p}_\psi^2 \tilde{\psi}^2}{c} \,,
\end{equation}
so we set $A=2\,,B=-2$.
Taking into account all contributions (note that $\partial F/\partial t = 0$ for this canonical transformation), we find a number of simplifications finally giving
\begin{equation}
    \tilde{\mathcal{H}}(\tilde{\psi},\tilde{p}_\psi) = \mathcal{H}^{(2)}(\tilde{\psi},\tilde{p}_\psi) \,.
\end{equation}
This agrees with the finding $\tilde{\tilde{U}}=0$ in Eq.~\eqref{eq: tilde tilde U toy model 2} of the in-in formalism with total time derivatives, that we had obtained after several integrations by part, uses of the linear equations of motion and careful considerations regarding cancellations between different vertex orders of the perturbation theory.
We find the method based on canonical transformations at the level of the full (rather than the interaction picture one) Hamiltonian much more straightforward.

Now, correlations functions of $\psi$ may be found from the ones of $\tilde{\psi}$ using $\psi = \tilde{\psi} - \tilde{\psi}^2 + 8 \tilde{\psi}^3/3$.
\paragraph{Bispectrum.} Going from real to Fourier space, it is straightforward to compute the expectation value:
\begin{align}
    \Braket{\mathcal{O}} = \Braket{\psi_{\vec{k}_1}\psi_{\vec{k}_2}\psi_{\vec{k}_3}} &= \Braket{\tilde{\psi}_{\vec{k}_1}\tilde{\psi}_{\vec{k}_2}\tilde{\psi}_{\vec{k}_3}} - \int \frac{\dd^3 \vec{q}}{(2\pi)^3}\Braket{\tilde{\psi}_{\vec{q}}\,\tilde{\psi}_{\vec{k}_1-\vec{q}}\,\tilde{\psi}_{\vec{k}_2}\tilde{\psi}_{\vec{k}_3}} + \text{2 perm.} + \ldots \nonumber \\
    &= (2\pi)^3\delta^{(3)}\left(\sum_{i=1}^3\vec{k}_i\right) \left[- 2P_\psi(k_1)P_\psi(k_2) + \text{2 perm.} \right]  + \ldots \,,
\end{align}
which exactly matches Eq.~\eqref{eq: bispectrum toy model 2}.
We have used that $\tilde{\psi}$ has itself a vanishing bispectrum and we have consistently neglected loop-level corrections to the bispectrum of $\psi$ from the following contributions
\begin{itemize}
    \item the connected piece of the four-point function $\braket{\tilde{\psi}_{\vec{q}}\,\tilde{\psi}_{\vec{k}_1-\vec{q}}\,\tilde{\psi}_{\vec{k}_2}\tilde{\psi}_{\vec{k}_3}}$ (if not vanishing); 
    \item three insertions in $\Braket{\psi_{\vec{k}_1}\psi_{\vec{k}_2}\psi_{\vec{k}_3}}$ of the quadratic term in the expression relating $\psi$ to $\tilde{\psi}$;
    \item the difference between $P_\psi(k)$ and $P_{\tilde{\psi}}(k)$;
    \item all other contributions leading to loop corrections.
\end{itemize}
In Appendix~\ref{app: diagrammatic rules} (see Eqs.~\eqref{eq: diagrammatic rules}--\eqref{eq: diagram for B} therein), we define some diagrammatic rules which we find useful to systematize the knowledge of the loop level at which canonical transformations may contribute.

\paragraph{Trispectrum.} Following the same set of rules for the trispectrum (see Eqs.~\eqref{eq: diagrammatic rules}--\eqref{eq: diagram for T} below), it is straightforward to find that there are two distinct contributions relevant at tree level, and there remains no other difficulty than counting all permutations consistently.
The two contributions correspond to the usual $\tau_\mathrm{NL}$ and $g_\mathrm{NL}$ terms in the trispectrum, and we find:
\begin{align}
    \Braket{\mathcal{O}} = \Braket{\psi_{\vec{k}_1}\psi_{\vec{k}_2}\psi_{\vec{k}_3}\psi_{\vec{k}_4}} = & \,  \int \frac{\dd^3 \vec{q}_1\dd^3\vec{q}_2}{(2\pi)^6}\Braket{\tilde{\psi}_{\vec{q}_1}\,\tilde{\psi}_{\vec{k}_1-\vec{q}_1}\,\tilde{\psi}_{\vec{q}_2}\,\tilde{\psi}_{\vec{k}_2-\vec{q}_2}\,\tilde{\psi}_{\vec{k}_3}\tilde{\psi}_{\vec{k}_4}} + \text{11 perm.}  \\
    & + \frac{8}{3}\int \frac{\dd^3 \vec{q}_1\dd^3\vec{q}_2}{(2\pi)^6}\Braket{\tilde{\psi}_{\vec{q}_1}\,\tilde{\psi}_{\vec{q}_2}\,\tilde{\psi}_{\vec{k}_1-\vec{q}_1-\vec{q}_2}\,\tilde{\psi}_{\vec{k}_2}\tilde{\psi}_{\vec{k}_3}\tilde{\psi}_{\vec{k}_4}} + \text{3 perm.} +\ldots \nonumber \\
    = & \,  (2\pi)^3\delta^{(3)}\left(\sum_{i=1}^4\vec{k}_i\right) \Big\{ 4  \left[P_\psi(k_1)P_\psi(k_2)P_\psi(k_{13})  + \text{11 perm.} \right] \nonumber \\
    &+ \frac{8}{3} \times 6 \times \left[P_\psi(k_1)P_\psi(k_2)P_\psi(k_3)  + \text{3 perm.} \right] \Big\} +\ldots \nonumber \,,
\end{align}
which exactly matches Eq.~\eqref{eq: trispectrum toy model 2} and where, once more, we consistently neglected loop corrections.

\paragraph{One-loop power spectrum.}
Following the set of rules for the power spectrum at one loop (see Eqs.~\eqref{eq: diagrammatic rules}--\eqref{eq: diagram for P-loop} below), it is straightforward to find that there are only two contributions for this toy model where $\tilde{\psi}$ is exactly free, and there remains no other difficulty than counting all permutations consistently:
\begin{align}
    \Braket{\mathcal{O}}^{1\mathrm{-loop}} = \Braket{\psi_{\vec{k}}\psi_{\vec{k}^\prime}}^{1\mathrm{-loop}} = & \,  \int \frac{\dd^3 \vec{q}_1\dd^3\vec{q}_2}{(2\pi)^6}\Braket{\tilde{\psi}_{\vec{q}_1}\,\tilde{\psi}_{\vec{k}-\vec{q}_1}\,\tilde{\psi}_{\vec{q}_2}\,\tilde{\psi}_{\vec{k}^\prime-\vec{q}_2}}  \\
    & + \frac{8}{3}  \int \frac{\dd^3 \vec{q}_1\dd^3\vec{q}_2}{(2\pi)^6}\Braket{\tilde{\psi}_{\vec{q}_1}\,\tilde{\psi}_{\vec{q}_2}\,\tilde{\psi}_{\vec{k}-\vec{q}_1-\vec{q}_2}\,\tilde{\psi}_{\vec{k}^\prime}} + (\vec{k}\leftrightarrow \vec{k}^\prime) \nonumber \\
    = & \,  (2\pi)^3\delta^{(3)}\left(\vec{k}+\vec{k}^\prime \right) \Big\{ 2 \times \int \frac{\dd^3 \vec{q}}{(2\pi)^3} P_\psi^\mathrm{tree}(q)P_\psi^\mathrm{tree}(|\vec{q}+\vec{k}|) \nonumber \\
    & + \frac{8}{3} \times 3 \times 2 \times P_\psi^\mathrm{tree}(k)  \int \frac{\dd^3 \vec{q}}{(2\pi)^3} P_\psi^\mathrm{tree}(q)\Big\}\nonumber \,,
\end{align}
where we used that at this loop order $P_{\tilde{\psi}}=P_{\tilde{\psi}}^\mathrm{tree}=P_\psi^\mathrm{tree}$.
How to compute the loops themselves is not the topic of this work.

\section{Single-field inflation}
\label{sec: single field inflation}

We apply the formalism with canonical transformations developed in the previous section to the case of single-field inflation.
This scenario is both the simplest and the most commonly explored, making it illustrative to observe the simplifications introduced by our formalism.
We start from the action of a single scalar field minimally coupled to gravity, which is given by:
\begin{equation}
    S=  \int {\rm d}^4x\sqrt{-g}\bigg[\frac{\Mp^2}{2}\,R(g)-\frac{1}{2}\partial^\mu\phi \partial_\mu \phi-V(\phi)\bigg]+S_{\rm GHY}\,,
\end{equation}
where we have supplemented the Einstein-Hilbert and scalar field action by the Gibbons-Hawking-York (GHY) boundary term that makes the initial problem well defined. 
As customary in the context of inflationary cosmology, we adopt the ADM form of the metric~\cite{Arnowitt:1962hi}:
\beq
\label{eq:ADM_metric}
{\rm d}s^2=-N^2 {\rm d}t^2+h_{ij}({\rm d}x^i+\N^i {\rm d} t)({\rm d}x^j+\N^j {\rm d}t)\,,
\eeq
where $N$ is the lapse function and $\N^i$ the shift vector, together defining a slicing of spacetime. 
Using the Gauss-Codazzi relation (see e.g.~\cite{Wald:1984rg}), the four-dimensional Ricci scalar can be rewritten as:\begin{equation}
 \frac{\Mp^2}{2}\int {\rm d}^4x\sqrt{-g} \, R(g) =   \frac{\Mp^2}{2}\int {\rm d}^4x\sqrt{-g}\,\left[ R^{(3)} + (K_{ij} K^{ij}-K^2)\right] -S_{\mathrm{GHY}}\,.
\end{equation}
In this expression, the tensor $K_{ij}$ is the so-called extrinsic curvature tensor and reads
\begin{equation}
    K_{ij}=\frac{1}{2N} \left(\dot{h}_{ij}-\nabla_i\N_j-\nabla_j\N_i\right) \,,
\end{equation}
where $\nabla_i$ denotes the spatial covariant derivative associated with 
the projection of the spacetime metric on these spatial hypersurfaces: $h_{ij}$.
Also, $R^{(3)}$ is the Ricci curvature on three-dimensional spatial hypersurfaces, calculated with $h_{ij}$ too. 
As a consequence of this decomposition, the GHY term cancels out in the total action which, upon plugging the ADM form of the metric,  reads
\beq\bal
S&=\frac{\Mp^2 }{2}\int {\rm d}t {\rm d}^3 x \sqrt{h}N\,\left[ R^{(3)} + (K_{ij} K^{ij}-K^2)\right] \\
&+ \frac{1}{2}\int {\rm d}t {\rm d}^3 x \sqrt{h}N\,\left[\frac{1}{N^2}\left(\dot \phi -\N^j \partial_j \phi\right)^2 - h^{ij} \partial_i \phi \partial_j \phi-2V \right] \,,
\label{eq:action-ADM}
\eal
\eeq
where $h=$ det$(h_{ij})$.

Actually, general relativity plus a scalar field, as we consider, is a fully constrained system as we are reminding in the following.
Indeed, the Lagrangian action of Eq.~\eqref{eq:action-ADM}, with $(\phi,h_{ij},N,\mathcal{N}^i)$ as fundamental position variables, defines the on-shell\footnote{What we mean by ``on-shell'' here and in the following is an equation that is valid in the sense of a weak equality, i.e. valid on the hypersurface of the phase space where constraints (to be derived right after) are identically verified.
We denote on-shell quantities by a bar.} canonical conjugate momenta:
\begin{align}
    \bar{p}_\phi &= \frac{\sqrt{h}}{N}\left( \dot{\phi} - \mathcal{N}^i \partial_i \phi \right) \,, \\
    \bar{p}^{ij} &= \frac{\sqrt{h}\Mp^2 }{2}\left(K h ^{ij} - K^{ij}\right)\,, \\
    \bar{p}_N &= 0 \,, \\
    \bar{p}_{\mathcal{N},i} &= 0 \,,
\end{align}
from which the Hamiltonian action of the theory is found by Legendre transform of the Lagrangian:
\begin{equation}
    I = \int \dd t \dd^3\vec{x} \left[p_\phi \dot{\phi} + p^{ij} \dot{h}_{ij} + p_N \dot{N}  + p_{\mathcal{N},i} \dot{\mathcal{N}}^i - \mathcal{H}\right]\,,
\end{equation}
where the Hamiltonian density reads 
\begin{align}
\label{eq:C=0}
   \mathcal{H} &= p_N \dot{N} + p_{\mathcal{N},i}\dot{\mathcal{N}}^i + N \mathcal{C}\left(\phi,p_\phi,h_{ij},p^{ij}\right) + \mathcal{N}^i \mathcal{C}_i\left(\phi,p_\phi,h_{ij},p^{ij}\right) \\ 
   \text{with}  \quad \mathcal{C} &= \frac{2}{\sqrt{h}\Mp^2}\left[p_{ij} p^{ij}- \frac{1}{2}(p_i^i)^2\right] - \frac{\sqrt{h}\Mp^2}{2}R^{(3)} + \frac{1}{2 \sqrt{h}} p_\phi^2 + \frac{\sqrt{h}h^{ij}}{2} \partial_i \phi \partial_j \phi + \sqrt{h} V  \,, \nonumber  \\ 
   \text{and}  \quad \mathcal{C}_i &= - 2 \sqrt{h} \nabla_j\left(\frac{p^j_i}{\sqrt{h}}\right) + p_\phi \partial_i \phi  \nonumber \,.
\end{align}
To find this expression, we had to evaluate on-shell the time derivatives $\dot{\phi}$ and $\dot{h}_{ij}$ in the Lagrangian as functions of the momenta $p_\phi$ and $p_h^{ij}$ which should themselves be considered unconstrained again, as could be seen from an explicit path integral approach with $\int \mathcal{D} \psi_a e^{i S[\psi_a]} =\int \mathcal{D} \psi_a  \int \mathcal{D} p_\psi^a e^{i I[\psi_a,p_\psi^a]}$, where $\psi_a$ collectively denotes all position variables, and $p_\psi^a$ all momentum variables.
Note however, that time derivatives of $N$ and $\mathcal{N}^i$ could not be replaced as their momentum is vanishing, so no inverse can be found.
Technically, this is because they do not appear with time derivatives in the Lagrangian.
Physically, this is because the foliation of spacetime in the ADM formalism is arbitrary and does not describe physically propagating degrees of freedom at this stage.
Therefore $N$ and $\mathcal{N}^i$ act as simple Lagrange multipliers, as in any constrained system.
Let us construct the on-shell, constrained subspace and its corresponding Hamiltonian $\bar{\mathcal{H}}$.

The so-called primary constraints, valid without invoking the equations of motion of the theory, consist in the vanishing momenta that we have already encountered:
\begin{align}
  \textbf{Primary constraints:} \quad \bar{p}_N &= 0 \,, \quad \bar{p}_{\mathcal{N},i} = 0 \quad \rightarrow \quad \mathcal{H}= N \mathcal{C} + \mathcal{N}^i \mathcal{C}_i\,.
\end{align}
This partially on-shell form of the Hamiltonian of general relativity plus a scalar field, is the one often quoted in the literature.
Secondary constraints are found by requiring that primary constrains are conserved as time passes.
Technically, this amounts to asking that their Poisson brackets with the Hamiltonian density is vanishing on-shell, $\overline{\{p_N(t,\vec{x}),H(t)\}}=0=\overline{\{p_{\mathcal{N},i}(t,\vec{x}),H(t)\}}$ with $H(t)=\int \dd^3\vec{y}\, \mathcal{H}(t,\vec{y}) $, which gives:
\begin{align}
  \textbf{Secondary constraints:} \quad \bar{\mathcal{C}} &= 0 \,, \quad \bar{\mathcal{C}}_i = 0 \quad \rightarrow \quad \bar{\mathcal{H}}= 0 \,,
\end{align}
where we used the fact that $\mathcal{C}$ and $\mathcal{C}_i$ do not depend on $N$ nor $\mathcal{N}^i$.
Remarkably, secondary constraints imply the vanishing, on shell, of the quantities $\mathcal{C}$ and $\mathcal{C}_i$.
Those equations, which are only functions of the field $\phi$ and the spatial metric $h_{ij}$, as well as their canonically conjugate momenta, are known as the energy and momentum constraints in general relativity.
Moreover, that the (now fully) on-shell Hamiltonian is exactly vanishing, means that general relativity plus a scalar field is a fully constrained system.
Now, an important property of these 4 primary and 4 secondary constraints is that they are first-class constraints.
Indeed, just like in pure general relativity, the Poisson brackets of the constraints are either trivially vanishing (like $\{p_N(t,\vec{x}),p_N(t,\vec{y})\}=0\,,\{p_N(t,\vec{x}),\mathcal{C}(t,\vec{y})\}=0$, etc.), or vanishing on-shell (like $\{\mathcal{C}(t,\vec{x}),\mathcal{C}(t,\vec{y})\}$ is a linear combination of $ \mathcal{C}_i(t,\vec{x})$ and $ \mathcal{C}_i(t,\vec{y})$~\cite{Prokopec:2010be}, so $\overline{\{\mathcal{C}(t,\vec{x}),\mathcal{C}(t,\vec{y})\}}=0$, etc.).
Defining $C_\alpha \equiv \left(\mathcal{C},\mathcal{C}_{i}, p_N, p_{\mathcal{N},i}\right)$, we therefore have:
\begin{align}
  \textbf{First-class constraints:} \quad \forall \alpha\,,\beta\,, \quad \overline{\left\{C_\alpha(t,\vec{x}),C_\beta(t,\vec{y})\right\}} = 0 \,.
\end{align}
Importantly, the fact that these are first-class constraints also closes the system of constraints, as no tertiary constraints can be found.
Indeed, Poisson brackets of the secondary constraints with the Hamiltonian are linear combinations of themselves, and thus vanishing on-shell.
These features are due to the invariance of the theory under spacetime diffeomorphisms, and they are crucial to correctly count degrees of freedom (d.o.f.) as we now recapitulate.
General relativity alone has 20 phase-space d.o.f. ($N, N_i, h_{ij}$ and their momenta), but there are 8 first-class constraints each counting double, leaving for 4 phase-space d.o.f. only: the two polarizations of gravitational waves and their conjugate momenta.
As we have seen, adding the scalar field and its momenta does not affect this picture, and simply adds 2 phase-space d.o.f.

On the other hand, the dynamical equations of the theory are given by the Hamilton equations, i.e. the Poisson brackets of the phase-space variables with non-trivial on-shell momenta.
The dynamical part of Einstein equations can be found from 
\begin{align}
\label{eq: dynamical Eisntein eq}
    \dot{h}_{ij} &=\frac{\delta}{\delta  p^{ij}}\int \dd^3 \vec{x} \,\mathcal{H} \,, \\
    \dot{p}^{ij} &=-\frac{\delta}{\delta h_{ij}}\int \dd^3 \vec{x} \, \mathcal{H} \nonumber \,,
\end{align}
but we will not need to write them explicitly at this stage.
Hamilton equations in the scalar sector read:
\begin{align}
\label{eq: dynamical scalar field eq}
    \dot{\phi}&=\frac{\delta}{\delta  p_\phi} \int \dd^3 \vec{x} \,\mathcal{H}  = \frac{N}{\sqrt{h}} p_\phi + \mathcal{N}^i \partial_i \phi \,, \\
    \dot{p}_\phi &=-\frac{\delta}{\delta \phi}\int \dd^3 \vec{x} \, \mathcal{H} = - \sqrt{h} N V^\prime(\phi) + \partial_i\left(\sqrt{h} N h^{ij} \partial_j \phi + \mathcal{N}^i p_\phi \right) \,. \nonumber
\end{align}

Although this fully constrained Hamiltonian formalism for general relativity plus a scalar field is very elegant, it is hard to use in practice, for it requires the manipulation of fully non-perturbative objects in real space like $\phi(t,\vec{x})$ and $h_{ij}(t,\vec{x})$.
What can be done, however, starting from this understanding, is to systematically expand the Hamiltonian density and the energy and momentum constraints up to a given order in perturbation theory.
Then, at each order, one can identify the physical degrees of freedom propagating on the constrained subset of the phase space, also quotient by the unphysical gauge degrees of freedom ubiquitous in constrained systems.
Gauge-invariant variables may be defined, and their dynamics can be studied, consistently order by order in perturbation theory.
To our knowledge, this procedure has been mainly used up to quadratic order in the Hamiltonian, and therefore for linear order cosmological fluctuations only~\cite{Langlois:1994ec} (see however~\cite{Domenech:2017ems} for an interesting first work at cubic order).
Instead of pursuing this direction, which avoids the \textit{ad hoc} fixing of a gauge, but requires switching already to a perturbation theory, we decide to take another route enabling us to make connection with the literature in the comoving gauge, although first in an innovative way valid at all orders in perturbation theory.

Before that, we quickly specify the background dynamics, assuming all phase-space variables to be homogeneous functions of time.
Specifically, we focus on a FLRW metric with $h_{ij} = a^2 \delta_{ij}$ and we fix the lapse and shift to $N=1\,,\, \mathcal{N}^i = 0$.
The constraint $\bar{\mathcal{C}}=0$ yields the first Friedmann equation:
\begin{equation}
    3 H^2 \Mp^2 = \frac{p_\phi^2}{2 a^6} + V(\phi) \,,
\end{equation}
while the second constraint vanishes trivially even off-shell, $\mathcal{C}^i = 0$, and therefore does not bring additional information.
Hamilton equations in the scalar field sector are
\begin{align}
    \dot{\phi} = \frac{p_\phi}{a^3} \,, \quad \dot{p}_\phi = - a^3 V^\prime(\phi) \,.
\end{align}
These three equations can be combined to give the homogeneous Klein-Gordon equation for a scalar field in a FLRW spacetime, and the usual form of the two Friedmann equations:
\begin{align}
    \ddot{\phi} + 3 H \dot{\phi} + V^\prime(\phi) &= 0 \,, \\
    3 H^2 \Mp^2 &=  \frac{1}{2} \dot{\phi}^2 + V(\phi) \,, \\
    -\frac{\dot{H}}{H^2} &= \frac{\dot{\phi}^2}{2 H^2 \Mp^2} \,.
\end{align}
The second Friedmann equation could also be found from the dynamical part of the Einstein equations in Eq.~\eqref{eq: dynamical Eisntein eq}.

\subsection{Hamiltonian in the comoving gauge}
\label{subsec: non-perturbative Hamiltonian}

As explained in the introduction, in this paper, we are interested in calculations in the comoving gauge.
Discarding tensor perturbations for the moment (we will come back to this matter in Sec.~\ref{subsec: tensors}), this choice corresponds to 
\begin{align}
    &\delta\phi^{{\rm comoving}}=0,\\
    &h^{{\rm comoving}}_{ij}=a^2e^{2\zeta}\delta_{ij} \,.
\end{align}
Instead of arriving to this non-perturbative definition for the curvature fluctuation by consistently requiring, order by order in perturbation theory, a gauge invariant variable in the physical subset of the phase space, we have ourselves decided what would be the gauge for performing calculations.
The price for picking the gauge by hand, is that such choice breaks the structure of the constrained system, as e.g. $\mathcal{C}$ and $\mathcal{C}_i$ are now functions of $N$ and $\mathcal{N}^i$.
Moreover, one does not know how to deal with the momenta $p_\phi$ and $p_h^{ij}$ in the Hamiltonian system, nor how to see the appearance of the momentum $p_\zeta$ that is canonically conjugate to the variable $\zeta$ defined above.
Instead, we first go back to the Lagrangian description in terms of $\phi$, $h_{ij}$ and their time derivatives, and we will go back to the Hamiltonian after a short pause.
In the comoving gauge, the Lagrangian density $\Lag$ takes the following form
\begin{align}
    \Lag=&\Mp^2\Biggl\{-\frac{6 a^3 e^{3 \zeta } H \dot{\zeta} }{N}+6 a^3 e^{3
   \zeta } H \dot{\zeta} -\frac{3 a^3 e^{3 \zeta }
   \dot{\zeta} {}^2}{N}+\frac{e^{-\zeta } \left(\partial
   _j\mathcal{N}_i\right){}^2}{4 a N}+\frac{e^{-\zeta }
   \mathcal{N}_i{}^2 \left(\partial_j\zeta
   \right){}^2}{a N}\notag\\&+\frac{e^{-\zeta } \partial
   _j\mathcal{N}_i \partial
   _i\mathcal{N}_j}{4 a N}-\frac{e^{-\zeta } \partial
   _i\mathcal{N}_i \partial
   _j\mathcal{N}_j}{2 a N}-\frac{e^{-\zeta }
   \mathcal{N}_i \partial_j\zeta  \partial
   _j\mathcal{N}_i}{a N}-\frac{e^{-\zeta }
   \mathcal{N}_i \partial_j\zeta  \partial
   _i\mathcal{N}_j}{a N}\notag\\&+\frac{2 a e^{\zeta } H
   \partial_i\mathcal{N}_i}{N}+\frac{2 a e^{\zeta } H
   \mathcal{N}_i \partial_i\zeta }{N}-\frac{a e^{\zeta
   } \dot{\zeta}  \left(\partial_i\zeta
   \right){}^2}{H}-\frac{2 a e^{\zeta } \partial_i\zeta 
   \partial_i\dot{\zeta} }{H}-a e^{\zeta } N
   \left(\partial_i\zeta \right){}^2\notag\\&-2 a e^{\zeta } N
   \partial^2\zeta +\frac{2 a e^{\zeta }
   \dot{\zeta}  \partial
   _i\mathcal{N}_i}{N}+\frac{2 a e^{\zeta }
   \mathcal{N}_i \dot{\zeta}  \partial
   _i\zeta }{N}-a e^{\zeta } \left(\partial_i\zeta
   \right){}^2-a e^{\zeta } \epsilon  \left(\partial_i\zeta
   \right){}^2\notag\\&+6 a^3 e^{3 \zeta } H^2-\frac{3 a^3 e^{3 \zeta } H^2}{N}-2 a^3 e^{3 \zeta } H^2
   \epsilon\notag +\frac{a^3 e^{3 \zeta } \dot\phi^2}{2 N \Mp^2}
 - \frac{a^3}{\Mp^2} e^{3 \zeta } N V(\phi )\\
   &+\frac{\dd}{\dd t}\left[-2 a^3 e^{3 \zeta } H\right]
   +\frac{\dd}{\dd t}\left[\frac{a e^{\zeta } \left(\partial_i\zeta \right){}^2}{H}\right]\Biggr\} .\label{eq:action_nonpert_withboundary}
\end{align}
Let us emphasize that this equation is valid at all orders in perturbations in the curvature perturbation $\zeta$. As we will see below, this will prove to be a significant simplification in deriving the Hamiltonian.
In particular, we have isolated in the last line above two total time derivatives at all orders in perturbation theory.

The on-shell conjugate momenta $p_X\equiv\delta L/\delta \dot{X}$ computed from the complete Lagrangian above (including the total time derivatives) are given by:
\begin{align}
\label{eq:momentum_zeta}
    \bar{p}_\zeta=&\Mp^2\left[-\frac{6 a^3 e^{3 \zeta } \dot{\zeta} }{N}+\frac{2 a
   e^{\zeta } \partial_i\mathcal{N}_i}{N}+\frac{2 a
   e^{\zeta } \mathcal{N}_i \partial_i\zeta
   }{N}-\frac{6 a^3 e^{3 \zeta } H}{N}\right]\\
   \label{eq:mom_N}
    \bar{p}_N=&0\\
   \label{eq:mom_Na}
    \bar{p}_{\N_,i}=&0\,.
\end{align}
The Hamiltonian can be easily computed via the Legendre transformation, and reads:
\begin{align}
    \mathcal{H}
    =&p_N \dot{N} + p_{\N,i}\dot{\N}^i + \frac{1}{\Mp^2}\Biggl[-\frac{e^{-3 \zeta } N p_{\zeta }^2}{12 a^3}-p_{\zeta } \left(H-\frac{e^{-2 \zeta } \mathcal{N}_i \partial
   _i\zeta }{a^2}\right)-\frac{e^{-2 \zeta }
   \mathcal{N}_i \partial_ip_{\zeta }}{3
   a^2}\notag\\&+a e^{\zeta } N \left(\partial
   _i\zeta \right){}^2+2 a e^{\zeta } N \partial
   _i\partial_i\zeta +3 a^3 e^{3 \zeta } H^2 N-a^3
   e^{3 \zeta } H^2 N \epsilon -\frac{a^3 e^{3 \zeta } H^2 \epsilon
   }{N}\notag\\&-\frac{e^{-\zeta } \mathcal{N}_i \partial_jN
   \partial_j\mathcal{N}_i}{4 a N^2}-\frac{e^{-\zeta }
   \mathcal{N}_i \partial_jN \partial
   _i\mathcal{N}_j}{4 a N^2}+\frac{e^{-\zeta }
   \mathcal{N}_i \partial_iN \partial
   _j\mathcal{N}_j}{6 a N^2}-\frac{e^{-\zeta }
   \mathcal{N}_i{}^2 \left(\partial_j\zeta
   \right){}^2}{a N}\notag\\&+\frac{3 e^{-\zeta } \mathcal{N}_i
   \partial_j\zeta  \partial_j\mathcal{N}_i}{4
   a N}+\frac{e^{-\zeta } \mathcal{N}_i \partial
   _j\partial_j\mathcal{N}_i}{4 a N}+\frac{3
   e^{-\zeta } \mathcal{N}_i \partial_j\zeta  \partial
   _i\mathcal{N}_j}{4 a N}-\frac{e^{-\zeta }
   \mathcal{N}_i \partial_i\zeta  \partial
   _j\mathcal{N}_j}{2 a N}\notag\\&+\frac{e^{-\zeta }
   \mathcal{N}_i \partial_i\partial
   _j\mathcal{N}_j}{12 a N}-\frac{e^{-\zeta }
   \mathcal{N}_i \mathcal{N}_j \partial_i\zeta 
   \partial_j\zeta }{3 a N}\notag\Biggr] \,,
\end{align}
where we have replaced instances of $\dot{\zeta}$ by inverting Eq.~\eqref{eq:momentum_zeta}.
Importantly, this replacement can be performed explicitly because, {\em before} solving the constraints, the form of $p_\zeta$ in Eq.~\eqref{eq:momentum_zeta} can be analytically inverted to express $\dot{\zeta}$ as a function of $p_\zeta$.
This is in contrast to more general situations where the relation $\dot{\zeta}=\dot{\zeta}(\zeta,\,\p)$ has to be inverted perturbatively, see e.g. Refs.~\cite{Wang:2013zva,Chen:2017ryl}.
In the following, we will focus on the partially on-shell Hamiltonian by imposing the vanishing of the on-shell momenta $\bar{p}_N$ and $\bar{p}_{\N,i}$, and we therefore drop the two first terms in the expression above.

The Hamiltonian $\mathcal{H}$ still contains contributions generated by the total time derivatives terms, see the last line of Eq.~\eqref{eq:action_nonpert_withboundary}. We now apply the general lesson learnt in Sec.~\ref{sec: canonical transformations} and we define the following type II canonical transformation to remove such undesired interactions:
\begin{align}
\label{eq: non linear canonical transf}
        \tilde{\zeta} & \equiv \zeta,\\
        p_\zeta & \equiv \tilde{p}_\zeta - 2\frac{ a}{H} e^{\zeta } \partial^2 \zeta - \frac{ a }{H}e^{\zeta } (\partial \zeta)^2 -6
   a^3 e^{3 \zeta } H \,,
\end{align}
with the corresponding generating function given by
\begin{equation}
\label{eq:first_canonical_transformation}
    F[{\tilde{p}_\zeta,\,\zeta,\,t}]=\tilde{p}_\zeta \zeta + \frac{a}{H}e^{\zeta }\left(\partial \zeta\right)^2 -2 a^3 e^{3 \zeta } H.
\end{equation}
Note that this canonical transformation does not affect $\zeta$, as we now expect from the general understanding of interactions expressed as total time derivatives of functions of fields only, and not their time derivatives.
Using the formulae developed in Section~\ref{sec: canonical transformations}, we can compute the new Hamiltonian:
\begin{align}\label{eq:Hamiltonian_final_nonpert}
\tilde{\mathcal{H}}\,\Mp^2=&-\frac{e^{-3 \tz } N \tilde{p}_\zeta ^2}{12 a^3}+\tilde{p}_\zeta  \left[\frac{e^{-2 \tz } }{ a^2 H}\left(\frac{N \left(\partial
   _i\tz \right){}^2}{6 }+\frac{ N \partial^2\tz
   }{3 }+ H \mathcal{N}_i
   \partial_i\tz\right)+H
   (N-1)\right]\notag\\&-\frac{e^{-2 \tz } \mathcal{N}_i
   \partial_i\tilde{p}_\tz }{3 a^2}+\frac{e^{-\tz } }{3 a
   H^2}\Biggl[-\frac{N}{4}\left(\partial_i\tz \partial_i\tz +2 \partial^2\tz\right)^2- 2 H
   \mathcal{N}_i\Biggl( \partial_i\tz 
   \left(\partial_j\tz \right){}^2\notag\\&-
   \partial_j\tz  \partial_i\partial
   _j\tz +2 \partial_i\tz 
   \partial^2\tz -   \partial_i\partial^2\tz \Biggr)\Biggr]+\frac{e^{-\tz }}{ a
   N}\Biggl[
   \frac{\mathcal{N}_i (3N\partial_j\tz-\partial_jN)\left(\partial
   _j\mathcal{N}_i+\partial
   _i\mathcal{N}_j\right)}{4 
   N}\notag\\&+\frac{ \mathcal{N}_i
   \partial_iN \partial
   _j\mathcal{N}_j}{6 
   N}- \left(\mathcal{N}_i
   \partial_j\tz \right)^2-\frac{ \mathcal{N}_i \partial
   _i\tz  \partial
   _j\mathcal{N}_j}{2}+\frac{ \mathcal{N}_i \partial^2\mathcal{N}_i}{4}+\frac{\mathcal{N}_i \partial
   _i\partial_j\mathcal{N}_j}{12
   }-\frac{ \left(\mathcal{N}_i
    \partial_i\tz\right)^2}{3 }\Biggr]\notag\\&+a e^{\tz } \left[\left(2+\epsilon\right) 
   \left(\partial_i\tz \right){}^2+ 
   2\partial^2\tz \right]+\left(2 a^3 e^{3 \tz }-a^3 e^{3
   \tz } N -\frac{a^3 e^{3 \tz }  }{N}\right) H^2 \epsilon
   \,.
\end{align}
We stress that this Hamiltonian, written in terms of the phase space variables, is valid to all orders in perturbation theory, and has been derived for the first time in this paper.
From now on, we will drop all tildes $\tilde{}$, both on $\mathcal{H}$ and on the new phase-space variables and simply take this Hamiltonian as the starting point for the remaining of the paper.
Also, the curvature fluctuation $\zeta$ of observational relevance has not been affected by the canonical transformation.
Would one seek the dynamics of the initial conjugate momentum of $\zeta$, one would need to perform the inverse canonical transformation shown in Eq.~\eqref{eq: non linear canonical transf} from the dynamics of the variables now called $(\zeta,p_\zeta)$.
We stress that there may be no particular reason to do, as the new $p_\zeta$ variable is by definition the canonical conjugate momentum of $\zeta$ associated to the Hamiltonian above, and is not less meaningful than the initial one.
Equation~\eqref{eq:Hamiltonian_final_nonpert} above is one of the main results of our paper, and we insist that it is valid at all orders in perturbation theory.
It is also in a form where unwanted interactions generated by total time derivatives in the Lagrangian have already been removed, thereby simplifying the calculation to come of interactions, at all orders in perturbation theory and at once.
It can be used as a starting point for many applications, including but not restricted to calculations in perturbation theory, to which we will turn in the next section.

Before that, we explain a conceptual but important point related to constraints and the notions of off-shell versus on-shell Hamiltonians.
We have already mentioned that because we picked a gauge before defining the Hamiltonian, we lost the nice geometrical structure of the complete description in terms of a fully constrained Hamiltonian.
In particular, we \textit{cannot} simply evaluate the constraint equations $\bar{\mathcal{C}} = 0 = \bar{\mathcal{C}}_i $ by replacing the moments in terms of their expressions in the comoving gauge.
Indeed, both $\mathcal{C}$ and $\mathcal{C}_i$ become functions of $N$ and $\mathcal{N}^i$ and the new constraint equations would instead read $ \bar{\mathcal{C}} + N \overline{\left(\frac{\delta \mathcal{C}}{\delta N}\right)} = 0$, etc.
This form is not particularly enlightening so we avoid using explicitly the quantities $\mathcal{C}$ and $\mathcal{C}_i$ in the following.
Instead, we simply write the constraint equations as
\begin{equation}
\label{eq: constraints any gauge}
    \left.\frac{\delta \mathcal{H}}{\delta N}\right|_{\bar{N},\,\bar{\mathcal{N}}_i} = 0\,, \quad       \left.\frac{\delta \mathcal{H}}{\delta \mathcal{N}^i}\right|_{\bar{N},\,\bar{\mathcal{N}}_i}  = 0 \,,
\end{equation}
which still hold true in any particular gauge.
Conceptually, this shows that the ADM slicing is no longer arbitrary once a gauge is chosen: $N$ and $\mathcal{N}_i$ are given in terms of other fluctuating quantities.
So, although the lapse and the shift are non-dynamical, they need to be solved for, in contrast to the fully constrained approach.
Conceptually, enforcing these constrains to hold in the Hamiltonian corresponds to evaluating it on-shell, i.e. on the region in phase space where physical processes happen classically.
When we compute cosmological observables, we do want to use the on-shell Hamiltonian where the lapse and the shift are assigned their physical values.
However, when we perform canonical transformations, we do so at the level of the off-shell Hamiltonian.
This is so because when we change phase-space variables $(\zeta, p_\zeta) \rightarrow (\tz,\tp)$, we implicitly leave invariant the remaining of the phase space made of $(N,\mathcal{N}^i,p_N,p_{\mathcal{N}, i})$.
Then, in the new phase space, constraints can still be derived and relate $N$ and $\mathcal{N}^i$ to the variables $(\tz,\tp)$ and their (inverse) spatial derivatives.
This conceptual subtlety makes no practical difference up to cubic order in phase-space variables, but not following this prescription would actually lead to mistakes starting at quartic order.
Note also that, strictly speaking, we should even consider the fully off-shell Hamiltonian including the $p_N \dot{N} + p_{\mathcal{N},i} \dot{\mathcal{N}}^i$ term, but it is easy to realize that it makes no practical difference at all to simply consider the partially on-shell Hamiltonian with $\bar{p}_N=0=\bar{p}_{\mathcal{N},i}$ that we have used above.

\subsection{Quadratic Hamiltonian}

We now expand the Hamiltonian~\eqref{eq:Hamiltonian_final_nonpert} perturbatively.
In that utility, we will assign a book-keeping perturbative parameter $\varepsilon$ (not to be confused with the slow-roll parameter $\epsilon$) to our fundamental canonical variables
$\zeta \rightarrow \varepsilon \zeta$ and $p_\zeta \rightarrow \varepsilon p_\zeta$.
We also expand the lapse and shift in powers of the phase-space variables, as\footnote{
The $1$ in $N$ corresponds to a physical choice for the time variable denoted as $t$ from the beginning, for which $t$ coincides with the cosmic time.
Other choices lead to $t$ coinciding with other time variables, like conformal time for $N^{(0)}= a$ the scale factor, or the number of $e$-folds for $N^{(0)}=H$ the Hubble parameter.
}
\begin{align}
\label{eq:perturbative_expansion_lapse_shift}
    N &=  1 +\varepsilon \,\alpha^{(1)}+ \varepsilon^2\, \alpha^{(2)}+\cdots  \\
    \N_i &= 0 + \varepsilon\left(\partial_i \theta^{(1)}+ b_i^{(1)} \right) + \varepsilon^2\left(\partial_i \theta^{(2)} + b_i^{(2)} \right) +\cdots
\end{align}
where $ b_i^{(n)}$  are divergence-free vectors, i.e. $\forall i\,, \partial^i b_i^{(n)}= 0 $.
The Hamiltonian itself can then be expanded consistently:
\begin{equation}
    \mathcal{H} = \mathcal{H}^{(0)}+\varepsilon\mathcal{H}^{(1)}+\varepsilon^2\mathcal{H}^{(2)}+\varepsilon^3\mathcal{H}^{(3)}+\varepsilon^4\mathcal{H}^{(4)}+\ldots
\end{equation}
In order to find the solutions for the lapse and the shift, we will vary the Hamiltonian with respect to them, as in Eq.~\eqref{eq: constraints any gauge}, at each order in perturbation variables\footnote{Interestingly, a solution for lapse function $N$ valid at all orders in perturbation theory can be found by varying Eq.~\eqref{eq:Hamiltonian_final_nonpert}
and reads $\bar{N}=\pm   a e^{\zeta} H\sqrt{f_1/f_2}$,
where we defined
\begin{align}
f_1=& 12 a^4 e^{4 \zeta } H^2 \epsilon 
 + 3 \left(\partial_j\mathcal{N}_i \right)^2
 + 3 \partial_j\mathcal{N}_i \partial_i\mathcal{N}_j
 - 2 \partial_i\mathcal{N}_i \partial_j\mathcal{N}_j
 - 12 \partial_j \zeta   \partial_j\mathcal{N}_i \mathcal{N}_i
\notag\\& - 12 \partial_j \zeta   \partial_i\mathcal{N}_j \mathcal{N}_i
 + 8 \partial_i \zeta   \partial_j\mathcal{N}_j \mathcal{N}_i
 + 12 \left(\partial_j \zeta \right)^2 \mathcal{N}_i {}^2
 + 4 \partial_i \zeta  \partial_j  \zeta   \mathcal{N}_i
\mathcal{N}_j\notag\\
f_2=&12 a^6 e^{6  \zeta  } H^4 \epsilon 
 - 12 a^3 e^{3  \zeta  } H^3 p_\zeta /\Mp^2
 + H^2 p_\zeta ^2
 - 2 a e^{ \zeta  } H p_\zeta  \left(\partial_i \zeta  \right)^2
 + a^2 e^{2  \zeta  } \left(\partial_i \zeta  \right)^2 \left(\partial_j \zeta  \right)^2
 \notag\\&- 4 a e^{ \zeta } H p_\zeta  \partial^2 \zeta 
 + 4 a^2 e^{2  \zeta } \left(\partial_i \zeta \right)^2 \
\partial^2\zeta 
 + 4 a^2 e^{2  \zeta  } \partial^2  \zeta   \partial^2 \zeta  . 
\end{align}
The solution with the $+$ sign is the one that matches the perturbative solution we will soon encounter.
Unfortunately, no solution valid at all orders for  the shift $\bar{\N}_i$ can be found. }.
As well known, we will only need first order constraints to compute the quadratic and cubic order interactions, whereas constraints of order $n\geq2$ contribute to interactions of order $\mathcal{H}^{(\geq2n)}$.

The $0^{\rm th}$ order Hamiltonian is simply proportional to the first Friedmann equation and therefore vanishes trivially.
Next, the first-order Hamiltonian reads:
\begin{align}
\mathcal{H}^{(1)} = &\left(
 \frac{1}{2} a^3 \dot\phi {}^2
 + a^3 V(\phi )- 3 a^3 H^2
\Mp^2 \right)\alpha^{(1)}
 \notag\\&+ \left(
  6 a^3 H^2 \epsilon \Mp^2- 9 a^3 H^2\Mp^2
 - \frac{3}{2} a^3 \dot\phi {}^2
 + 3 a^3 V(\phi )\right)\zeta   \,,
\end{align}
which also vanishes upon using the two Friedmann equations.
Seen differently, by varying $\mathcal{H}^{(1)}$ with respect to $\alpha^{(1)}$ and $\zeta$, we would find respectively the first and second Friedmann equations.
In the following, we always enforce the Friedmann equations to hold, which sets $\mathcal{H}^{(0)}=0=\mathcal{H}^{(1)}$ even for off-shell calculations.

Next, we consider the quadratic Hamiltonian, which is given by:
\begin{align} \label{eq:H2_not_solved}
    \mathcal{H}^{(2)} =& 
-\frac{\p ^2}{12 a^3 \Mp^2}
 + H \p  \alpha^{(1)}
 - a^3 H^2 \epsilon  \Mp^2 \left(\alpha^{(1)}\right)^2 
 + a \Mp^2 \epsilon  \left(\partial \zeta \right)^2+ \frac{\p   }{3 a^2 }\left(\partial_i \mathcal{N}^{(1)}_i+\frac{  \partial^2\zeta }{ H}\right)
\notag\\& 
 - \frac{\Mp^2 \partial^2\zeta}{3 a H}\left( \frac{\partial^2\zeta}{H}+2\partial_i \mathcal{N}^{(1)}_i\right)
 - \frac{\Mp^2 \left(\partial_i\mathcal{N}^{(1)}_j \right)^2}{4 a}  - \frac{\Mp^2 \left(\partial_i\mathcal{N}^{(1)}_i\right)^2}{12 a}.
\end{align}
Varying it with respect to $\alpha^{(1)}$ and $\mathcal{N}^{(1)}_i$, we obtain the linear order constraints, which read:
\begin{align}
&H p_\zeta  -2 a^3 \Mp^2 \bar{\alpha}^{(1)} H^2 \epsilon=0,\\
&-\frac{\partial_i p_\zeta  }{3 a^2 \Mp^2}+\frac{2 \partial_i \partial^2\zeta }{3 a H}+\frac{\partial_i \partial^2\bar{\theta}^{(1)}}{3 a}+\frac{\partial^2 \bar{b}^{(1)}_i}{4
   a}=0.
\end{align}
Recalling that $b^{(1)}_{i}$ is by definition traceless, we can take the divergence of the second equation to arrive at:
\begin{align}
\label{eq: linear constraints}
    \bar{\alpha}^{(1)}=\,&\frac{p_\zeta }{2 a^3\Mp^2 H \epsilon },\\
    \bar{b}_i^{(1)}=\,&0,\\
    \bar{\theta}^{(1)}=\,&\chi-\frac{\zeta }{H},
\end{align}
where we have defined 
\begin{equation}
\label{eq:chi}\partial^2\chi\equiv p_\zeta /2 a\Mp^2.
\end{equation}
Inserting the solutions for the linear constraints into Eq.~\eqref{eq:H2_not_solved}, we get the on-shell quadratic Hamiltonian
\begin{equation}
\label{eq:H2_solved}
    \bar{\mathcal{H}}^{(2)} = \frac{1}{\Mp^2}\frac{p_\zeta
   ^2}{4 a^3 \epsilon }+a \epsilon \Mp^2  \left(\partial_i\zeta \right){}^2,
\end{equation}
which is the standard form of the quadratic Hamiltonian for single-field inflation.
In perturbation theory, the interaction picture is often defined with this quadratic Hamiltonian as the free Hamiltonian dictating equations of motion for the interaction picture field $\zeta^I$ and momentum $p_\zeta^I=2 \Mp^2 a^3 \epsilon \dot{\zeta}^I$.

\subsection{Cubic Hamiltonian} 
\label{sec:cubic_hamiltonian}
We now consider cubic interactions. Expanding the Hamiltonian~\eqref{eq:Hamiltonian_final_nonpert} up to third order in perturbations, we get:
\begin{align}
    \mathcal{H}^{(3)}=&\Mp^2\Biggl\{ \left(3\zeta -\alpha^{(1)}\right)\left(\frac{\p ^2}{12 a^3 \Mp^4}
 - a^3 H^2 \epsilon \left(\alpha^{(1)}\right)^2  
 + \frac{\left(\partial^2\theta^{(1)}\right)^2}{6 
a}\right)+(\alpha^{(1)}
 - 2 \zeta ) \frac{p_\zeta  \partial^2\zeta }{3 a^2 H \Mp^2}\notag\\
&+\left(3\alpha^{(1)}
 - \zeta \right) \frac{\left(\partial_i\partial_j\theta^{(1)}\right)^2}{6 a}-\left(\alpha^{(1)}
 - \zeta \right)\frac{ (\partial^2\zeta)^2 }{3 a 
H^2}+ \frac{a \epsilon }{3}\left(\partial\zeta \right)^2 \left(\frac{\p }{2 a^3 H \epsilon\Mp^2}
 +  3\zeta 
\right) \notag\\& -\frac{1}{3 a^2 \Mp^2}\zeta  \partial_i\p  \partial_i\theta^{(1)}+ \frac{2 }{3 a H}\zeta 
\partial_i\partial^2\zeta  \partial_i\theta^{(1)}
 - \frac{2 }{3 a}\partial^2\zeta  \left(\partial\theta^{(1)}\right)^2
 - \frac{1 }{3  a H^2}\left(\partial\zeta \right)^2 \partial^2\zeta\notag\\&-\left(\frac{\p  \zeta }{a^2 \Mp^2}
 +\frac{\left(\partial_i\zeta \right)^2}{3 a H}
 - \frac{4 \zeta  \partial^2\zeta }{3 a H}\right)\partial^2\theta^{(1)}\Biggr\}\,.
\label{eq:H3_interaction_initial}
\end{align}

As well known, the lapse and shift at second order  do not appear in the cubic Hamiltonian, as interactions involving them can be recast into total spatial derivatives.
The expression of $\mathcal{H}^{(3)}$ in Eq.~\eqref{eq:H3_interaction_initial} is quite cumbersome,
but its on-shell evaluation can be written in a more compact form:
\begin{align}
    \bar{\mathcal{H}}^{(3)}=&-\Mp^2\Biggl\{\left(3\zeta-\frac{p_\zeta}{2 a^3 \epsilon H \Mp^2}\right)\Biggl[\frac{p_\zeta^2}{4 a^3 \epsilon  \Mp^4}+\frac{1}{2a}\left(\partial_i\partial_j\bar{\theta}^{(1)}\partial_i\partial_j\bar{\theta}^{(1)}-\left(\partial^2\bar{\theta}^{(1)}\right)^2\right)\Biggr]
 \notag\\&-a\epsilon\,\zeta (\partial \zeta)^2-\frac{2}{a}\,\partial_i \zeta\,\partial_i \bar{\theta}^{(1)}\,\partial^2\bar{\theta}^{(1)}\Biggr\}\,.
\label{eq:H3_interaction}
\end{align}
It can be easily checked that, upon writing this expression in terms of the interaction picture fields and momenta,  Eq.~\eqref{eq:H3_interaction}  
matches exactly the starting point for the third order bulk Lagrangian derived in~\cite{Garcia-Saenz:2019njm}, up to an overall minus sign, as expected.
Note however the absence of the total time derivatives denoted $\mathcal{D}_0$ in that reference, which instead have already been fully evacuated (at all orders in perturbation theory) by the canonical transformation Eq.~\eqref{eq: non linear canonical transf}.

Clearly, the interactions in Eq.~\eqref{eq:H3_interaction} are not all sufficiently slow-roll suppressed, even when plugging interaction picture momenta $p_\zeta \rightarrow p_\zeta^I\propto \epsilon \dot{\zeta}_I$.
In order to make the suppression manifest, we thus seek a canonical transformation as outlined in Section~\ref{sec: canonical transformations}.
As we have already explained, we should strictly speaking perform them at the level of the off-shell quadratic and cubic Hamiltonians in Eqs.~\eqref{eq:H2_not_solved}--\eqref{eq:H3_interaction_initial}.
However, it can be checked that using the on-shell ones gives a wrong answer starting at quartic order only, so we do so for simplicity in this section and we will come back to this issue in Sec.~\ref{sec:quartic_Hamiltonian}.
Let us look at a subset of the total cubic Hamiltonian to illustrate these simplifications:
\begin{equation}
\label{eq: contrib Ham A}
    \mathcal{H}^{(3)}(\zeta,p_\zeta) \supset  \mathcal{H}^{(3)}_A (\zeta,p_\zeta) = -\frac{1}{4 a^3 \epsilon  \Mp^2} \left(3\zeta-\frac{p_\zeta}{2 a^3 \epsilon H \Mp^2}\right) p_\zeta^2 \,.
\end{equation}
Upon plugging interaction picture fields and momenta, this contribution ``A'' reduces to $-\mathcal{L}^{(3)}_A(\zeta_I,\dot{\zeta}_I)$ of the Lagrangian approach that we have already encountered, see Eq.~\eqref{eq: 2 contrib to L3}.
We want to incorporate the dominant $p_\zeta^3$ interaction into the kinetic term for a new momentum variable $p_\zeta \rightarrow \tilde{p}_\zeta$, i.e.
\begin{equation}
    \frac{1}{\Mp^2}\frac{p_\zeta
   ^2}{4 a^3 \epsilon }+
   \frac{p_\zeta ^3}{8 a^6 H \epsilon ^2\Mp^4} \rightarrow  \frac{1}{\Mp^2}\frac{\tp
   ^2}{4 a^3 \epsilon } \,.
\end{equation}
Inverting this equation to find the desired $p_\zeta$ as a function of $\tp$, we find the following expression to remove the large cubic interaction:
\begin{equation}
\label{eq: old p Ham A}
    p_\zeta \equiv \tilde{p}_\zeta - \frac{1}{4 a^3 \epsilon H \Mp^2} \tp^2,
\end{equation}
from which, using $p_\zeta=\delta F/\delta \zeta$, we get the corresponding type II generating function:
\begin{equation}
    F\left(\z , \tp, t\right) = \z\tp - \frac{1}{4 a^3 \epsilon H \Mp^2} \tp^2 \z + f(\tp,t)\,.
\end{equation}
We can now deduce the consistent transformation law for the position variable $\zeta$, in order for the new phase-space variables to be canonically conjugate one to the other: 
\begin{equation}
     \tilde{\zeta}=  \frac{\delta F}{\delta \tilde{p}_\zeta} = \z - \frac{\tp \z}{2 a^3 \epsilon H \Mp^2} + \frac{\partial f}{\partial \tilde{p}_\zeta} \,. 
\end{equation}
Since we are free to choose $f=0$ for simplicity, we do so.
We can then invert perturbatively this last equation, to find:
\begin{equation}
\label{eq: old z Ham A}
    \zeta = \tz +  \frac{\tp \tz}{2 a^3 \epsilon H \Mp^2} + \ldots
\end{equation}
with dots denoting terms of order three or more in the new phase-space variables.
Note that, following this procedure, the expression $p_\zeta(\z,\tp)$ is exact while the one for $\z(\tz,\tp)$---and therefore also the one $p_\zeta(\tz,\tp)$ in the general case, although not relevant here---is only perturbative.
This needs not be a problem as long as calculations are performed consistently at a given order in perturbation theory.
Another contribution is given by the time partial derivative of the generating function,
\begin{equation}
   \frac{\partial F}{\partial t}=\frac{1}{4 a^3 \Mp^2}\left(-1
 + \frac{3}{\epsilon }
 + \frac{\eta }{ \epsilon }
  \right) \tp^2 \z.
\end{equation}
Inserting Eqs.~\eqref{eq: old p Ham A} and~\eqref{eq: old z Ham A} into Eqs.~\eqref{eq:H2_solved} and~\eqref{eq: contrib Ham A}, and adding $\partial F/\partial t$ given above as we should, we get a new Hamiltonian $\tilde{\mathcal{H}} = \tilde{\mathcal{H}}^{(2)} + \tilde{\mathcal{H}}^{(3)}_A + \ldots$ with
\begin{align}
    \tilde{\mathcal{H}}^{(2)}(\tz,\tp) &= \mathcal{H}^{(2)}(\tz,\tp) \\
    \tilde{\mathcal{H}}^{(3)}_A(\tz,\tp) &= \frac{\eta-\epsilon}{4a^3 \epsilon \Mp^2} \tp^2\tz - \frac{1}{2 a^2 H} \tz \tp \partial^2 \tz \,.
\end{align}
A few remarks are in order.
First, $p_{\zeta}^3$ interactions have been removed and the explicit size of the $p_\zeta^2 \zeta$ interaction has been reduced from order $\epsilon$ to order $\epsilon(\epsilon-\eta)$ (after inserting interaction picture fields and momenta), and is therefore further suppressed by slow-roll parameters.
Second, in stark contrast with the approach with integrations by parts in the Lagrangian, we have not generated neither total time derivatives nor terms proportional to the linear equations of motion, but simply defined a canonical transformation $(\zeta, p_\zeta) \rightarrow (\tz,\tp)$.
Third and finally, we have generated a new cubic interaction with the operator $\tz \tp \partial^2 \tz$, but just like in the Lagrangian approach (see Eq.\eqref{eq: combine with LB}) it can be combined with another contribution to perform an additional canonical transformation and simplify further the Hamiltonian.

We perform canonical transformations step by step to simplify all types of cubic Hamiltonian interactions.
The final type II generating function is
\begin{align}
    F\left(\z , \tp, t\right)  &= \z\tp - \frac{1}{4 a^3 \epsilon H \Mp^2} \tp^2 \z - \frac{a \epsilon \Mp^2 }{H} \z (\partial \z)^2 - \frac{\Mp^2}{6a H^3 }\z \left[\z_{ij}\z_{ij}-(\partial^2  \z)^2\right] \notag\\
&
    + \frac{\Mp^2}{2 a H^2 } \z \left[\z_{ij} \tilde{\chi}_{ij} - \partial^2 \z \partial^2 \tilde{\chi} \right]
    - \frac{\Mp^2}{2 a H} \z \left[\tilde{\chi}_{ij} \tilde{\chi}_{ij} - (\partial^2 \tilde{\chi})^2 \right] \,, 
\label{eq:generating_function_H3}
\end{align}
with $\tilde{\chi}$ given by Eq.~\eqref{eq:chi}, but applied to $\tp$ instead of $\p$.
From the generating function, we can derive
\begin{align}
\label{eq:canonical_transf_SF_z}
\tilde{\zeta}=&\z - \frac{\tp \z}{2 a^3 \epsilon H \Mp^2}  + \frac{1}{4 a^2 H^2 }  \left[\partial^{-2}\partial_i\partial_j\left(\z\,\z_{ij}\right)  - \zeta\partial^2 \z \right]
    - \frac{1}{2 a^2 H}  \left[\partial^{-2}\partial_i\partial_j \left(\z \tilde{\chi}_{ij}\right) - \zeta\partial^2 \tilde{\chi} \right],\\
        p_\zeta=&\tilde{p}_\zeta - \frac{1}{4 a^3 \epsilon H \Mp^2} \tp^2  + \frac{a \epsilon \Mp^2 }{H}  (\partial \z)^2 +\frac{2a \epsilon \Mp^2 }{H} \z \partial^2 \z- \frac{\Mp^2}{6a H^3 } \left[\z_{ij}\z_{ij}-(\partial^2  \z)^2\right]\notag\\&- \frac{\Mp^2}{3a H^3 } \left[\partial_i\partial_j\left(\z\,\z_{ij}\right)-\partial^2(\zeta \,\partial^2  \z)\right]
    + \frac{\Mp^2}{2 a H^2 }  \left[\z_{ij} \tilde{\chi}_{ij} - \partial^2 \z \partial^2 \tilde{\chi} \right]- \frac{\Mp^2}{2 a H}  \left[\tilde{\chi}_{ij} \tilde{\chi}_{ij} - (\partial^2 \tilde{\chi})^2 \right]\notag\\
\label{eq:canonical_transf_SF_p}&
    + \frac{\Mp^2}{2 a H^2 }  \left[\partial_i\partial_j\left(\z\,\tilde{\chi}_{ij}\right) - \partial^2 \left(\z \partial^2 \tilde{\chi}\right) \right].
\end{align}
As mentioned above, although the generating function is cubic in $(\zeta,\tilde{p}_\zeta)$ and therefore the above expressions for $(\tilde{\zeta},p_\zeta)$ are exactly quadratic in the same variables, inverting the relation for $\zeta$ to express old variables in terms of new variables can only be done perturbatively, schematically as
\begin{align}
\label{eq:perturbative_expansion_ct}    \zeta & \equiv \zeta^{(1)}\left(\tilde{\zeta},\,\tilde{p}_\zeta\right) + \zeta^{(2)}\left(\tilde{\zeta},\,\tilde{p}_\zeta\right)+ \zeta^{(3)}\left(\tilde{\zeta},\,\tilde{p}_\zeta\right)+\cdots\\
    p_\zeta & \equiv p_\zeta^{(1)}\left(\tilde{\zeta},\,\tilde{p}_\zeta\right) + p_\zeta^{(2)}\left(\tilde{\zeta},\,\tilde{p}_\zeta\right)+ p_\zeta^{(3)}\left(\tilde{\zeta},\,\tilde{p}_\zeta\right)+\cdots 
\end{align}
where $\zeta^{(1)}=\tilde{\zeta}$ and $p_\zeta^{(1)} = \tilde{p}_\zeta$. 
Several effects must be taken into account.
Let us list them consistently.

First, expressing the old Hamiltonian $\mathcal{H}=\mathcal{H}^{(2)}+\mathcal{H}^{(3)}+\ldots$ in terms of the new variables leads to corrections at cubic order and higher ones.
The old quadratic Hamiltonian gives 
\begin{equation}
    \mathcal{H}^{(2)}\left(\zeta(\tilde{\zeta},\tilde{p}_\zeta),p_\zeta(\tilde{\zeta},\tilde{p}_\zeta)\right) =   \mathcal{H}^{(2)}\left(\tilde{\zeta},\tilde{p}_\zeta\right) + \Delta \tilde{\mathcal{H}}^{(3)\,\mathrm{from}\,(2)} +  \ldots
\end{equation}
with
\begin{equation}
\label{eq:H3from2}
    \Delta \tilde{\mathcal{H}}^{(3)\,\mathrm{from}\,(2)} =\sum_A\left.\frac{\delta \mathcal{H}^{(2)}}{\delta z^A}\right|_{\tilde{z}}\left(z^A\right)^{(2)},
\end{equation}
where $A,\,B=1,\,2$, $z^A =(\z,\,\p)$ and $\tilde{z}^A =(\tz,\,\tp)$. \footnote{In a more general settings with multiple fields, or including tensors, $z^A$ would also contain them as well as their associated momenta.}
We have not written terms of order four and more.
For notation simplicity, we have denoted by the symbol $\delta/\delta ( \zeta,p_\zeta)$ the functional derivative including the correct factors of (inverse) spatial derivatives operators.
The old cubic Hamiltonian gives
\begin{equation}
    \mathcal{H}^{(3)}\left(\zeta(\tilde{\zeta},\tilde{p}_\zeta),p_\zeta(\tilde{\zeta},\tilde{p}_\zeta)\right) =  \mathcal{H}^{(3)}\left(\tilde{\zeta},\tilde{p}_\zeta\right)  + \ldots
\end{equation}
with corrections starting at quartic order only.

The generating function also contributes to the new Hamiltonian via its partial derivative with respect to time.
Explicitly, the correct contribution to $\tilde{\mathcal{H}}(\tilde{\zeta},\tilde{p}_\zeta,t)$ is
\begin{equation}
    \left.\frac{\partial F(\zeta,\tilde{p}_\zeta,t)}{\partial t}\right|_{\zeta\left(\tilde{\zeta},\,\tilde{p}_\zeta\right),\tilde{p}_\zeta} =  \Delta \tilde{\mathcal{H}}^{(3)\,\mathrm{from}\,F} + \ldots=\left.\frac{\partial F(\zeta,\tilde{p}_\zeta,t)}{\partial t}\right|_{\tilde{z}} + \ldots,
\end{equation}
where it is important to take the partial time derivative \textit{before} plugging in the expression for $\zeta$ in terms of the new variables, and where once more higher orders are not written explicitly. 

An important consequence of having only cubic corrections to the trivial term $\zeta \tp$ in the generating function is that our canonical transformation affects the Hamiltonian only starting at cubic order.
It therefore leaves invariant the background theory, as it should, and does not create tadpoles with linear order terms.
Moreover, it leaves invariant the quadratic Hamiltonian, so that in the corresponding perturbation theory, the free theory is invariant under the canonical transformation:
\begin{equation}
\tilde{\mathcal{H}}^{(2)}\left(\tz,\, \tp \right)=\mathcal{H}^{(2)}\left(\tz,\, \tp \right) \,.
\end{equation}
In particular, the linear equations of motion for interaction picture fields defined from this quadratic Hamiltonian are not affected.
However, non-linear interactions are affected.
Focusing for now on the cubic order terms in the Hamiltonian, we have:
\begin{equation}
\tilde{\mathcal{H}}^{(3)}\left(\tz,\, \tp \right)=\mathcal{H}^{(3)}\left(\tz,\, \tp \right)+\Delta \tilde{\mathcal{H}}^{(3)\,\mathrm{from}\,(2)}+\Delta \tilde{\mathcal{H}}^{(3)\,\mathrm{from}\,F}
\end{equation}
After some straightforward simplifications, the canonically transformed cubic Hamiltonian, on shell, becomes:
\begin{equation}
\label{eq:H3_sr_suppressed}
\bar{\tilde{\mathcal{H}}}^{(3)} = \frac{\eta-\epsilon}{4a^3 \epsilon \Mp^2} \tp^2\tz - \epsilon(\epsilon+\eta) a \Mp^2 \tz (\partial \tz)^2 +\frac{\Mp^2}{a}(2-\epsilon/2) (\partial \tz)(\partial \tilde{\chi}) \partial^2 \tilde{\chi} - \frac{\epsilon \Mp^2}{4a} \partial^2 \tz (\partial \tilde{\chi})^2 \,.
\end{equation}
Upon inserting the interaction picture fields and momenta, this equation reduces to the bulk part of the cubic scalar Lagrangian after all simplifications in the traditional approach, see Eq.~\eqref{eq:cubic action scalars}.
The important point, however, is that no integration by parts has been performed in this Hamiltonian approach, but rather a canonical transformation to new variables.
The consequence is the absence of total time derivative interactions in the interaction Hamiltonian of the corresponding perturbation theory, leading to simpler in-in formulas.
Our canonical transformation generalizes Maldacena's change of variable~\cite{Maldacena:2002vr} to the whole phase space without invoking cancellations of total time derivatives or terms proportional to the equations of motion.
From Eqs.~\eqref{eq:canonical_transf_SF_z} and~\eqref{eq:canonical_transf_SF_p}, the old variables are connected to the new ones via the following canonical transformation:
\begin{align}
\label{eq:z_of_tz}
    \zeta= &\, \tz + \frac{\tp \tz}{2 a^3 \epsilon H \Mp^2}  - \frac{1}{4 a^2 H^2 }  \left[\partial^{-2}\partial_i\partial_j\left(\tz\,\tz_{ij}\right)  - \tz\partial^2 \tz \right]
    + \frac{1}{2 a^2 H}  \left[\partial^{-2}\partial_i\partial_j \left(\tz \tilde{\chi}_{ij}\right) - \tz \partial^2 \tilde{\chi} \right] \nonumber \\
     &\, + \ldots  \\
    \label{eq:p_of_tz}
    p_\zeta= &\, \tilde{p}_\zeta - \frac{1}{4 a^3 \epsilon H \Mp^2} \tp^2  + \frac{a \epsilon \Mp^2 }{H}  (\partial \tz)^2 +\frac{2a \epsilon \Mp^2 }{H} \tz \partial^2 \tz- \frac{\Mp^2}{6a H^3 } \left[\tz_{ij}\tz_{ij}-(\partial^2  \tz)^2\right]  \\ 
    &\, - \frac{\Mp^2}{3a H^3 } \left[\partial_i\partial_j\left(\tz\,\tz_{ij}\right)-\partial^2(\tz \,\partial^2  \tz)\right]
    + \frac{\Mp^2}{2 a H^2 }  \left[\tz_{ij} \tilde{\chi}_{ij} - \partial^2 \tz \partial^2 \tilde{\chi} \right]- \frac{\Mp^2}{2 a H}  \left[\tilde{\chi}_{ij} \tilde{\chi}_{ij} - (\partial^2 \tilde{\chi})^2 \right] \nonumber \\
    &\, 
    + \frac{\Mp^2}{2 a H^2 }  \left[\partial_i\partial_j\left(\tz\,\tilde{\chi}_{ij}\right) - \partial^2 \left(\tz \partial^2 \tilde{\chi}\right) \right] +\ldots \nonumber
\end{align}
where we have consistently included only quadratic corrections.
Indeed, although our canonical transformations technically include cubic and higher-order corrections to these equations, those terms only contribute to correlation functions that also get affected by quartic and higher order interactions that we have not consistently kept track of.

In particular, correlation functions of $\z$ (and similarly for those of $\p$ or mixed correlators) can be related to those of $\tz$ and $\tp$ via Eqs.~\eqref{eq:z_of_tz} and~\eqref{eq:p_of_tz}, which are computed using the new $\tilde{\mathcal{H}}^{(3)}$.
For illustration, let us relate the late-time bispectrum of $\zeta$, i.e. its Fourier space 3-point correlation function on super-horizon scales, defined as:
\beq
\underset{\forall i\,, - k_i \tau \rightarrow 0}{\mathrm{lim}} \,\,
\langle
\zeta_{\boldsymbol{k}_1}
\zeta_{\boldsymbol{k}_2} \zeta_{\boldsymbol{k}_3} \rangle (\tau) \equiv (2\pi)^3 \delta\left( \boldsymbol{k}_1 + \boldsymbol{k}_2 + \boldsymbol{k}_3\right) B_{\zeta}(k_1,k_2,k_3)\,,
\label{eq:bispectrum}
\eeq
to the one of $\tilde{\zeta}$, $B_{\tz}$ that can be computed from the Hamiltonian~\eqref{eq:H3_sr_suppressed} or equivalently from the Lagrangian path integral approach,
soon after horizon crossing for these modes.
Indeed, a subtlety comes about here.
In general, we are only able to compute correlation functions with the in-in formalism in a given range of $e$-folds where we consider all background quantities to be slowly varying.
When we compute correlation functions for $\zeta$ directly, this makes little difference as $\zeta$ generally reaches an adiabatic limit on super-horizon scales, at least in attractor scenarios of single-field inflation.
In that case, it is enough to compute the evolution of the correlation functions from deep inside the Hubble radius to slightly after Hubble crossing when the fluctuations freeze.
However, when the variable used differs from $\zeta$, like $\tz$ here, we need to be careful and in practice we are only able to relate correlation functions of $\zeta$ to those of $\tz$ soon after horizon crossing.
The other option, perfectly valid, consists in following exactly the (slow) evolution of correlators of $\tz$ until the end of inflation on super-horizon scales, and only there relate them to those of $\zeta$.
Here we follow the first route, better adapted to analytical calculations.

By using the Fourier transform of Eq.~\eqref{eq:z_of_tz} at a time $\tau_\star$ slightly after horizon crossing for all three wavenumbers, i.e. $\forall i\,, \, - k_i \tau_\star \ll 1$, we find:
\begin{align}
    B_\z = B_\z^\star = B_{\tz}^\star&- 2\, \b(-\k_1,-\k_2,-\k_3) P_\tz^\star(k_1) P_\tz^\star(k_2)\,\, +2\, {\rm perm.}\nonumber\\&-
2\, \h(-\k_1,-\k_2,-\k_3) P_\tz^\star(k_3) P_{\tz \tp}^\star(k_2) \,\,+5\, {\rm perm.}\,,
\label{eq:bispectrum_old_new}
\end{align}
where we have defined
\bea
\b(\k_1,\k_2,\k_3) &=&\frac{1}{8 a^2 H_\star^2} \left(k_1^2+k_2^2-(\k_1 \cdot \hat{\k}_3)^2-(\k_2 \cdot \hat{\k}_3)^2 \right) \\
\h(\k_1,\k_2,\k_3) &=&\frac{1}{8 a^3 \epsilon_\star H_\star \Mp^2} \left[-2+\epsilon_\star \,(1-(\hat{\k}_1 \cdot \hat{\k}_2)^2) \right]\,,
\eea
a starred quantity being evaluated around the time $\tau_\star$,
and where $\hat{\k}_i$ is the unit vector $\k_i/k_i$.
Note that $P_{\tz \tp}(k)=\frac12 \left(\langle \tz(\k) \tp(-\k) \rangle^{'}+\langle \tp(\k) \tz(-\k) \rangle^{'} \right)$ is the real cross-spectrum of $\tz$ and $\tp$, and $P_\tz(k)$ is the power spectrum of $\tz$.
At tree-level, those power spectra are equal to those of $\z,\p$.
The result for the bispectrum obtained with our formalism exactly matches the one computed in~\cite{Garcia-Saenz:2019njm} obtained using the in-in formalism with boundary terms outlined in Section~\ref{sec:in-in}, and we have intentionally adopted their same notation. 

Given this canonical transformation, we can see that correlation functions of $\zeta$ will be equal to those of $\tz$, plus correction terms either proportional to correlation functions including the momentum $\tp$ or spatial gradients.
Therefore, in single-field inflation with an attractor phase, for cosmological scales of observational relevance, one can conclude that the above canonical transformation used to simplify the cubic Hamiltonian, will only produce negligible corrections.
This conclusion is much easier to arrive at using the canonical transformations approach rather than the IBP one, for in the latter case we have seen that the in-in perturbation theory quickly becomes cumbersome with cancellations to seek for between different vertex orders.

\subsection{Scalar and tensor interactions}
\label{subsec: tensors}
So far, we have exclusively examined scalar interactions, but it is instructive to apply our formalism to tensor interactions as well. Therefore, we extend the results of the previous section to encompass tensor perturbations. We present the cubic Hamiltonian for tensors and mixed scalar-tensor interactions. Our findings can be used for calculating correlation functions involving tensor perturbations, such as in computations related to scalar-induced gravitational waves~\cite{Domenech:2021ztg} and gravitational wave anisotropies~\cite{Dimastrogiovanni:2022afr}. Additionally, we highlight an intriguing aspect of our approach—a nontrivial contribution of tensor perturbations to scalar interactions and vice versa, beginning at quartic order in interactions. 

Including tensor perturbations, the spatial metric in the comoving gauge takes the following form:
\begin{align}
	h_{ij}&=a^2e^{2\zeta}(e^\gamma)_{ij} \label{eq:spatial_metric_with_tensors},
\end{align}
where the tensor perturbation $\gamma_{ij}$ is transverse  and traceless, i.e. $\partial_i\gamma_{ij}=\gamma_{ii}=0$.
Let us note that~\eqref{eq:spatial_metric_with_tensors} is just a shorthand and it should be understood as follows
\begin{equation}
 h_{ij}=a^2e^{2\zeta}(e^\gamma)_{ij} = a^2e^{2\zeta} \left(\delta_{ij}+\gamma_{ij}+\frac{1}{2}\gamma_{i l}\gamma_{lj}+\frac{1}{6}\gamma_{i l}\gamma_{lm}\gamma_{mj}+\dots\right).~\label{eq:spatial_metric_with_tensors_expanded}
\end{equation}
Using the exponential of $\gamma_{ij}$ to carry out an analysis for the tensor sector valid at all orders in perturbation theory is not as straightforward as for scalar perturbations, so that we need to use the expression~\eqref{eq:spatial_metric_with_tensors_expanded} instead. Plugging it into the action~\eqref{eq:action-ADM} we get:
\begin{equation}
\label{eq:lagrangian_tensors_tot}
  \mathcal{L}=\mathcal{L}_{\rm scalars}  + \mathcal{L}_{\gamma} + \mathcal{L}_{\gamma\gamma} + \mathcal{L}_{\gamma\gamma\gamma} + \mathcal{L}_{\gamma\gamma\gamma\gamma}+\cdots,
\end{equation}
where $\mathcal{L}_{\rm scalars}$ is given by the scalar Lagrangian in Eq.~\eqref{eq:action_nonpert_withboundary}, $\mathcal{L}_{\gamma},\, \mathcal{L}_{\gamma\gamma},\, \mathcal{L}_{\gamma\gamma\gamma},\,\mathcal{L}_{\gamma\gamma\gamma\gamma}$ contain 1, 2, 3 and 4 powers of $\gamma_{ij}$ or its spatial or time derivatives as well as other powers of the scalar perturbation $\zeta$.
Explicit expressions for each term in Eq.~\eqref{eq:lagrangian_tensors_tot} are provided in the Appendix~\ref{app:tensors_L_and_H}. 
 From such expressions, we can derive
the on-shell conjugate momenta $\bar{p}_X\equiv\delta L/\delta \dot{X}$:

\begin{align}
\label{eq:momentum_zeta_with_tensors}
    \bar{p}_\zeta=\Mp^2\Biggl(&-\frac{6 a^3 e^{3 \zeta } \dot{\zeta} }{N}+\frac{2 a
   e^{\zeta } \partial_i\mathcal{N}_i}{N}+\frac{2 a
   e^{\zeta } \mathcal{N}_i \partial_i\zeta
   }{N}-\frac{6 a^3 e^{3 \zeta } H}{N}-\frac{2 a e^{\zeta } \partial_j\mathcal{N}_i \,\gamma
_{ij}}{N }\Biggr)+\cdots,\\
\label{eq:momentum_gamma_with_tensors}
    \bar{p}_{i j}=\Mp^2\Biggl(&\frac{a^3 e^{3 \zeta } \dot\gamma _{ij}}{4 N}
 - \frac{a e^{\zeta }  \mathcal{N}_i \partial_jN}{2 N^2}
 + \frac{3 a e^{\zeta }   \mathcal{N}_i \partial_j\zeta}{2 N}
 - \frac{a e^{\zeta }  \mathcal{N}_k \partial_k\gamma_{ij}}{4 N}
 + \frac{a e^{\zeta }  \mathcal{N}_k \partial_j\gamma_{ik}}{2 N}
  \notag\\ &+ \frac{a e^{\zeta } \partial_k\mathcal{N}_j  \gamma_i{}_k}{4 N}
+ \frac{a e^{\zeta } \partial_j\mathcal{N}_k \gamma_i{}_k}{4 N}\Biggr)_{\rm symm}+\cdots,
\end{align} where the subscript ``${\rm symm}$'' denotes symmetrization with respect to the indices $i,\,j$, i.e. $A_{ij}\rvert_{\rm symm}\equiv(A_{ij}+A_{ji})/2$,
and the conjugate momenta to $N$ and $\mathcal{N}_i$ are still zero.
The dots denote cubic order terms in the expressions of  the momenta $\bar{p}_\zeta$ and $\bar{p}_{ij}$, and only become relevant for the computation of quintic or higher order interactions, which we are not concerned with in this paper.

We immediately notice that the momentum $\bar{p}_\zeta$ differs from that in Eq.~\eqref{eq:momentum_zeta} by the term $-2 a \Mp^2 e^{\zeta } \partial_j\mathcal{N}_i \,\gamma_{ij}\, / N $. Furthermore, the tensor momentum $\bar{p}_{ij}$ receives contributions from products of scalar perturbations.
This fact will play a crucial role in deriving the quartic Hamiltonian in phase space.
For example, the kinetic terms $\sim\dot{\zeta}^2$ and $\sim(\dot{\gamma}_{ij})^2$ will also generate purely tensor and scalar quartic interactions respectively, as well as mixed scalar-tensor quartic interactions.
We will come back to this matter in Sec.~\ref{sec:quartic_Hamiltonian}.

Using Eqs.~\eqref{eq:lagrangian_tensors_tot}, ~\eqref{eq:momentum_zeta_with_tensors} and~\eqref{eq:momentum_gamma_with_tensors}, we can perform the Legendre transform, followed by the canonical transformation~\eqref{eq:first_canonical_transformation} arrive at the Hamiltonian.
Up to cubic order, the Hamiltonian consists in the following contributions:

\begin{equation}
    \mathcal{H} = \mathcal{H}^{(2)}_{\rm scalars}+\mathcal{H}^{(2)}_{\rm tensors}+\mathcal{H}^{(3)}_{\zeta\zeta\zeta}+\mathcal{H}^{(3)}_{\zeta\zeta\gamma}+\mathcal{H}^{(3)}_{\zeta\gamma\gamma}+\mathcal{H}^{(3)}_{\gamma\gamma\gamma}
\end{equation}
where the purely scalar interactions $\mathcal{H}^{(2)}_{\rm scalars}$ and $\mathcal{H}^{(3)}_{\zeta\zeta\zeta}$ are given by Eqs.~\eqref{eq:H2_solved} and ~\eqref{eq:H3_interaction} respectively.
The evolution of the interaction picture tensor degrees of freedom is often set to be governed by their quadratic Hamiltonian, which reads:
\begin{equation}
   \mathcal{H}^{(2)}_{\rm tensors}=\frac{2 p_{ij}^2}{a^3 \Mp^2}
 + \frac{\Mp^2}{8} a \left(\partial_k\gamma _{ij} \right)^2 \,,
\end{equation}
in which case $p_{ij}^I =  a^3 \Mp^2 \dot{\gamma}^I_{ij}$.
The cubic interactions involving tensors take the following form
\begin{align}
 \mathcal{H}^{(3)}_{\zeta\zeta\gamma}=&
\frac{6 \zeta  p_{ij} \partial_i\partial_j\theta^{(1)}}{a^2}
 - \frac{p_\zeta  p_{ij} \partial_i\partial_j\theta^{(1)}}{a^5 H 
\epsilon \Mp^2}
 - \frac{\Mp^2}{2 a}\theta^{(1)} \partial_i\partial_j\partial_k\theta^{(1)} \
\partial_k\gamma _{ij}
 - a \Mp^2\zeta  \partial_i\partial_j\zeta  \gamma _{ij}
\notag\\& - \frac{p_\zeta  \partial_i\partial_j\zeta  \gamma _{ij}}{a^2 H \
\epsilon }\\ 
 \mathcal{H}^{(3)}_{\zeta\gamma\gamma}=&-\frac{6 \zeta  p_{ij}^2}{a^3 \Mp^2}
 + \frac{p_\zeta  p_{ij}^2}{a^6 H \epsilon \Mp^4 }
 + \frac{p_{ij} \partial_k\theta^{(1)} \partial_k\gamma _{ij}}{a^2}
 + \frac{\Mp^2}{8} a \zeta  \left(\partial_k\gamma _{ij} \right)^2
 + \frac{p_\zeta  \left(\partial_k\gamma _{ij} \right)^2}{16 a^2 H \epsilon }
 \\
 \mathcal{H}^{(3)}_{\gamma\gamma\gamma}=&- \frac{a \Mp^2}{4}  \partial_l\gamma_{ik}\, \partial_k\gamma_{jl} \,
\gamma _{ij}
 + \frac{a \Mp^2}{8}\,  \partial_k\partial_l\gamma _{ij}\, \gamma_{ij} \,
\gamma_{kl}.
\end{align}
As before,  we need to perform a canonical transformation to simplify the cubic Hamiltonian and make the slow-roll suppression manifest.
We arrive at the following generating function: 
\begin{align}
\label{eq:generating_function_tensors}
    F[\tilde{p}_\zeta,\,\zeta,\,\tilde{p}_{ij},\,\gamma_{ij},\,t]=&\tilde{p}_\zeta \zeta+\tilde{p}_{ij}\,\gamma_{ij} \\&- \frac{1}{4 a^3 \epsilon H \Mp^2} \tp^2 \z - \frac{a \epsilon \Mp^2 }{H} \z (\partial \z)^2 - \frac{\Mp^2}{6a H^3 }\z \left[\z_{ij}\z_{ij}-(\partial^2  \z)^2\right] \notag\\
&
    + \frac{\Mp^2}{2 a H^2 } \z \left[\z_{ij} \tilde{\chi}_{ij} - \partial^2 \z \partial^2 \tilde{\chi} \right]
    - \frac{\Mp^2}{2 a H} \z \left[\tilde{\chi}_{ij} \tilde{\chi}_{ij} - (\partial^2 \tilde{\chi})^2 \right]\notag \\& - \frac{a  \Mp^2}{H}\partial_i\zeta  \partial_j\zeta  \gamma _{ij}+\frac{1}{a^2 H^2} \partial_i\zeta  \partial_j\zeta  \tilde{p} _{ij}+ \frac{2 }{a^2 H \Mp^2}\tilde{\chi}  \partial_i\partial_j\zeta  \tilde{p} _{ij}
	\notag\\&- \frac{a \Mp^2}{8 H}\zeta  \left(\partial_l\gamma _{ij} \right)^2-\frac{2 }{a^3 H \Mp^2}\zeta  \tilde{p} _{ij}^2.\notag
\end{align}
The second and third lines of the generating functions are the same as in Eq.~\eqref{eq:generating_function_H3}, and only affect $\mathcal{H}^{(3)}_{\zeta\zeta\zeta}$, recasting it in the form~\eqref{eq:H3_sr_suppressed}. Terms in the fourth and fifth lines are new. Like in the scalar case, when evaluated on interaction picture fields, the fourth line turns out to be equal to the argument of total time derivative terms in the Lagrangian, as can be easily checked by comparing with Ref.~\cite{Ning:2023ybc}, where total time derivative tensor interactions were highlighted for the fist time.

 From the generating function, we can derive the relations between the new and the old variables:
\begin{align}
    \label{eq:transf_z_w_tensors}
    \tilde{\zeta}=&\,{\rm Eq.~\eqref{eq:canonical_transf_SF_z}} -\frac{1}{a^3 H}\partial^{-2}\left(\partial_i\partial_j \zeta\,\tilde{p}_{ij}\right)\\
    \p=&\,{\rm Eq.~\eqref{eq:canonical_transf_SF_p}} + \frac{2 a \Mp^2}{H}\partial_i\partial_j\zeta\,\gamma_{ij}-\frac{2}{a^2 H^2}\partial_j\left(\partial_i\zeta\,\tilde{p}_{ij}\right)+\frac{2}{a^2 H  \Mp^2}\partial_i\partial_j\left(\tilde{\chi}\,\tilde{p}_{ij}\right)\notag\\
    \label{eq:transf_p_w_tensors}
    &-\frac{a  \Mp^2}{8 H}\left(\partial_k\,\gamma_{ij}\right)^2-\frac{2}{a^3 H  \Mp^2}\tilde{p}_{ij}^2,\\
    \label{eq:transf_gammaij}
    \tilde{\gamma}_{ij}=&\gamma_{ij}+ \frac{  \Mp^2}{a^2 H^2    }\partial_i\zeta\partial_j\zeta+ \frac{2   }{a^2 H}\tilde{\chi}  \partial_i\partial_j\zeta
	-\frac{4  }{a^3 H}\zeta  \tilde{p} _{ij},\\
 \label{eq:transf_pij}
    p_{ij}=&\tilde{p}_{ij}-\frac{a \Mp^2}{H}\partial_i\zeta\partial_j\zeta+\frac{a \Mp^2}{4H}\partial_k\left(\zeta\partial_k\gamma_{ij}\right).
\end{align}
We can invert the relations above  perturbatively  to find the old variables in terms of the new one, and compute the new Hamiltonian $\tilde{\mathcal{H}}$. As previously announced, we see that the canonical transformations mix scalar and tensor perturbations, which implies that correlation functions of the old scalar {\em or} tensor variables will be found from {\em both} new scalar {\em and} tensor variables. One may derive diagrammatic rules following the procedure outlined in Appendix~\ref{app: diagrammatic rules} to find the precise relations, although we do not do it here. 

Using the transformations above, we simplify the tensor interactions:
\begin{align}
\tilde{\mathcal{H}}_{\zeta\zeta\gamma}&=-\Mp^2\left[-\frac{2\epsilon}{a^2 \Mp^2}  \tilde{\chi}  \partial_i\partial_j\tilde{\zeta}  \tilde{p}_{ij}
	+ \frac{1}{4a\Mp^2}\partial_i\tilde{\chi}  \partial_j\tilde{\chi}  \partial^2\tilde{\gamma}_{ij}
	+ a \epsilon  \partial_i\tilde{\zeta}  \partial_j\tilde{\zeta}  \tilde{\gamma}_{ij}\right] \ ,\label{eq:H2zeta1gamma_bulk} \\
 &-\frac{1}{a^2}\left[\frac{\epsilon}{H }\partial_j\left(\tilde{\zeta}^2\right) -\frac{2}{\Mp^2}\partial_j\left(3\tilde{\zeta}-\frac{\tilde{p}_\zeta}{2 a^3 H \epsilon \Mp^2}\right)\,\tilde{\chi}\right]\partial_i \tilde{p}_{ij}\ ,\notag\\
\tilde{\mathcal{H}}_{\zeta\gamma\gamma}&=-\Mp^2\left[\frac{2  \epsilon}{a^3 \Mp^4}  \tilde{\zeta}  \tilde{p} _{ij} {}^2
	- \frac{1}{a^2 \Mp^4}\partial_l\tilde{\chi}  \tilde{p} _{ij} \partial_l\tilde{\gamma}_{ij}
	+ \frac{1}{8} a \epsilon  \tilde{\zeta}  \left(\partial_l\tilde{\gamma} _{ij} \right)^2\right]\ \label{eq:H1zeta2gamma_bulk}\\&+\frac{2}{a^2}\partial_k\tilde{\theta}^{(1)}\,\gamma_{jk}\,\partial_i\tilde{p}_{ij},\notag\\
\tilde{\mathcal{H}}_{\gamma\gamma\gamma}&=-\Mp^2\left[\frac{1}{4} a \partial_m\tilde{\gamma}_i{}_l \partial_l\tilde{\gamma}_{jm} 
	\tilde{\gamma} _{ij}
	+ \frac{1}{8} a \partial_i\tilde{\gamma}_{lm} \partial_j\tilde{\gamma}_{lm} 
	\tilde{\gamma}_{ij}\right] \label{eq:H3gamma_bulk} .
\end{align}
Upon inserting the interaction picture fields $\tilde{\gamma}^I_{ij}$ and $\tilde{p}^I_{ij}=\Mp^2 a^3 \dot{\tilde{\gamma}}^I _{ij}$, we have that  $\partial_i\tilde{p}^I_{ij}=0$ and the equation above simply reduces to the bulk interactions found in Ref.~\cite{Ning:2023ybc}, as expected.

One may want to seek for an additional canonical transformation to remove the interactions proportional to $\partial_i\tilde{p}_{ij}$ so that the Hamiltonian do not contain at all such terms, even without evaluating it on interaction picture fields.  This can be done by modifying the generating function as follows:
\begin{align}
\label{eq:gen_func_transverse_pij}
    F[\tilde{p}_\zeta,&\,\zeta,\,\tilde{p}_{ij},\,\gamma_{ij},\,t]=\,{\rm Eq.~\eqref{eq:generating_function_tensors}}\\
  & \notag +\frac{a\Mp^2}{4}\gamma_{ij}\,\partial_i\,\Biggl\{2\partial_k\left[\left(\frac{\tilde{\chi}}{\Mp^2}-\frac{\z}{H}\right) \gamma_{jk}\right]-\left[\partial_j\left(\frac{\epsilon \z^2}{H}\right)-2\partial_j\left(3\zeta-\frac{\tilde{p}_\zeta}{2 a^3 H \epsilon\Mp^2}\right)\right]\frac{\tilde{\chi}}{\Mp^2}  \Biggr\}.
\end{align}
Adding this term only modifies the canonical transformation law for the tensor conjugate momentum as
\begin{align}
\label{eq:transf_pij_final}
   & p_{ij}=\,{\rm Eq.~\eqref{eq:transf_pij}}\\\notag&-\frac{a\Mp^2}{4}\,\Biggl\{-2\partial_i\partial_k\left[\left(\frac{\tilde{\chi}}{\Mp^2}-\frac{\z}{H}\right) \gamma_{jk}\right]+\partial_i\left[\partial_j\left(\frac{\epsilon \z^2}{H}\right)-2\partial_j\left(3\zeta-\frac{\tilde{p}_\zeta}{2 a^3 H \epsilon\Mp^2}\right)\right]\frac{\tilde{\chi}}{\Mp^2}  \Biggr\}
\end{align}
and does not induce any additional transformations for $\zeta,\,\p$ and $\gamma_{ij}$. Furthermore,  Eq.~\eqref{eq:gen_func_transverse_pij} is easily seen to be a total spatial derivative, and so is its partial time derivative. Therefore, the sole effect of adding this term is the removal of interactions proportional to $\partial_i\tilde{p}_{ij}$ in Eqs.~\eqref{eq:H2zeta1gamma_bulk} and ~\eqref{eq:H3gamma_bulk}, and does not introduce additional terms.

\subsection{Quartic Hamiltonian}
\label{sec:quartic_Hamiltonian}

Let us finally comment on the quartic Hamiltonian. Quartic vertices are relevant for the calculations of tree-level trispectra and loop-corrections to primordial correlators.
The procedure to arrive at a manifestly slow-roll suppressed Hamiltonian is conceptually the same as outlined in the previous sections.
However, the calculation is significantly more involved, due to the large number of SR unsuppressed terms that need to be removed via canonical transformations. For this reason, here we will simply outline the steps of the calculation, leaving a full simplification for future works. To show how the procedure works in practice, we apply it to a simplified situation in which the size of only a subset of interactions is much larger than the other ones, so that a simplification becomes feasible.

\subsubsection{All contributions to the quartic Hamiltonian}

Let us remind the reader that the quartic Hamiltonian is not obtained by simply expanding the original Hamiltonian up to fourth order in perturbations. Indeed, following the notation of Section~\ref{sec:cubic_hamiltonian}, the quartic Hamiltonian  receives several contributions that look schematically as follows:
\begin{equation}
\label{eq:H4_starting_point}
\tilde{\mathcal{H}}^{(4)}=\mathcal{H}^{(4)}\left(\tz,\,\tp,\,\tilde{\gamma}_{ij},\,\tilde{p}_{ij}\right)   +\Delta \tilde{\mathcal{H}}^{(4)\,\mathrm{from}\,(2)}+\Delta \tilde{\mathcal{H}}^{(4)\,\mathrm{from}\,(3)}+   \Delta \tilde{\mathcal{H}}^{(4)\,\mathrm{from}\,F}  .
\end{equation}
We now explain the origin of each contribution in this equation, and clarify some subtleties associated to their calculation.  
\begin{itemize}
    \item  $\mathcal{H}^{(4)}\left(\tz,\,\tp,\,\tilde{\gamma}_{ij},\,\tilde{p}_{ij}\right)$. This is the standard contribution from the Legendre transform, expanded at fourth order in perturbations, and written in terms of the new variables.  Had we not performed the canonical transformation~\eqref{eq:gen_func_transverse_pij}, this would be the final result for $\mathcal{H}^{(4)}$, and the starting point for subsequent simplifications. As well known, it is in general not equal to  $-\mathcal{L}^{(4)}$~\cite{Chen:2006dfn}. Computing this term is already quite demanding as, in general, the relation between the time derivative of the coordinate and its conjugate momentum is  not always analytically invertible. One could need to invert it perturbatively, as explained in e.g.~\cite{Wang:2013zva,Chen:2017ryl}. However, in our case we get a simple analytical expression of  $\dot{\zeta}$ and $\dot{\gamma}_{ij}$ in terms of their momenta~\eqref{eq:momentum_zeta_with_tensors} and~\eqref{eq:momentum_gamma_with_tensors}, thereby significantly simplifying its calculation. 

Let us note that varying the quartic Hamiltonian with respect to the second order lapse and shift we can get the second order constraint equations needed for this calculation.
Actually, terms involving $\alpha^{(2)},$ and shift $\mathcal{N}^{(2)}_i$ only appear in the first term in Eq.~\eqref{eq:H4_starting_point}, as all other ones are generated by the canonical transformation~\eqref{eq:gen_func_transverse_pij}.
Before moving on to the next contribution, let us finally stress that in order to obtain the correct $\mathcal{H}^{(4)}$ we have to take into account both scalar and tensor perturbations, as the conjugate scalar and tensor momenta in~\eqref{eq:momentum_zeta_with_tensors} and~\eqref{eq:momentum_gamma_with_tensors} contain second order tensor and scalar contributions, respectively.

\item  $\Delta \tilde{\mathcal{H}}^{(4)\,\mathrm{from}\,(2)}$.  Inserting  the expression for the old variables in terms of the new ones in the quadratic Hamiltonian we get a correction to $\tilde{\mathcal{H}}^{(4)}$, which can be calculated as:
\begin{equation}
    \label{eq:H4from2}
    \Delta \tilde{\mathcal{H}}^{(4)\,\mathrm{from}\,(2)} = \sum_A \left.\frac{\delta \mathcal{H}^{(2)}}{\delta z^A} \right|_{\tilde{z}}\left(z^A\right)^{(3)} \,+\,\left.\frac{1}{2} \sum_{A,\,B}\frac{\delta^2 \mathcal{H}^{(2)}}{\delta z^A \delta z^B}\right|_{\tilde{z}}\left(z^A\right)^{(2)}\left(z^B\right)^{(2)}   \,.
\end{equation}
 To derive the cubic contributions $\left(z^A\right)^{(3)}$, Eqs.~\eqref{eq:transf_z_w_tensors},~\eqref{eq:transf_p_w_tensors},~\eqref{eq:transf_gammaij} and~\eqref{eq:transf_pij_final} need to be expanded to third order in the new~$\,\tilde{}\,$~variables. Let us also emphasize once more that it is crucial not to plug the solutions for the linear constraints into the quadratic Hamiltonian before performing the canonical transformation. The first term in Eq.~\eqref{eq:H4from2} yield the same contribution to $\Delta\mathcal{H}^{(4)\,\mathrm{from}\,(2)}$ irrespective of whether one starts from  the off-shell quadratic Hamiltonian~\eqref{eq:H2_not_solved} or the on-shell one~\eqref{eq:H2_solved}. However, this consistency does not extend to the second term in~\eqref{eq:H4from2}. Indeed, it can be easily checked that quartic interactions made up of products of two quadratic terms in the perturbative expansion~\eqref{eq:perturbative_expansion_ct} differ depending on whether one uses the off-shell or the on-shell Hamiltonians as starting points for the canonical transformation.

\item  $\Delta \tilde{\mathcal{H}}^{(4)\,\mathrm{from}\,(3)}$.  Similarly, inserting  the expression for the old variables in terms of the new ones in the cubic Hamiltonian we get the following correction to $\tilde{\mathcal{H}}^{(4)}$:
\begin{equation}
\label{eq:H4from3}
 \Delta \tilde{\mathcal{H}}^{(4)\,\mathrm{from}\,(3)} = \left. \sum_A \frac{\delta \mathcal{H}^{(3)}}{\delta z^A} \right|_{\tilde{z}}\left(z^A\right)^{(2)}.
\end{equation}
Also in this case, the starting point for the simplification should be \textit{a priori} the off-shell cubic Hamiltonian~\eqref{eq:H3_interaction_initial}, before any on-shell solution for the linear constraints is enforced.

\item $\Delta \tilde{\mathcal{H}}^{(4)\,\mathrm{from}\,F}$. Finally, there is a fourth contribution from plugging the canonical transformation into $\partial F/\partial t$:
\begin{equation}
     \Delta \tilde{\mathcal{H}}^{(4)\,\mathrm{from}\,F} = \left.\left(\frac{\delta}{\delta \zeta} \frac{\partial F(\zeta,\,\gamma_{ ij},\,\tilde{p}_\zeta,\tilde{p}_{ij},t)}{\partial t}\zeta^{(2)}+\frac{\delta}{\delta \gamma_{ij}} \frac{\partial F(\zeta,\,\gamma_{ ij},\,\tilde{p}_\zeta,\tilde{p}_{ij},t)}{\partial t}\gamma_{ij}^{(2)}\right)\right|_{\tilde{z}^A} .
\end{equation}
We remind the reader that our type II generating function is already a function of the new momenta, so only the new coordinates needs to be expanded perturbatively to calculate this contribution.
It is important to first take the partial time derivative, and then only expand in terms of the new phase-space variables.

\end{itemize}

The process of extracting manifestly slow-roll suppressed quartic interactions from $\tilde{\mathcal{H}}^{(4)}$ is the same as in Section~\ref{sec:cubic_hamiltonian}. The difference lies in the much larger number of initial interactions to be simplified, which makes the simplification computationally demanding.
Rather than performing all such kinds of simplifications, which goes beyond the scopes of this paper, we will focus on a subset of interactions to show how the algorithm described at the end of Section~\ref{sec:cubic_hamiltonian} works in practice for quartic interactions.

\subsubsection{Dominant interactions with a large $\eta$} 

In this section, we focus on a subset of the quartic interactions, that dominate whenever $\eta$ and its derivatives are large.
This may be relevant for inflationary scenarios with a transient stage during which the usual slow-roll approximation breaks down, with higher order slow-roll parameters become of order one ($\epsilon_{i>1}\sim\mathcal{O}(1)$) while $\epsilon$ is still remaining small.
We already encountered some interactions proportional to $\epsilon_2=\eta$ in the simplified cubic Hamiltonian~\eqref{eq:H3_interaction_initial}.
Those are indeed the dominant cubic interactions in this regime, and we can estimate their sizes as, for example,\footnote{We estimate the size of interactions around the time of Hubble crossing using $\dd t\sim1/H$, $ a \sim k/H$, $\partial \zeta \sim k \zeta$, $p_\zeta \sim \epsilon a^3 \Mp^2 \dot{\zeta}$ and $\zeta\sim H/ 
(k^{3/2}\epsilon^{1/2} \Mp)$.
How to relate $\dot{\zeta}$ to $\zeta$ depends on the details of this non-slow-roll phase.
}
\begin{equation}
\label{diag:large_eta_cubic}
\text{Large $\eta$ cubic vertex:} \quad
\vcenter{\hbox{
\begin{tikzpicture}[line width=1. pt, scale=2]
\vspace*{-2cm}
\draw[pygreen] (0.4, 0) -- (0.6, 0);
\draw[pygreen] (0.6, 0.) -- (0.7, 0.173) ;
\draw[pygreen]  (0.6, 0.)  -- (0.7, -0.173);
\draw[fill=pygreen] (0.6, 0) circle (.04cm);
\end{tikzpicture}}} \,\,\, /\, \tz
\,\sim \left[ \epsilon \eta a \Mp^2 \tz (\partial \tz)^2 \right] / \,\tz \sim \eta \,.
\end{equation}
The other cubic interaction with $\eta$, schematically of the form $(\eta/\epsilon)\tp^2 \tz$, contributes roughly the same order, although the precise relative size of these two contributions depends also on the details of the hierarchies between higher order slow-roll parameters $\epsilon_{i>1}$ and one, as encoded in the ratio $\dot{\zeta}/(H \zeta)$.
We also stress already that in addition to these two leading interactions, the canonical transformation used to simplify the cubic order Hamiltonian may lead to non-negligible contributions in scenarios with large $\eta$ where the momentum $\tp$ does not necessarily decay on super-horizon scales, in contrast to the usual slow-roll attractor scenario.

The lowest order correlation functions to which quartic interactions contribute are the tree-level trispetrum, and the 1-vertex correction to the 1-loop power spectrum. The same correlation functions will receive a contribution from two insertions of~\eqref{diag:large_eta_cubic}.
We can use this to estimate the typical size of a quartic interactions as important as, or more important than, the cubic ones, as follows:
\begin{equation}
\label{diag:large_eta_quartic}
\text{Large $\eta$ quartic vertex:} \quad
\vcenter{\hbox{
\begin{tikzpicture}[line width=1. pt, scale=2]
\vspace*{-2cm}
\draw[pygreen] (0.4, 0.21) -- (0.6, 0);
\draw[pygreen] (0.4, -0.21) -- (0.6, 0);
\draw[pygreen] (0.6, 0.) -- (0.8, 0.21) ;
\draw[pygreen]  (0.6, 0.)  -- (0.8, -0.21);
\draw[fill=pygreen] (0.6, 0) circle (.04cm);
\end{tikzpicture}}}\,\,\, / \, \tz^2\,\gtrsim\, 
\left(\vcenter{\hbox{
\begin{tikzpicture}[line width=1. pt, scale=2]
\vspace*{-2cm}
\draw[pygreen] (0.4, 0) -- (0.6, 0);
\draw[pygreen] (0.6, 0.) -- (0.7, 0.173) ;
\draw[pygreen]  (0.6, 0.)  -- (0.7, -0.173);
\draw[fill=pygreen] (0.6, 0) circle (.04cm);
\end{tikzpicture}}} \,\,\, / \,\tz \,\right)^2
\,\sim\eta^2\,.
\end{equation}
For example, a quartic Hamiltonian interaction of the form $ a \epsilon \eta^2 \Mp^2 \tz^2 (\partial \tz)^2 $ would parametrically contribute as much as the leading cubic interactions.
Quartic interactions with an estimated size smaller than~\eqref{diag:large_eta_quartic} would provide a subdominant contribution during a stage of $\eta\sim\mathcal{O}(1)$.

Let us now isolate such dominant interactions in $\tilde{\mathcal{H}}^{(4)}$.  The Lagrangian for single-field inflation~\eqref{eq:action-ADM} does not contain any interactions with $\epsilon_{i>1}$ in its original form and the only way such interactions can appear  in the Hamiltonian is by performing canonical transformations. There are therefore only  two possibilities. Either such interactions are generated by the canonical transformation in~\eqref{eq:generating_function_tensors} to simplify the cubic interactions, or they are generated by a new canonical transformation to render manifest the true size of quartic interactions.  

Let us start by the former possibility, i.e. by inspecting the canonical transformations derived from the generating function~\eqref{eq:generating_function_tensors}. They only depend on the background quantities $(a,\, H,\, \epsilon)$, so that interactions involving $\eta$ can only be generated through the partial time derivative of the generating function, i.e. $\Delta \tilde{\mathcal{H}}^{(4)\,\mathrm{from}\,F}$. Among such interactions, the 
ones least suppressed by $\epsilon \ll 1$ 
are given by:

\begin{align}    \label{eq:Hamiltonian_eta_leading}
\tilde{\mathcal{H}}^{(4)}_{\eta,\,{\rm dominant}}=&
\frac{  \eta }{8 a^6 H \Mp^4\epsilon ^2}\tp^3 \tz
 + \frac{  \eta }{2 a^2 H}\tp\tz \left(\partial\tz \right)^2
 + \frac{\eta}{a^2 H}\tp\tz ^2 \partial^2\tz+\mathcal{H}^{(4)}_{\eta,\,{\rm \,inv.deriv.}}\,,
\end{align}
with $\mathcal{H}^{(4)}_{\eta,\,{\rm \,inv.deriv.}}$ involving interactions of the form $\epsilon \eta \mathcal{D}( \tz^4)$ and $(\eta/\epsilon) \mathcal{D} ( \tz^2 \tp^2 )$ with $\mathcal{D}$ a complicated operator involving inverse spatial derivatives.
Interactions above only contain one power of $\eta$, thus providing an apparently smaller size than our estimate in~\eqref{diag:large_eta_quartic} when $\eta$ is large.
However, as in Section~\ref{sec:cubic_hamiltonian}, we can perform a canonical transformation to incorporate the  interactions not involving inverse spatial derivatives in~\eqref{eq:Hamiltonian_eta_leading} into the kinetic term for a new momentum variable.
This will result in interactions of the form $\zeta^4$ or $\p^2 \z^2$ but with additional factors of $\eta$ and its derivatives, which would decrease the size of the interaction in slow roll, while here it increases it.
In practice, we define the following generating function:
\begin{equation}
\label{eq:generating_function_eta}
   F_\eta[\tz , \ttp, t] = \ttp \tz 
 - \frac{ \eta }{8 a^3 H \Mp^2\epsilon }\ttp^2 \tz^2
 - \Mp^2\frac{a \epsilon  \eta    }{6 H}\tz ^3\partial^2\tz,
\end{equation} 
from which we can derive the expressions relating old and new variables as follows
\begin{align}
        \tp=&\ttp
 - \frac{ \eta }{4 a^3 H \Mp^2\epsilon }\,\ttp^2 \tz 
 - \Mp^2\frac{a \epsilon   \eta }{H} \tz  \left(\partial \tz \right)^2
 - \Mp^2\frac{a  \epsilon  \eta   }{H}\,\tz^2\,\partial^2\tz\\
 \label{eq: CT relevant for large eta}
 \ttz=&\tz 
 + \frac{ \eta }{8 a^3 H \Mp^2\epsilon }\,\ttp \tz ^2.
\end{align}
Let us note that, being the generating function quartic (besides the trivial term), not only does this canonical transformation leave the quadratic action invariant, but also the cubic one.
Computing the new quartic Hamiltonian, and keeping only terms quadratic in $\eta$ and its logarithmic derivative we get:
\begin{equation}
\label{eq:H4_large_eta}
\tilde{\tilde{\mathcal{H}}}^{(4)}_{\eta^2,\,{\rm dominant}}=\frac{\ttp^2 \ttz ^2 \eta ^2}{8 a^3 \Mp^2 \epsilon }
 - \frac{\ttp^2 \ttz ^2 \eta  \eta_2}{8 a^3 \Mp^2 \epsilon }
 + \frac{\Mp^2}{2} a \epsilon  \ttz ^2 \eta ^2 \left(\partial\ttz \right)^2
 + \frac{\Mp^2 }{2} a \epsilon  \ttz ^2 \eta  \eta_2 \left(\partial\ttz \right)^2,
\end{equation}
where we have defined $\eta_2\equiv  \dot{\eta}/H\eta$.
We have also consistently neglected $\mathcal{H}^{(4)}_{\eta,\,{\rm \,inv.deriv.}}$ as it is less boosted by $\epsilon_{i>1}$ parameters.
Interestingly, this expression exactly matches the dominant $\eta$ interactions derived using the EFT of inflation in Ref.~\cite{Firouzjahi:2023aum}.
However, to be consistent when computing a given observable generated by such large quartic interactions, one should also consider contributions given by the canonical transformation relating correlation functions of $\z$ to those of $\ttz$.
One can check that the only term containing $\eta$ is the one shown in Eq.~\eqref{eq: CT relevant for large eta}, leading to $\zeta \simeq \ttz -\eta/(8a^3 \epsilon H \Mp^2) \ttp \ttz^2 $.
Whether it can give a contribution to correlation functions of $\zeta$ as sizable as the dominant cubic and quartic interactions should be decided depending on the details of the modelling of the large $\eta$ phase, and in particular depending on the behaviour of the momenta which are super-horizon and the time of evaluation of the correlation functions (see discussion around Eq.~\eqref{eq:bispectrum}).

Next, let us explore the second potential source of large $\eta$ contributions, namely, additional canonical transformations aimed at elucidating the magnitude of the entire quartic Hamiltonian.
As mentioned earlier, we do not undertake this entire calculation, as it is beyond the scope of this work.
Instead, our goal is to present a general argument explaining why such transformations are unlikely to yield dominant contributions to the large $\eta$ interactions.
Let $\mathcal{O}^{(i)}_{\p^n\,\zeta^m}$ denote an interaction consisting solely of operators with $n$ powers of $\p$, $m$ powers of $\zeta$, and $i$ spatial derivatives, suitably contracted. Inverse Laplacian operators contribute with a factor of $-2$ to the sum for computing the index $i$. We find the leading contribution to our on-shell quartic Hamiltonian~\eqref{eq:H4_starting_point} to be superficially of slow-roll order $\epsilon^0$, and of the form:
\begin{equation}
\label{eq:H4_leading}
    \tilde{\mathcal{H}}^{(4)}_{\epsilon^0}= \frac{1}{a^7 H^4\epsilon^2 \Mp^2}\mathcal{O}^{(4)}_{\tp^2\,\tz^2} + \frac{1}{a^6 H^5\epsilon}\mathcal{O}^{(6)}_{\tp\,\tz^3}+ \frac{ \Mp^2}{a^4 H^3\epsilon}\mathcal{O}^{(4)}_{\tp\,\tz^3}+ \frac{ \Mp^2}{a^3 H^4}\mathcal{O}^{(6)}_{\tz^4}+ \frac{ \Mp^2}{a H^2}\mathcal{O}^{(4)}_{\tz^4}.
\end{equation}
One can follow the usual procedure and determine the canonical transformation to simplify the interactions above by incorporating the first two terms in~\eqref{eq:H4_leading} into the quadratic kinetic term. Only the generating function corresponding to 
the first term above includes factors of $\epsilon$ and eventually leads to an $\eta$ dependent term upon taking the partial time derivative.
The resulting $\eta$-dependent contribution is of the form $\eta\, \mathcal{F}^{(4)}_{\ttp\,\ttz^3}/a^4 H^3 \epsilon$, and is superficially of slow-roll order $\sim\epsilon$. 

After having properly chosen the generating function so as to cancel out the leading order Hamiltonian~\eqref{eq:H4_leading}, the remaining interactions are superficially of slow-roll order $\sim\epsilon$, and we find their form to be:
\begin{align}
    \tilde{\tilde{\mathcal{H}}}^{(4)}_{\epsilon^1}=& \frac{1}{a^8 H^3\epsilon^2\Mp^4}\mathcal{O}^{(2)}_{\ttp^3\,\ttz}+ \frac{1}{a^6 H\epsilon^2\Mp^4}\mathcal{O}^{(0)}_{\ttp^3\,\ttz} + \frac{1}{a^5 H^2\epsilon\Mp^2}\mathcal{O}^{(2)}_{\ttp^2\,\ttz^2} + \frac{1}{a^3 \epsilon\Mp^2}\mathcal{O}^{(0)}_{\ttp^2\,\ttz^2} \notag\\& + \frac{1}{a^7 H^4\epsilon\Mp^2}\mathcal{O}^{(4)}_{\ttp^2\,\ttz^2}+\frac{1}{a^4 H^3 }\mathcal{O}^{(4)}_{\ttp\,\ttz^3}+\frac{1}{a^2 H }\mathcal{O}^{(2)}_{\ttp\,\ttz^3}+\frac{\epsilon\Mp^2}{a H^2}\,\mathcal{O}^{(4)}_{\ttz^4}-\frac{\epsilon\Mp^2}{a^3 H^4}\,\mathcal{O}^{(6)}_{\ttz^4}\notag\\&\label{eq:H4_next_to_leading}+\frac{1}{a^4 H^3 }\frac{\eta}{\epsilon}\mathcal{F}^{(4)}_{\ttp\,\ttz^3},
\end{align}
where the last term is precisely the one mentioned above, produced by the first canonical transformation. As in the previous step, one could then find another canonical transformation to reduce the size of the interactions. It is easy to see check that of all the terms in Eq.~\eqref{eq:H4_next_to_leading}, only the first two terms and the last one  can possibly produce interactions proportional to $\eta$ through this procedure.
The former generate a term of the form $\eta \mathcal{F}^{(2)}_{\tilde{\tilde{\tilde{p}}}_\zeta^2\,\tilde{\ttz}^2}/a^5 H^2  \Mp^2$, which is, however, subdominant with respect to those in the leading interactions in the large $\eta$ regime derived in Eq.~\eqref{eq:H4_large_eta}.
On the  other hand, simplifying the last term in Eq.~\eqref{eq:H4_next_to_leading}, may lead to an interaction of the form $\eta( \eta_2-H)\Mp^2\mathcal{F}^{(4)}_{\tilde{\tilde{\tilde{\z}}}^4}/a H^2\Mp^2$.
The interaction $\propto \eta \eta_2$  would be of the same order as those in Eq.~\eqref{eq:H4_large_eta}.
However, the one $\propto \eta$ would be of slow-roll order $\epsilon^1$, and cannot be
cancelled by any other contribution.
This would be problematic, as such interaction would lead to a large trispectrum of order unity even during standard slow-roll inflation, while it is known to start at the next order in this regime~\cite{Seery:2006vu,Seery:2008ax}.
Therefore, we conclude that the complicated operator $\mathcal{F}^{(4)}_{\tilde{\tilde{\tilde{\z}}}^4}$ must vanish upon careful computation of its various contributions and uses of spatial integration by parts.
Based on this argument, we conclude that the dominant interactions in the large $\eta$ limit take the form  that we have derived in Eq.~\eqref{eq:H4_large_eta}.
Moreover, the canonical transformations needed to reach the stage with the operator $\mathcal{F}^{(4)}_{\tilde{\tilde{\tilde{\z}}}^4}$ do not introduce $\eta$-dependent terms so we can consider $\tilde{\tilde{\tilde{\zeta}}} = \ttz$ in the large $\eta$ regime, and the only relevant canonical transformation is still given by Eq.~\eqref{eq: CT relevant for large eta} alone.

This example serves primarily for illustration, providing a simplified setting to demonstrate the method for manipulating  $\mathcal{H}^{(4)}$ using canonical transformations. However, let us mention that our exercise  has already  interesting applications on its own. In particular, a heated discussion has recently emerged regarding whether models with a transient stage of large $\eta$ can produce large loop-corrections to the spectrum of $\zeta$ (see Refs.~\cite{Cheng:2021lif,Inomata:2022yte,Kristiano:2022maq,Riotto:2023hoz,Choudhury:2023vuj,Motohashi:2023syh,Franciolini:2023lgy,Tasinato:2023ukp,Fumagalli:2023loc,Maity:2023qzw,Davies:2023hhn,Iacconi:2023ggt,Inomata:2024lud} for an incomplete list of papers). One of the controversies in this context is related to how to deal with total time derivative interactions terms, recently discussed in Refs.~
\cite{Fumagalli:2023hpa,Tada:2023rgp,Firouzjahi:2023bkt}. We hope that our identification of the dominant interactions in the large $\eta$ regime, together with our thorough explanation of how to deal with total time derivatives in the approach with integrations by parts, our equivalently with canonical transformations at the level of the full Hamiltonian in phase space, can contribute to solving this matter.

\subsubsection{Order $n$ Hamiltonian}

Before concluding, let us thus generalize the learnt lesson, and provide  a prescription to simplify interactions at order $n$ in perturbations. To be fully general we reintroduce a collection of multiple fields $\psi^A$ and associated momenta $p_\psi^A$ with $A=1,\,\ldots,\,N$.  Starting from a Hamiltonian with interactions of superficially slow-roll size $\epsilon_i^{m}$ ($m$ powers of either of the slow-roll parameters) after introducing interaction picture fields, we propose the following prescription to make explicit the true size of interactions:
\begin{itemize}
    \item Take the interactions with the highest powers of momenta, and superficial size $\epsilon_i^{m}$. At order $n$ this would be $\left(p_\psi^A\right)^n$.
    \item Define new momenta $\tilde{p}_\psi^A$ to incorporate this interaction into the kinetic term for this new momentum.
    \item Find the corresponding generating function and new position variable $\tilde{\psi}^A$, possibly perturbatively, and derive the new Hamiltonian $\tilde{\mathcal{H}}$ for the new phase-space variables. The generating function is of order $n$, and lower order interactions of order $< n$ are not modified.
    \item Repeat this procedure for the interaction of order $\left(p_\psi^A\right)^{n-1} \psi^A$, etc. The last step  is the order $p_\psi^A \left(\psi^A\right)^{n-1}$, after which one is left with only terms of the form $\left(\psi^A\right)^{n}$. At this point there are three possibilities. (i) the remaining interactions of the form  $\left(\psi^A\right)^{n}$ do not cancel out among each others: the {\em true} size of the Hamiltonian is $\epsilon_i^{m}$. (ii)  the remaining interactions of the form  $\left(\psi^A\right)^{n}$ cancel out among each others: the {\em true} size of the Hamiltonian is $<\epsilon_i^{m}$, and  we will have to repeat the procedure above until we reach the true size of the interaction Hamiltonian at order $n$.
\end{itemize}
The perturbative canonical transformations obtained to simplify the Hamiltonian of order $n$ way will also generate higher order interactions, which will add to other interactions of the same order. One can then go to the order $n+1$ and apply again the algorithm we have just proposed.

\section{Conclusions}
\label{sec: conclusions}

	Despite the established use of in-in perturbation theory for computing cosmological correlation functions, unresolved ambiguities were persisting in handling total time derivative interactions in the Lagrangian. In the context of inflation, these interactions primarily emerge from integrations by parts designed to reveal the correct slow-roll suppression of non-linear interactions in the comoving gauge. Although the importance of not overlooking total time derivatives was already emphasized, a systematic approach to such terms beyond the single vertex order was still lacking in the literature. Our work precisely clarifies the impact of total time derivatives and terms proportional to the linear equations of motion in the in-in perturbation theory and formalizes their analysis.

	In this work, we systematize the treatment of total time derivative interactions in the interaction Hamiltonian of the in-in perturbation theory and we elucidate their possible contributions to correlation functions.
    We show that they lead to i) boundary terms in the form of equal-time nested commutators with the external operator $\mathcal{O}$ whose correlation functions are sought for; ii) modifications of the usual bulk interactions that need to be integrated over time, in the form of effective interactions. 
    Only the former contribution was investigated before our work, while we show that the latter is also crucial to cancel superficially large contributions from higher-order bulk interactions in the Hamiltonian generated from the Legendre transformation of the Lagrangian that includes total time derivative interactions.
    We provide a general formula valid at any order in perturbation theory that enables to recast all contributions from total time derivatives in the interaction Hamiltonian as either a redefinition of the external operator $\mathcal{O}$ or of the bulk interactions.
    Explicitly expanding the formula up to second order, as required for practical calculations of the most common primordial correlation functions, bispectrum, contact and exchange trispectrum and 1-loop corrections to the power spectrum, we demonstrate its practical use beyond the single-vertex order.
    For completeness and sake of comparison, we also explored how the Lagrangian path integral approach of the in-in formalism performs in dealing with total time derivatives and terms proportional to linear equations of motion, and showcased concrete calculations proving equivalence with the Hamiltonian operator approach. 
    Although perfectly suitable for performing calculations, these two versions of the in-in perturbation theory with total time derivative interactions that we have developed could be judged cumbersome, with either fine cancellations to be sought for between different vertex-orders or interactions with complicated operators to take into account.
    We thus propose another route to the calculation of primordial correlation functions, that avoids altogether the generation of total time derivative interactions.

    Instead of performing integrations by parts in the Lagrangian to reveal the correct size of interactions, our proposed method operates directly in phase space, where the Hamiltonian is defined.
    By utilizing canonical transformations, we are able to simplify interactions without generating total time derivatives, so that the canonically transformed Hamiltonian exclusively comprises bulk interactions.
    The canonical transformations establish a relationship between correlation functions of old and new variables, and we explain how to retrieve the former from the knowledge of the latter that can be computed using the transformed Hamiltonian.
    With this approach the interaction Hamiltonian is significantly simpler, which greatly trivializes the computation of correlation functions beyond the single-vertex order.
    We demonstrate the equivalence of this new method with the in-in perturbation theory that includes total time derivatives for the two toy models that we had already introduced.
    In order to build intuition on the possible contributions from the inverse canonical transformations relating correlation functions of the old phase-space variables to those of the new ones, we also introduce diagrammatic rules of thumb.

    As an application, we utilized our formalism to compute the cubic Hamiltonian for scalar and tensor perturbations in canonical single-field inflation.
    Using canonical transformations, we are able to extract the correct slow-roll order suppression of the phase-space interactions without introducing total time derivatives nor terms proportional to the linear equations of motion.
    Importantly, this cubic Hamiltonian is relevant beyond interaction picture calculations, and can also be used for Schrödinger picture non-linear evolutions.
    When evaluated on interaction picture fields and momenta though, it exactly reduces to the cubic interaction Hamiltonian found from the cubic scalar and tensor Lagrangians derived in earlier works, with the exception of total time derivative interactions which are now absent.
    In comparison to the more traditional approach with integration by parts, our method proves simpler to compute the primordial bispectrum and illuminates the connection between the use of either boundary terms and field redefinitions a la Maldacena, the latter representing simply a subset of the canonical transformations proposed in our approach.
    We also discuss the generalization of our formalism to quartic and higher-order interactions.
    To illustrate practicality, we applied our formalism to recover the subset of phase-space quartic interactions that are dominant in the large $\eta$ limit.
    When evaluated on interaction picture fields and momenta, this quartic interaction picture Hamiltonian exactly coincides with findings using the Lagrangian of the EFT of inflation beyond the slow-varying background approximation.
    However, we additionally provide with the crucial information about the inverse canonical transformations to be performed in order to retrieve correlation functions of the exact primordial curvature perturbation seeding the large-scale structures of our universe.
    
    Our formalism can be extended and applied in several directions.
    An evident example is a comprehensive computation of quartic interactions in single-field inflation in the comoving gauge, a task that remains unexplored in the existing literature, even with the more traditional approach at the level of the Lagrangian and using integrations by parts.
    Although this computation will be demanding, our work establishes all the essential tools for undertaking such an endeavor.
    Another intriguing avenue consists in the simplification of the phase-space Hamiltonian interactions in multifield models of inflation that encompass both the adiabatic curvature perturbation that we have considered in this work, and a set of entropic---or isocurvature---fluctuations.
    We also anticipate that our insights into large $\eta$ interactions will contribute to clarifying the recent debate surrounding loop corrections to the primordial power spectrum arising from a temporary non-attractor stage of inflation.
    Lastly, we stress again that the phase-space interactions elucidated in this work can be applied beyond interaction picture calculations of primordial correlation functions. 
    For instance, they may prove useful in studying the quantum properties of cosmological perturbations beyond the linear order.
    Concrete applications could range from the calculation of self-decoherence in single-field inflation from non-linear interactions between different scales, to Schrödinger picture non-linear evolutions.
    We plan to explore ourselves some of these directions in future works.

\begin{acknowledgments}

We are grateful to
Xingang Chen,
David Langlois,
Jérôme Martin,
Sébastien Renaux-Petel,
David Seery and
Denis Werth
for useful discussions at various stages of this project.
We also thank 
Xingang Chen, Guillem Domenech, Jacopo Fumagalli, Sebastian Garcia-Saenz,
Sébastien Renaux-Petel and
Denis Werth
for comments on a draft of this paper. Some calculations in this paper were performed with the Mathematica software \texttt{MathGR}~\cite{Wang:2013mea}. 
We would like to acknowledge CERN and the EuCAPT Consortium for the 2023 EuCAPT International Travel Award that helped foster this work.
L.P. acknowledges funding support from the Initiative Physique des Infinis (IPI), a research training program of the Idex SUPER at Sorbonne Université.
L.P. would also like to thank New York University for hospitality during the latest stages of this project.

\end{acknowledgments}

\appendix

\section{Canonical quantization and quantum anomalies.}
\label{app:quantum_anomalies}
In most cosmological applications, a semi-classical and perturbative treatment is used.
Semi-classical because only vacuum fluctuations around a homogeneous background are quantified.
Perturbative because the theory is first assumed to be free, and therefore the Hamiltonian is truncated at quadratic order before canonical quantization.
Then only, non-linear interactions expressed in terms of the quantized free fields are added to the theory in a perturbative scheme.
We will therefore focus here on a free quadratic Hamiltonian, for simplicity still with a single degree of freedom and no mass term as in Eq.~\eqref{eq: free Hamiltonian toy model}.
Roughly speaking, canonical quantization amounts to promoting the phase-space variables $(\psi,p_\psi)$ in $\mathcal{H}_\mathrm{free}$ to quantum operators verifying the canonical commutation relations inherited from the Poisson brackets in the corresponding classical theory:
\begin{align}
    \psi(t,\vec{x}) &\rightarrow \hat{\psi}(t,\vec{x}) \,,  \\
    p_\psi(t,\vec{x}) &\rightarrow \hat{p}_\psi(t,\vec{x}) \,, \nonumber \\ 
    \forall t \,, \quad \left\{\psi(t,\vec{x}), p_\psi(t,\vec{y})\right\} = \delta^{(3)}\left(\vec{x}-\vec{y}\right) & \rightarrow \left[\hat{\psi}(t,\vec{x}) , \hat{p}_\psi(t,\vec{y}) \right] = i \hbar \, \delta^{(3)}\left(\vec{x}-\vec{y}\right) \,, \nonumber
\end{align}
where we kept the $\hbar$-factor exceptionally.
Then, one must define the vacuum state of the theory, $\ket{0}$, often found by minimization of the energy $\braket{0|\mathcal{H}|0}$.
The creation and annihilation quantum operators are used to construct the corresponding Fock space made of $n$-particles states, e.g. $\ket{1_{\vec{k}}}=\hat{a}^\dagger_{\vec{k}}\ket{0}$, etc.
The creation and annihilation operators must verify the commutation relations
\begin{equation}
    \left[\hat{a}_{\vec{k}},\hat{a}^\dagger_{\vec{k}^\prime} \right]=(2\pi)^3 \delta^{(3)}\left(\vec{k}-\vec{k}^\prime\right) \,.
\end{equation}
Since they form a complete set of quantum states, the position and momentum operators in Fourier space can be decomposed in this basis, and the coefficients of the decomposition are the mode functions $\left(\psi_k(t),p_{\psi,k}(t)\right)$:
\begin{align}
    \hat{\psi}(t,\vec{x})&=\int \frac{\dd^3 \vec{k}}{(2\pi)^3} e^{i \vec{k} \cdot \vec{x}} \left( \psi_k(t) \hat{a}_{\vec{k}} +  \psi_k^*(t) \hat{a}^\dagger_{-\vec{k}} \right) \,, \\
    \hat{p}_\psi(t,\vec{x})&=\int \frac{\dd^3 \vec{k}}{(2\pi)^3} e^{i \vec{k} \cdot \vec{x}} \left( p_{\psi,k}(t) \hat{a}_{\vec{k}} +  p_{\psi,k}^{*}(t) \hat{a}^\dagger_{-\vec{k}} \right) \,.
\end{align}
From the canonical commutation relations in real space, those for the creation and annihilation operators, and  the above expressions, the mode functions must verify the Wronskian condition: $\forall t\,, \,\, \psi_k p_{\psi,k}^{*} -  \psi_k^* p_{\psi,k} = i \hbar $.

A quantum canonical transformation must leave invariant all these properties.
In practice, the difference with classical mechanics is the presence of non-commutating operators.
This implies for example that the order in which quantum canonical transformations are performed now matters, and that the inverse transformations must be defined with care.
Also, a canonical transformation should a priori be assigned a given operator ordering, e.g. $\hat{\tilde{p}}_\psi=\hat{p}_\psi + \hat{\psi} \hat{p}_\psi $ is different from $\hat{\tilde{p}}_\psi=\hat{p}_\psi + \hat{p}_\psi \hat{\psi} $, etc., see Ref.~\cite{Anderson:1992lns} in the context of quantum mechanics.
One way to go around the operator ordering ambiguities is to consider a path-integral approach where all instances of the fields and their conjugate momenta are simply dummy variables of integration, instead of operators: $\mathcal{Z}=\int \mathcal{D} \psi \mathcal{D} p_\psi e^{i I[\psi, p_\psi]}$.
However, then, equivalent subtleties arise at the level of the discrete time path.
The main point is that, for quantum path integrals, the ``jump'' between two values $\psi_{j}$ and $\psi_{j-1}$ (resp. $p_\psi^j$ and $p_\psi^{j-1}$) may not be proportional to the time elapsed $t_{j}-t_{j-1}\equiv \Delta t$ during the jump, but to its square root $\sqrt{\Delta t}$.
This property, ubiquitous in classical but stochastic processes, result in so-called quantum anomalies in the context of canonical transformations.
\begin{itemize}
    \item Given a generating function $F$ for a canonical transformation at the time $t_j$, the relation between old and new variables is affected by corrections proportional to higher-order derivatives of $F$ and to the jumps $\Delta \psi_j \equiv \psi_j-\psi_{j-1}\,,\Delta p_\psi^j \equiv p_\psi^{j+1}- p_\psi^j$.
    For example, for a type III generating function, Eq.~\eqref{eq: type III generating function} would now read~\cite{Swanson:1994jn}:
    \begin{align}
        \psi_j &= - \frac{F_3(p_\psi^{j+1} , \tilde{\psi}_j,t_j)-F_3(p_\psi^{j} , \tilde{\psi}_{j},t_j)}{\Delta p_{\psi}^j} \\
        &=-  \frac{\partial F_3}{\partial p_{\psi}^{j}}(p_\psi^j , \tilde{\psi}_j,t_j) - \frac{1}{2} \frac{\partial^2 F_3}{\partial p_{\psi}^{j\,2}}(p_\psi^j , \tilde{\psi}_j,t_j) \Delta p_{\psi}^j +\ldots \nonumber \\
        \tilde{p}_\psi^j &= - \frac{F_3(p_\psi^j , \tilde{\psi}_j,t_j)-F_3(p_\psi^j , \tilde{\psi}_{j-1},t_j)}{\Delta \tilde{\psi}_j} \\
        &=-  \frac{\partial F_3}{\partial \tilde{\psi}_j}(p_\psi^j , \tilde{\psi}_j,t_j) + \frac{1}{2} \frac{\partial^2 F_3}{\partial \tilde{\psi}_j^2}(p_\psi^j , \tilde{\psi}_j,t_j) \Delta \tilde{\psi}_j +\ldots \nonumber
    \end{align}
    where usually one discards the contributions explicitly proportional to the jumps in the limit $\Delta t \rightarrow 0$.
    However in general in a quantum theory those contribute finite corrections, though suppressed by an additional factor of $\hbar$, not explicit here.
    In practice, this has the consequence that the new Hamiltonian that describes the system in terms of the new phase-space variables is not only a function of the fields and momenta, but also of their time derivatives---the equivalent of the ``jumps'' in the continuous time description---$\tilde{H}\left(\tilde{\psi},\tilde{p}_\psi, \dot{\tilde{\psi}},\dot{\tilde{p}}_\psi\right)$.
    \item A second, related anomaly concerns the Jacobian of the transformation in the path integral, taking into account all canonical transformations at the discrete times $t_j$.
    Indeed, the equivalent of Eq.~\eqref{eq: Jac condition} now becomes~\cite{Swanson:1994jn}:
    \begin{equation}
        J = \prod_{j=0}^{\infty}\left[1+ A_j \Delta \tilde{\psi}_j + B_j \Delta \tilde{p}_{\psi}^j\right] \neq 1\,,
    \end{equation}
    with contributions proportional to the jumps at each time step, $A_j$ and $B_j$ being given by third derivatives of the generating function.
    In the continuous limit, one can exponentiate the Jacobian and finds $J \rightarrow \exp\{\int \dd t  A(t) \dot{\tilde{\psi}}(t) + B(t) \dot{\tilde{p}}_\psi (t) \}$, contributing as external source terms for the time derivatives.
\end{itemize}
Ref.~\cite{Swanson:1994jn} shows examples of physical systems and canonical transformations including all relevant corrections in a quantum system, and proves that the quantum anomalies cancel each other.
In those cases, one can therefore consider a Hamiltonian path integral in terms of the new phase-space variables and independent of the time derivatives of the fields and momenta.
Although definitely a fascinating direction for future work, we defer the study of such quantum anomalies in the cosmological context.
In practice, we will therefore overlook the possibility of having quantum anomalies and use canonical transformations at the level of the classical Hamiltonian, before canonical quantization.
It remains to be known whether quantum anomalies may affect the calculation of loop-level diagrams or purely quantum effects like entanglement, etc., in the context of quantum canonical transformations.

\section{Diagrammatic rules}
\label{app: diagrammatic rules}

We present here some diagrammatic rules of thumb that we found convenient to develop, in order to classify the possible contributions to an observable $\braket{\mathcal{O}(\psi)}$, where $\psi$ is related to a variable $\tilde{\psi}$ via a canonical transformation.
These are only approximate diagrammatic rules in the sense that we pay no attention to numerical factors, permutations, etc.
Moreover, for simplicity, we overlook the presence of momenta $\tilde{p}_\psi$ in the expression for $\psi$, as well as the possibility to compute expectation values of an operator including $p_\psi$.
More generally, as in the rest of this section, we only consider explicilty a single degree of freedom $\psi$ instead of a collection $\{\psi^a\}_{a=1,\ldots, N}$.
However, we already propose a straight generalization of these diagrammatic rules to a multifield setup in phase space: by the inclusion of indices $A$ such that $\{\psi^A\}_{A=1,\ldots,2N} = \{\psi^a, p_\psi^a \}_{a=1,\ldots,N}$, with both propagators and vertices mixing them.

For definiteness, we now focus on a transformation $\psi = g_2 \tilde{\psi}+g_3  \tilde{\psi}^2+ g_4  \tilde{\psi}^3 + \ldots$ where we wrote explicitly only the relevant contributions for the calculations of the tree-level bispectrum and trispectrum, as well as the one-loop power spectrum.
In the diagrams below, a white dot simply corresponds to the transition $\psi \rightarrow g_2 \tilde{\psi}$ (quadratic vertex), while a black dot represents a non-linear transition $\psi \rightarrow g_3 \tilde{\psi}^2$  (cubic vertex), $\psi \rightarrow g_4 \tilde{\psi}^3$  (quartic vertex), etc. 
A larger and red dot represents a non-linear interaction for the $\tilde{\psi}$ field itself, so any diagram containing this kind of vertex will be proportional to correlation functions of $\tilde{\psi}$ beyond its power spectrum.
For $n$-point connected correlation functions of $\psi$, only the exact same $n$-point connected correlation functions of $\tilde{\psi}$ can contribute at tree level.
Blue lines represent external propagators, but they actually only appear as amputated after a white or black vertex, so they do not contribute and they are only shown for illustration purposes.
Red lines represent internal propagators and may result either in power spectra (if connected to white and black dots only) or higher-order correlation functions of $\tilde{\psi}$ (if connected to a single red dot and to another either white or black dot).
Note however that red lines cannot connect to two red dots, as this amounts to dissecting the different contributions to a given correlation function of $\tilde{\psi}$.
Diagrammatic rules now follow.
\begin{align}
\label{eq: diagrammatic rules}
&\text{Amputated vertices:}\quad
\vcenter{\hbox{
\begin{tikzpicture}[line width=1. pt, scale=2]
\vspace*{-2cm}
\draw[pyblue] (0.4, 0) -- (0.5, 0);
\draw[pyred] (0.5, 0) -- (0.6, 0);
\draw[fill=white] (0.5, 0) circle (.03cm);
\end{tikzpicture}}} 
\,\sim\,
g_2\,, \quad
\vcenter{\hbox{
\begin{tikzpicture}[line width=1. pt, scale=2]
\vspace*{-2cm}
\draw[pyblue] (-0.1, 0) -- (0, 0);
\draw[pyred] (-0.05, 0.08660) -- (0, 0);
\draw[pyred] (-0.05, -0.08660) -- (0, 0);
\draw[densely dotted][pyred] (0.05, -0.08660) -- (0, 0);
\draw[densely dotted][pyred] (0.05, +0.08660) -- (0, 0);
\draw[pyred] (0, 0) -- (0.1, 0);
\draw[fill=black] (0, 0) circle (.03cm);
\end{tikzpicture}}} 
\,\sim\,
g_{3,4,\ldots,n} \,. \\
&\text{Internal 2-point functions:} \quad
\vcenter{\hbox{
\begin{tikzpicture}[line width=1. pt, scale=2]
\vspace*{-2cm}
\draw[pyblue] (0.4, 0) -- (0.5, 0);
\draw[pyred] (0.5, 0) -- (0.8, 0);
\draw[pyblue] (0.8, 0) -- (0.9, 0);
\draw[fill=white] (0.5, 0) circle (.03cm);
\draw[fill=white] (0.8, 0) circle (.03cm);
\end{tikzpicture}}} 
\,\sim\,
\vcenter{\hbox{\begin{tikzpicture}[line width=1. pt, scale=2]
\vspace*{-2cm}
\draw[pyblue] (0.4, 0) -- (0.5, 0);
\draw[pyred] (0.5, 0) -- (0.8, 0);
\draw[pyred] (0.8, 0) -- (0.9, 0);
\draw[fill=white] (0.5, 0) circle (.03cm);
\draw[fill=black] (0.8, 0) circle (.03cm);
\end{tikzpicture}}}
\,\sim\,
\vcenter{\hbox{\begin{tikzpicture}[line width=1. pt, scale=2]
\vspace*{-2cm}
\draw[pyred] (0.4, 0) -- (0.5, 0);
\draw[pyred] (0.5, 0) -- (0.8, 0);
\draw[pyred] (0.8, 0) -- (0.9, 0);
\draw[fill=black] (0.5, 0) circle (.03cm);
\draw[fill=black] (0.8, 0) circle (.03cm);
\end{tikzpicture}}} 
\,\sim\,
P_{\tilde{\psi}}(k)\,, \nonumber \\
&\text{Internal $n$-point functions:} \quad
\vcenter{\hbox{
\begin{tikzpicture}[line width=1. pt, scale=2]
\vspace*{-2cm}
\draw[pyred] (0.4, 0) -- (0.6, 0);
\draw[pyred] (0.6, 0.) -- (0.7, 0.173) ;
\draw[pyred]  (0.6, 0.)  -- (0.7, -0.173);
\draw[fill=pyred] (0.6, 0) circle (.04cm);
\end{tikzpicture}}} 
\,\sim\,
B_{\tilde{\psi}}(k_1,k_2,k_3)\,,
\quad 
\vcenter{\hbox{
\begin{tikzpicture}[line width=1. pt, scale=2]
\vspace*{-2cm}
\draw[pyred] (0.4, 0) -- (0.6, 0);
\draw[pyred] (0.6, 0.) -- (0.6, 0.2) ;
\draw[pyred] (0.6, 0.) -- (0.6, -0.2) ;
\draw[pyred]  (0.6, 0.)  -- (0.8, 0);
\draw[fill=pyred] (0.6, 0) circle (.04cm);
\end{tikzpicture}}} 
\,\sim\,
T_{\tilde{\psi}}(\vec{k}_1,\vec{k}_2,\vec{k}_3,\vec{k}_4)\,, \quad \ldots \nonumber
\\
& \text{Internal loops:} 
\vcenter{\hbox{
\begin{tikzpicture}[line width=1. pt, scale=2]
\vspace*{-2cm}
\draw [pyred, xshift=0.6cm, domain=180:360] plot(\x:0.2);
\draw [pyred, xshift=0.6cm, domain=180:360] plot(-\x:0.2);
\end{tikzpicture}}} 
\,\sim\,
\int \dd^3 \vec{q}\,, \quad
\vcenter{\hbox{
\begin{tikzpicture}[line width=1. pt, scale=2]
\vspace*{-2cm}
\draw [pyred, xshift=0.6cm, domain=180:360] plot(\x:0.2);
\draw[pyred] (0.6, 0.2) -- (0.6, -0.2) ;
\draw [pyred, xshift=0.6cm, domain=180:360] plot(-\x:0.2);
\end{tikzpicture}}} 
\,\sim\,
\vcenter{\hbox{
\begin{tikzpicture}[line width=1. pt, scale=2]
\vspace*{-2cm}
\draw [pyred, xshift=0.6cm, domain=180:360] plot(\x:0.2);
\draw[pyred] (0.4, 0) -- (0.8, 0) ;
\draw [pyred, xshift=0.6cm, domain=180:360] plot(-\x:0.2);
\end{tikzpicture}}} 
\,\sim\,
\vcenter{\hbox{
\begin{tikzpicture}[line width=1. pt, scale=2]
\vspace*{-2cm}
\draw [pyred, xshift=0.6cm, domain=180:360] plot(\x:0.2);
\draw [pyred, xshift=0.6cm, domain=180:360] plot(-\x:0.2);
\draw [pyred, xshift=1cm, domain=180:360] plot(\x:0.2);
\draw [pyred, xshift=1cm, domain=180:360] plot(-\x:0.2);
\end{tikzpicture}}} 
\,\sim\,
\int \dd^3 \vec{q}_1 \dd^3 \vec{q}_2
\,, \quad\ldots \nonumber
\end{align}
We now show a few examples below, with all leading-loop order contributions and a few corrections at one higher order in loops, for connected diagrams only.
The bispectrum of $\psi$ is given by:
\begin{align}
\label{eq: diagram for B}
B_\psi(k_1,k_2,k_3) = & \,
\vcenter{\hbox{
\begin{tikzpicture}[line width=1. pt, scale=2]
\vspace*{-2cm}
\draw[pyblue] (0.1, 0) -- (0.4, 0);
\draw[pyred] (0.4, 0) -- (0.6, 0);
\draw[pyblue] (0.7, 0.173) -- (0.85, 0.433);
\draw[pyred] (0.6, 0.) -- (0.7, 0.173) ;
\draw[pyblue] (0.7, -0.173) -- (0.85,- 0.433);
\draw[pyred]  (0.6, 0.)  -- (0.7, -0.173);
\draw[fill=white] (0.4, 0) circle (.03cm);
\draw[fill=white] (0.7, 0.173) circle (.03cm);
\draw[fill=white] (0.7, -0.173) circle (.03cm);
\draw[fill=pyred] (0.6, 0) circle (.04cm);
\end{tikzpicture}}} 
\,+\,
\vcenter{\hbox{
\begin{tikzpicture}[line width=1. pt, scale=2]
\vspace*{-2cm}
\draw[pyblue] (0.3, 0) -- (0.6, 0);
\draw[pyblue] (0.7, 0.173) -- (0.85, 0.433);
\draw[pyred] (0.6, 0.) -- (0.7, 0.173) ;
\draw[pyblue] (0.7, -0.173) -- (0.85,- 0.433);
\draw[pyred]  (0.6, 0.)  -- (0.7, -0.173);
\draw[fill=white] (0.7, 0.173) circle (.03cm);
\draw[fill=white] (0.7, -0.173) circle (.03cm);
\draw[fill=black] (0.6, 0) circle (.03cm);
\end{tikzpicture}}} 
\,+\,
\vcenter{\hbox{
\begin{tikzpicture}[line width=1. pt, scale=2]
\vspace*{-2cm}
\draw[pyblue] (0.1, 0) -- (0.4, 0);
\draw [pyred, xshift=0.6cm, domain=180:360] plot(\x:0.2);
\draw [pyred, xshift=0.6cm, domain=180:360] plot(-\x:0.2);
\draw[pyblue] (0.7, 0.173) -- (0.85, 0.433);
\draw[pyblue] (0.7, -0.173) -- (0.85, -0.433);
\draw[fill=black] (0.4, 0) circle (.03cm);
\draw[fill=black] (0.7, 0.173) circle (.03cm);
\draw[fill=black] (0.7, -0.173) circle (.03cm);
\end{tikzpicture}}}
\,+\,
\vcenter{\hbox{
\begin{tikzpicture}[line width=1. pt, scale=2]
\vspace*{-2cm}
\draw[pyblue] (0.1, 0) -- (0.4, 0);
\draw [pyred, xshift=0.6cm, domain=180:360] plot(\x:0.2);
\draw [pyred, xshift=0.6cm, domain=180:360] plot(-\x:0.2);
\draw[pyred] (0.8, 0) -- (0.95, 0.260);
\draw[pyred] (0.8, 0) -- (0.95, -0.260);
\draw[pyblue] (0.95, 0.260) -- (1.20, 0.3);
\draw[pyblue] (0.95, -0.260) -- (1.20, -0.3);
\draw[fill=black] (0.4, 0) circle (.03cm);
\draw[fill=pyred] (0.8, 0) circle (.04cm);
\draw[fill=white] (0.95, 0.260) circle (.03cm);
\draw[fill=white] (0.95, -0.260) circle (.03cm);
\end{tikzpicture}}}
+ \ldots 
\\
\sim & \,\, g_2^3 \times B_{\tilde{\psi}}(k_1,k_2,k_3) + g_2^2 \times g_3 \left[ P_{\tilde{\psi}}(k_2) P_{\tilde{\psi}}(k_3)  \text{+ perm.}\right] \nonumber \\
&+ g_3^3  \times  \int \dd^3  \vec{q} \,\, \left[
P_{\tilde{\psi}}(q) 
P_{\tilde{\psi}}(|\vec{q}+\vec{k_1}|) 
P_{\tilde{\psi}}(|\vec{q}-\vec{k_3}|)
\text{+ perm.}\right]  \nonumber \\
&
+ g_2^2 \times g_3 \times  \int \dd^3  \vec{q} \,\, \left[T_{\tilde{\psi}}(\vec{q},\vec{k_1}-\vec{q},\vec{k}_2,\vec{k}_3)  \text{+ perm.}\right]  
+ \ldots\nonumber\,,
\end{align}
and where, at leading-order in the loop expansion, one has $P_{\tilde{\psi}}=P_\psi\,,\,B_{\tilde{\psi}}=B_\psi\,,\,T_{\tilde{\psi}}=T_\psi\,,\ldots$
However, to be consistent, the leading loop correction to the bispectrum of $\psi$ is given both by the loop terms starting from the second line above (plus other ones not written), and by the quantities appearing in the first line at the one-loop order, for which $P_{\tilde{\psi}}\neq P_\psi$ for example.
The trispectrum of $\psi$ is given by:
\begin{align}
\label{eq: diagram for T}
T_\psi(\vec{k}_1,\vec{k}_2,\vec{k}_3,\vec{k}_4) = & \,
\vcenter{\hbox{
\begin{tikzpicture}[line width=1. pt, scale=2]
\vspace*{-2cm}
\draw[pyred] (0.4, 0) -- (0.7, 0);
\draw[pyred] (0.7, 0.) -- (0.7, 0.3);
\draw[pyred] (0.7, 0.) -- (0.7, -0.3);
\draw[pyred]  (0.7, 0.)  -- (1, 0);
\draw[pyblue] (0.2, 0) -- (0.4, 0);
\draw[pyblue] (0.7, 0.3) -- (0.7, 0.5);
\draw[pyblue] (0.7, -0.3) -- (0.7, -0.5);
\draw[pyblue]  (1, 0.)  -- (1.2, 0);
\draw[fill=pyred] (0.7, 0) circle (.04cm);
\draw[fill=white]  (0.4, 0) circle (.03cm);
\draw[fill=white]  (0.7, 0.3) circle (.03cm);
\draw[fill=white]  (1, 0) circle (.03cm);
\draw[fill=white]  (0.7, -0.3) circle (.03cm);
\end{tikzpicture}}} 
\,+\,
\vcenter{\hbox{
\begin{tikzpicture}[line width=1. pt, scale=2]
\vspace*{-2cm}
\draw[pyred] (0.4, 0) -- (0.7, 0);
\draw[pyred] (0.4, 0.) -- (0.4, 0.3) ;
\draw[pyred] (0.7, 0.) -- (0.7, -0.3) ;
\draw[pyblue] (0.2, 0) -- (0.4, 0);
\draw[pyblue] (0.4, 0.3) -- (0.4, 0.5) ;
\draw[pyblue] (0.7, -0.3) -- (0.7, -0.5) ;
\draw[pyblue] (0.7, 0) -- (0.9, 0) ;
\draw[fill=black] (0.4, 0) circle (.03cm);
\draw[fill=black]  (0.7, 0) circle (.03cm);
\draw[fill=white]  (0.4, 0.3) circle (.03cm);
\draw[fill=white]  (0.7, -0.3) circle (.03cm);
\end{tikzpicture}}} 
\,+\,
\vcenter{\hbox{
\begin{tikzpicture}[line width=1. pt, scale=2]
\vspace*{-2cm}
\draw[pyred] (0.4, 0) -- (0.7, 0);
\draw[pyred] (0.4, 0.) -- (0.4, 0.3) ;
\draw[pyred] (0.4, 0.) -- (0.4, -0.3) ;
\draw[pyblue] (0.2, 0) -- (0.4, 0);
\draw[pyblue] (0.4, 0.3) -- (0.4, 0.5) ;
\draw[pyblue] (0.4, -0.3) -- (0.4, -0.5) ;
\draw[pyblue] (0.7, 0) -- (0.9, 0) ;
\draw[fill=black] (0.4, 0) circle (.03cm);
\draw[fill=white]  (0.7, 0) circle (.03cm);
\draw[fill=white]  (0.4, 0.3) circle (.03cm);
\draw[fill=white]  (0.4, -0.3) circle (.03cm);
\end{tikzpicture}}} 
\,+\,
\vcenter{\hbox{
\begin{tikzpicture}[line width=1. pt, scale=2]
\vspace*{-2cm}
\draw [pyred, xshift=0.6cm, domain=180:360] plot(\x:0.2);
\draw [pyred, xshift=0.6cm, domain=180:360] plot(-\x:0.2);
\draw[pyblue] (0.2, 0) -- (0.4, 0);
\draw[pyblue] (0.8, 0) -- (1.0, 0);
\draw[pyblue] (0.6, 0.2) -- (0.6, 0.4);
\draw[pyblue] (0.6, -0.2) -- (0.6, -0.4);
\draw[fill=black] (0.4, 0) circle (.03cm);
\draw[fill=black] (0.8, 0) circle (.03cm);
\draw[fill=black] (0.6, 0.2) circle (.03cm);
\draw[fill=black] (0.6, -0.2) circle (.03cm);
\end{tikzpicture}}}
+
\ldots 
\\
\sim & \,\, g_2^4 \times T_{\tilde{\psi}}(\vec{k}_1,\vec{k}_2,\vec{k}_3,\vec{k}_4) + g_2^2 g_3^2 \times \left[P_{\tilde{\psi}}(k_1)P_{\tilde{\psi}}(k_2)P_{\tilde{\psi}}(|\vec{k}_1+\vec{k}_3|) + \text{perm.} \right] \nonumber
\\ 
&+ g_2^3 g_4 \times \left[P_{\tilde{\psi}}(k_1)P_{\tilde{\psi}}(k_2)P_{\tilde{\psi}}(k_3) + \text{perm.} \right] \nonumber
\\ 
&+ g_2^4 \times \int \dd^3 \vec{q} \left[
P_{\tilde{\psi}}(q)
P_{\tilde{\psi}}(|\vec{q}+\vec{k}_1|)
P_{\tilde{\psi}}(|\vec{q}+\vec{k}_1+\vec{k}_2|)
P_{\tilde{\psi}}(|\vec{q}-\vec{k}_4|)
+ \text{perm.} \right] \nonumber
\\ 
& + \ldots \nonumber
\end{align}
The one-loop power spectrum of $\psi$ is given by:
\begin{align}
\label{eq: diagram for P-loop}
P^{1\mathrm{-loop}}_\psi(k) = & \, \Big[ P_\psi(k) - P^{\mathrm{tree}}_\psi(k)\Big]^{1\mathrm{-loop}} \\
= & \,
\Big[
\vcenter{\hbox{
\begin{tikzpicture}[line width=1. pt, scale=2]
\vspace*{-2cm}
\draw[pyblue] (0.2, 0) -- (0.4, 0);
\draw[pyred] (0.4, 0) -- (0.7, 0);
\draw[pyblue] (0.7, 0) -- (0.9, 0);
\draw[fill=white] (0.4, 0) circle (.03cm);
\draw[fill=white] (0.7, 0) circle (.03cm);
\end{tikzpicture}}}
-P^{\mathrm{tree}}_{\tilde{\psi}}(k)\Big]^{1\mathrm{-loop}} 
\,+\,
\vcenter{\hbox{
\begin{tikzpicture}[line width=1. pt, scale=2]
\vspace*{-2cm}
\draw[pyblue] (0.2, 0) -- (0.4, 0);
\draw [pyred, xshift=0.6cm, domain=180:360] plot(\x:0.2);
\draw [pyred, xshift=0.6cm, domain=180:360] plot(-\x:0.2);
\draw[pyblue] (0.8, 0) -- (1, 0);
\draw[fill=black] (0.4, 0) circle (.03cm);
\draw[fill=black] (0.8, 0) circle (.03cm);
\end{tikzpicture}}}
\,+\,
\vcenter{\hbox{
\begin{tikzpicture}[line width=1. pt, scale=2]
\vspace*{-2cm}
\draw [pyred, xshift=0.6cm, domain=180:360] plot(\x:0.2);
\draw [pyred, xshift=0.6cm, domain=180:360] plot(-\x:0.2);\draw[pyblue] (0.1, 0.26) -- (0.4, 0);
\draw[pyred] (0.2, -0.173) -- (0.4, 0);
\draw[pyblue] (0.1, -0.26) -- (0.2, -0.173);
\draw[fill=black] (0.4, 0) circle (.03cm);
\draw[fill=white] (0.2, -0.173) circle (.03cm);
\end{tikzpicture}}}
\,+\,
\vcenter{\hbox{
\begin{tikzpicture}[line width=1. pt, scale=2]
\vspace*{-2cm}
\draw[pyblue] (0.2, 0) -- (0.4, 0);
\draw [pyred, xshift=0.6cm, domain=180:360] plot(\x:0.2);
\draw [pyred, xshift=0.6cm, domain=180:360] plot(-\x:0.2);
\draw[pyred] (0.8, 0) -- (1.1, 0);
\draw[pyblue] (1.1, 0) -- (1.3, 0);
\draw[fill=black] (0.4, 0) circle (.03cm);
\draw[fill=red] (0.8, 0) circle (.04cm);
\draw[fill=white] (1.1, 0) circle (.03cm);
\end{tikzpicture}}} \nonumber
\\
= & \,\, P^{1\mathrm{-loop}}_{\tilde{\psi}}(k) 
+ 2  g_3^2  \times  \int \frac{\dd^3 \vec{q}}{(2\pi)^3} \,\,
P^{\mathrm{tree}}_{\tilde{\psi}}(q) 
P^{\mathrm{tree}}_{\tilde{\psi}}(|\vec{q}+\vec{k}|)
\nonumber \\
& \,+ 6 g_4 \times  \int \frac{\dd^3 \vec{q}}{(2\pi)^3} \,\, P^{\mathrm{tree}}_{\tilde{\psi}}(q) P^{\mathrm{tree}}_{\tilde{\psi}}(k)  + 2 g_3 \times \int \frac{\dd^3 \vec{q}}{(2\pi)^3} \,\, B^\mathrm{tree}_{\tilde{\psi}}(q,|\vec{k}+\vec{q}|,k) 
\,, \nonumber
\end{align}
where $\Big[\ldots\Big]^{1\mathrm{-loop}}$ means ``truncated at one loop''.
This expression for the one-loop power spectrum is exact, though we used that $g_2 = 1$ explicitly.
We see that the one-loop power spectrum of $\psi$ is given both by the one-loop power spectrum of $\tilde{\psi}$ as well as the one-loop corrections from the non-linear canonical transformation $\psi(\tilde{\psi})$.

\section{Complete expressions for the scalar and tensor Lagrangian}
\label{app:tensors_L_and_H}

\begin{align}
    \frac{\mathcal{L}_\gamma}{\Mp^2}=&-\frac{a e^{\zeta } \partial_jN \dot\gamma_{ij} \
\mathcal{N}_i}{2 N^2}
 + \frac{3 a e^{\zeta } \partial_j\zeta  \dot\gamma_{ij} \
\mathcal{N}_i}{2 N}
 - \frac{e^{-\zeta } \partial_k\mathcal{N}_j \partial_k\gamma_{ij} \mathcal{N}_i}{2 a N}
 - \frac{e^{-\zeta } \partial_k\mathcal{N}_j \partial_j\gamma_{ik} \mathcal{N}_i}{2 a N}
 \notag\\&+ \frac{e^{-\zeta } \partial_k\mathcal{N}_j \partial_i\gamma{}_j{}_k \mathcal{N}_i}{2 a N}
 + \frac{e^{-\zeta } \partial_k\zeta  \partial_k\gamma_{ij} \
\mathcal{N}_i \mathcal{N}_j}{a N}
 + a e^{\zeta } N \partial_i\zeta  \partial_j\zeta  \gamma_{ij}
 + 2 a e^{\zeta } N \partial_i\partial_j\zeta  \gamma_{ij}\notag\\&
 - \frac{2 a e^{\zeta } H \partial_j\mathcal{N}_i \gamma_{ij}}{N}
 - \frac{2 a e^{\zeta } \dot\zeta  \partial_j\mathcal{N}_i \gamma_{ij}}{N}
 - \frac{e^{-\zeta } \partial_k\mathcal{N}_i \
\partial_k\mathcal{N}_j \gamma_{ij}}{4 a N}
 - \frac{e^{-\zeta } \partial_k\mathcal{N}_i \
\partial_j\mathcal{N}_k \gamma_{ij}}{2 a N}
\notag\\& - \frac{e^{-\zeta } \partial_i\mathcal{N}_k
\partial_j\mathcal{N}_k \gamma_{ij}}{4 a N}
 + \frac{e^{-\zeta } \partial_j\mathcal{N}_i \
\partial_k\mathcal{N}_k \gamma_{ij}}{a N}
 - \frac{2 a e^{\zeta } H \partial_j\zeta  \mathcal{N}_i \gamma_{ij}}{N}
 - \frac{2 a e^{\zeta } \dot\zeta  \partial_j\zeta  \mathcal{N}_i \
\gamma_{ij}}{N}
\notag\\& + \frac{e^{-\zeta } \partial_k\zeta  \partial_k\mathcal{N}_j \
\mathcal{N}_i \gamma_{ij}}{a N}
 + \frac{e^{-\zeta } \partial_k\zeta  \partial_j\mathcal{N}_k \
\mathcal{N}_i \gamma_{ij}}{a N}
 + \frac{e^{-\zeta } \partial_j\zeta  \partial_k\mathcal{N}_i \
\mathcal{N}_i \gamma_j{}_k}{a N}
 \notag\\&+ \frac{e^{-\zeta } \partial_j\zeta  \partial_i\mathcal{N}_k \
\mathcal{N}_i \gamma_j{}_k}{a N},
\end{align}
\begin{align}
    \frac{\mathcal{L}_{\gamma\gamma}}{\Mp^2}=&\frac{a^3 e^{3 \zeta } \dot\gamma{}_i{}_j {}^2}{8 N}
 - \frac{1}{8} a e^{\zeta } N \left(\partial_k\gamma{}_i{}_j \right)^2
 - \frac{1}{4} a e^{\zeta } N \partial_k\gamma{}_i{}_j \
\partial_j\gamma{}_i{}_k
 + \frac{1}{4} a e^{\zeta } N \partial_k\gamma{}_i{}_j \
\partial_i\gamma{}_j{}_k
 \notag\\&+ \frac{a e^{\zeta } \partial_k\gamma{}_i{}_j \dot\gamma{}_j{}_k \
\mathcal{N}_i}{2 N}
 - \frac{a e^{\zeta } \dot\gamma{}_j{}_k \partial_i\gamma{}_j{}_k \
\mathcal{N}_i}{4 N}
 + \frac{e^{-\zeta } \partial_l\gamma{}_i{}_k \partial_l\gamma
{}_j{}_k \mathcal{N}_i \mathcal{N}_j}{4 a N}
\notag\\& + \frac{e^{-\zeta } \partial_l\gamma{}_i{}_k \partial_k\gamma
{}_j{}_l \mathcal{N}_i \mathcal{N}_j}{4 a N}
 - \frac{e^{-\zeta } \partial_l\gamma{}_i{}_k \partial_j\gamma
{}_k{}_l \mathcal{N}_i \mathcal{N}_j}{2 a N}
 + \frac{e^{-\zeta } \partial_i\gamma{}_k{}_l \partial_j\gamma
{}_k{}_l \mathcal{N}_i \mathcal{N}_j}{8 a N}
\notag\\& - a e^{\zeta } N \partial_k\zeta  \partial_j\gamma{}_i{}_k \gamma
{}_i{}_j
 + \frac{a e^{\zeta } \partial_k\mathcal{N}_i \dot\gamma{}_j{}_k \
\gamma{}_i{}_j}{4 N}
 + \frac{a e^{\zeta } \partial_i\mathcal{N}_k \dot\gamma{}_j{}_k \
\gamma{}_i{}_j}{4 N}
\notag\\& - \frac{a e^{\zeta } \partial_k\zeta  \dot\gamma{}_j{}_k \
\mathcal{N}_i \gamma{}_i{}_j}{2 N}
 + \frac{e^{-\zeta } \partial_l\mathcal{N}_k \partial_l\gamma
{}_j{}_k \mathcal{N}_i \gamma{}_i{}_j}{4 a N}
 + \frac{e^{-\zeta } \partial_l\mathcal{N}_k \partial_k\gamma
{}_j{}_l \mathcal{N}_i \gamma{}_i{}_j}{4 a N}
\notag\\& - \frac{e^{-\zeta } \partial_l\mathcal{N}_k \partial_j\gamma
{}_k{}_l \mathcal{N}_i \gamma{}_i{}_j}{2 a N}
 + \frac{a e^{\zeta } H \partial_k\gamma{}_i{}_j \mathcal{N}_i \
\gamma{}_j{}_k}{N}
 + \frac{a e^{\zeta } \dot\zeta  \partial_k\gamma{}_i{}_j \
\mathcal{N}_i \gamma{}_j{}_k}{N}
\notag\\& - \frac{e^{-\zeta } \partial_l\mathcal{N}_l \partial_k\gamma
{}_i{}_j \mathcal{N}_i \gamma{}_j{}_k}{2 a N}
 - \frac{a e^{\zeta } \partial_j\zeta  \dot\gamma{}_i{}_k \
\mathcal{N}_i \gamma{}_j{}_k}{2 N}
 + \frac{e^{-\zeta } \partial_l\mathcal{N}_j \partial_l\gamma
{}_i{}_k \mathcal{N}_i \gamma{}_j{}_k}{4 a N}
 \notag\\&+ \frac{e^{-\zeta } \partial_j\mathcal{N}_l \partial_l\gamma
{}_i{}_k \mathcal{N}_i \gamma{}_j{}_k}{4 a N}
 + \frac{e^{-\zeta } \partial_l\mathcal{N}_j \partial_k\gamma
{}_i{}_l \mathcal{N}_i \gamma{}_j{}_k}{2 a N}
 + \frac{e^{-\zeta } \partial_j\mathcal{N}_l \partial_k\gamma
{}_i{}_l \mathcal{N}_i \gamma{}_j{}_k}{2 a N}
\notag\\& - \frac{e^{-\zeta } \partial_l\mathcal{N}_j \partial_i\gamma
{}_k{}_l \mathcal{N}_i \gamma{}_j{}_k}{4 a N}
 - \frac{e^{-\zeta } \partial_j\mathcal{N}_l \partial_i\gamma
{}_k{}_l \mathcal{N}_i \gamma{}_j{}_k}{4 a N}
 - \frac{1}{2} a e^{\zeta } N \partial_i\zeta  \partial_k\zeta  \
\gamma{}_i{}_j \gamma{}_j{}_k
 \notag\\&- a e^{\zeta } N \partial_i\partial_k\zeta  \gamma{}_i{}_j \gamma
{}_j{}_k
 + \frac{a e^{\zeta } H \partial_k\mathcal{N}_i \gamma{}_i{}_j \
\gamma{}_j{}_k}{N}
 + \frac{a e^{\zeta } \dot\zeta  \partial_k\mathcal{N}_i \gamma
{}_i{}_j \gamma{}_j{}_k}{N}
 \notag\\&+ \frac{e^{-\zeta } \partial_l\mathcal{N}_i \
\partial_l\mathcal{N}_k \gamma{}_i{}_j \gamma{}_j{}_k}{8 a N}
 + \frac{e^{-\zeta } \partial_l\mathcal{N}_i \
\partial_k\mathcal{N}_l \gamma{}_i{}_j \gamma{}_j{}_k}{4 a N}
 + \frac{e^{-\zeta } \partial_i\mathcal{N}_l \
\partial_k\mathcal{N}_l \gamma{}_i{}_j \gamma{}_j{}_k}{8 a N}\notag\\&
 - \frac{e^{-\zeta } \partial_k\mathcal{N}_i \
\partial_l\mathcal{N}_l \gamma{}_i{}_j \gamma{}_j{}_k}{2 a N}
 + \frac{a e^{\zeta } H \partial_k\zeta  \mathcal{N}_i \gamma
{}_i{}_j \gamma{}_j{}_k}{N}
 + \frac{e^{-\zeta } \partial_k\mathcal{N}_i \
\partial_l\mathcal{N}_j \gamma{}_i{}_j \gamma{}_k{}_l}{4 a N}\notag\\&
 - \frac{e^{-\zeta } \partial_j\mathcal{N}_i \
\partial_l\mathcal{N}_k \gamma{}_i{}_j \gamma{}_k{}_l}{2 a N}
 + \frac{e^{-\zeta } \partial_k\mathcal{N}_i \
\partial_j\mathcal{N}_l \gamma{}_i{}_j \gamma{}_k{}_l}{4 a N},
\end{align}
\begin{align}
    \frac{\mathcal{L}_{\gamma\gamma\gamma}}{\Mp^2}=&\frac{1}{12} a e^{\zeta } N \partial_l\gamma{}_a{}_k \
\partial_k\gamma{}_j{}_l \gamma{}_i{}_j
 - \frac{1}{12} a e^{\zeta } N \partial_l\gamma{}_i{}_k \
\partial_j\gamma{}_k{}_l \gamma{}_i{}_j
 + \frac{1}{8} a e^{\zeta } N \partial_i\gamma{}_k{}_l \
\partial_j\gamma{}_k{}_l \gamma{}_i{}_j
 \notag\\&- \frac{a e^{\zeta } \partial_l\gamma{}_j{}_k \dot\gamma{}_k{}_l \
\mathcal{N}_i \gamma{}_i{}_j}{4 N}
 + \frac{a e^{\zeta } \dot\gamma{}_k{}_l \partial_j\gamma{}_k{}_l \
\mathcal{N}_i \gamma{}_i{}_j}{4 N}
 - \frac{a e^{\zeta } \partial_j\gamma{}_i{}_l \dot\gamma{}_k{}_l \
\mathcal{N}_i \gamma{}_j{}_k}{4 N}
 \notag\\&+ \frac{1}{3} a e^{\zeta } N \partial_l\zeta  \partial_k\gamma
{}_i{}_l \gamma{}_i{}_j \gamma{}_j{}_k
 - \frac{a e^{\zeta } \partial_l\mathcal{N}_i \dot\gamma{}_k{}_l \
\gamma{}_i{}_j \gamma{}_j{}_k}{12 N}
 - \frac{a e^{\zeta } \partial_i\mathcal{N}_l \dot\gamma{}_k{}_l \
\gamma{}_i{}_j \gamma{}_j{}_k}{12 N}
 \notag\\&+ \frac{1}{12} a e^{\zeta } N \partial_j\partial_l\gamma{}_i{}_k \
\gamma{}_i{}_j \gamma{}_k{}_l
 + \frac{1}{6} a e^{\zeta } N \partial_k\zeta  \partial_j\gamma
{}_i{}_l \gamma{}_i{}_j \gamma{}_k{}_l
 + \frac{1}{6} a e^{\zeta } N \partial_i\zeta  \partial_l\gamma
{}_j{}_k \gamma{}_i{}_j \gamma{}_k{}_l
\notag\\& - \frac{a e^{\zeta } \partial_k\mathcal{N}_i \dot\gamma{}_j{}_l \
\gamma{}_i{}_j \gamma{}_k{}_l}{12 N}
 - \frac{a e^{\zeta } H \partial_l\gamma{}_j{}_k \mathcal{N}_i \
\gamma{}_i{}_j \gamma{}_k{}_l}{3 N}
 - \frac{a e^{\zeta } H \partial_l\gamma{}_i{}_j \mathcal{N}_i \
\gamma{}_j{}_k \gamma{}_k{}_l}{3 N}
 \notag\\&- \frac{a e^{\zeta } H \partial_l\mathcal{N}_i \gamma{}_i{}_j \
\gamma{}_j{}_k \gamma{}_k{}_l}{6 N}
 - \frac{a e^{\zeta } H \partial_i\mathcal{N}_l \gamma{}_i{}_j \
\gamma{}_j{}_k \gamma{}_k{}_l}{6 N}
 + \frac{1}{3} a e^{\zeta } N \partial_i\partial_k\zeta  \gamma
{}_i{}_j \gamma{}_j{}_l \gamma{}_k{}_l,
\end{align}
\begin{align}
    \frac{\mathcal{L}_{\gamma\gamma\gamma\gamma}}{\Mp^2}=&\frac{a^3 e^{3 \zeta } \dot\gamma{}_i{}_l \dot\gamma{}_k{}_l \gamma
{}_i{}_j \gamma{}_j{}_k}{48 N}
 - \frac{1}{48} a e^{\zeta } N \partial_m\gamma{}_i{}_l \
\partial_m\gamma{}_k{}_l \gamma{}_i{}_j \gamma{}_j{}_k
 + \frac{1}{24} a e^{\zeta } N \partial_m\gamma{}_i{}_l \
\partial_k\gamma{}_l{}_m \gamma{}_i{}_j \gamma{}_j{}_k
 \notag\\&- \frac{1}{16} a e^{\zeta } N \partial_i\gamma{}_l{}_m \
\partial_k\gamma{}_l{}_m \gamma{}_i{}_j \gamma{}_j{}_k
 - \frac{a^3 e^{3 \zeta } \dot\gamma{}_i{}_k \dot\gamma{}_j{}_l \
\gamma{}_i{}_j \gamma{}_k{}_l}{48 N}
 + \frac{1}{48} a e^{\zeta } N \partial_m\gamma{}_i{}_k \
\partial_m\gamma{}_j{}_l \gamma{}_i{}_j \gamma{}_k{}_l
  \notag\\&+ \frac{1}{8} a e^{\zeta } N \partial_m\gamma{}_i{}_k \
\partial_l\gamma{}_j{}_m \gamma{}_i{}_j \gamma{}_k{}_l
 - \frac{1}{48} a e^{\zeta } N \partial_j\gamma{}_i{}_m \
\partial_l\gamma{}_k{}_m \gamma{}_i{}_j \gamma{}_k{}_l
\notag\\& - \frac{5}{24} a e^{\zeta } N \partial_m\gamma{}_i{}_k \
\partial_j\gamma{}_l{}_m \gamma{}_i{}_j \gamma{}_k{}_l
  + \frac{1}{24} a e^{\zeta } N \partial_k\gamma{}_i{}_m \
\partial_j\gamma{}_l{}_m \gamma{}_i{}_j \gamma{}_k{}_l
 \notag\\&+ \frac{1}{48} a e^{\zeta } N \partial_l\partial_m\gamma{}_i{}_k \
\gamma{}_i{}_j \gamma{}_j{}_m \gamma{}_k{}_l
 + \frac{1}{3} a e^{\zeta } N \partial_k\partial_m\zeta  \gamma
{}_i{}_j \gamma{}_i{}_l \gamma{}_j{}_m \gamma{}_k{}_l
 \notag\\& - \frac{1}{16} a e^{\zeta } N \partial_j\partial_m\gamma{}_i{}_k \
\gamma{}_i{}_j \gamma{}_k{}_l \gamma{}_l{}_m
 - \frac{5}{12} a e^{\zeta } N \partial_i\partial_k\zeta  \gamma
{}_i{}_j \gamma{}_j{}_m \gamma{}_k{}_l \gamma{}_l{}_m.
\end{align}

\bibliographystyle{JHEP}
\bibliography{biblio}

\end{document}